\documentclass[]{beilstein}

\usepackage{xspace}
\usepackage[dvipsnames,table]{xcolor}
\usepackage{graphicx}% Include figure files
\usepackage{dcolumn}% Align table columns on decimal point
\usepackage{bm}% bold math
\usepackage{subfig} % for \subfloat
\usepackage{hyperref}
\usepackage{xr-hyper}

\hypersetup{
    colorlinks=true,
    citecolor=PineGreen,
    urlcolor=RoyalBlue,
    linkcolor=PineGreen,
    pdftitle={Microscopic contributions to the deviation from Amontons friction law}
    }

\usepackage{pdfpages}

\begin{document}
\nolinenumbers
%%%%%%%%%%%%%%%%%%%%%%%%%%%%%%%%%%%%%%%%%%%%%%%%%%%%%%%%%%%%%%%%%%%%%
\title{Microscopic contributions to the deviation from Amontons friction law}
%%%\sititle{} % when using manuscript=suppinfo
\author*{Suresh Ravisankar}{ravissur@fel.cvut.cz}
\affiliation{Department of Control Engineering, Faculty of Electrical Engineering, Czech Technical University in Prague, Technicka 2, 16627 Prague 6, Czech Republic}
\author*[1]{Ravikant Kumar}{kumarra2@fel.cvut.cz}
\author[1]{Antonio Cammarata}
\author{Thilo Glatzel}
\affiliation{Department of Physics, University of Basel, 4056 Basel, Switzerland}
\author[1]{Tomas Polcar}
\maketitle
%%%%%%%%%%%%%%%%%%%%%%%%%%%%%%%%%%%%%%%%%%%%%%%%%%%%%%%%%%%%%%%%%%%%%
\begin{abstract} 
We investigate the nanoscale friction behaviour of MX$_2$ monolayers (M = Mo, W; X = S, Se) on Au(111) and Ag(111) substrates with a silicon tip using classical molecular dynamics simulations with machine-learning-based force fields. This approach enables an accurate description of tip-surface interactions and friction mechanisms at the atomic scale. We observe a pronounced nonmonotonic dependence of the friction force on the applied normal load, indicating a breakdown of Amontons's law at the nanoscale. Analysis of lateral force signals and their spatial Fourier transforms reveals the coexistence of multiple sliding modes, including longitudinal sliding, lateral slip, and zig-zag motions. We show that the overall friction response is governed by the relative contributions of these motions. While the qualitative features of friction are largely substrate-independent, both the magnitude of friction and the balance between sliding modes depend sensitively on the substrate-monolayer combination. In particular, Au/MoSe$_2$/Si exhibits significantly reduced friction due to suppression of lateral slip motion. Our results indicate that the method is broadly applicable for probing nanoscale friction in related heterostructures.
\end{abstract}
%%%%%%%%%%%%%%%%%%%%%%%%%%%%%%%%%%%%%%%%%%%%%%%%%%%%%%%%%%%%%%%%%%%%%
%%%%%%%%%%%%%%%%%%%%%%%%%%%%%%%%%%%%%%%%%%%%%%%%%%%%%%%%%%%%%%%%%%%%%
\section{Introduction}
\label{intro}
Friction arises whenever two contacting surfaces move relative to each other, leading to energy dissipation and material wear in mechanical systems.
Unfortunately, liquid lubricants are ineffective in reducing friction under extreme conditions, such as high normal load, high temperature and ultra-high vacuum \cite{Liquid}.
In recent years, two-dimensional layered materials such as graphene, boron nitride and transition metal dichalcogenides (TMDs) have emerged as promising solid lubricants in achieving superlubricity on rough and worn surfaces in engineering applications \cite{Nano_to_Meso, 2D_Tribology_review, 2D_Nanoscale_Review, 2D_Friction_Wear_Review, Sliding_2D_Review, Review_general_framework}. 
Among them, TMDs stand out for their layered structure and weak van der Waals bonding between the layers, \cite{Superlubricity_MoS2, MoS2_Review}
which facilitates interlayer shear, resulting in exceptionally low frictional properties in vacuum.
However, the dynamic behaviour of the TMD layers with atomic detail during sliding is subject of continuous study, due to the complex phenomena occurring concurrently in tribological conditions.
The most used and simple theoretical model of nanoscale friction is the Prandtl-Tomlinson model \cite{PT_model_1, PT_model_2}.
However, atomistic observation of lateral slips, dynamic deformation and thermal fluctuations is not well captured by this model;
indeed, advanced analysis based on the Fourier Transform of the friction force helps to quantify these lateral movements properties \cite{Ftrans, Ftrans_2}.

The interaction between TMD layers and metallic substrates, together with the influence of a scanning probe tip, determines the overall nanoscale friction response.
The atomically flat nature of metallic substrates helps to suppress out-of-plane fluctuations of the TMD layer, resulting in a more stable sliding interface and more reliable investigation of frictional mechanisms \cite{out_of_plane_floppiness}.
First-principles based calculations have provided valuable insights into the electronic and structural properties of these interfaces \cite{DFT_Friction_1, DFT_Friction_2};
however, their application to large-scale friction simulations is limited due to high computational cost.
Compared with first-principles calculations, classical molecular dynamics (MD) is a technique capable of simulating dynamic and kinetic problems such as sliding processes and frictional behavior at larger scale and reduced computational load \cite{TMDs_MD_1, TMD_MD_2, TMD_MD_3_REBO}.
MD simulations are often employed using empirical interatomic potentials, such as ReaxFF \cite{TMD_MD_5_ReaxFF, TMD_MD_6_ReaxFF} and reactive empirical bond-order (REBO) potentials \cite{TMD_MD_3_REBO, REBO_MD, TMD_MD_4_REBO};
nevertheless, the parametrisation of such potentials is challenging and time demanding.
Fortunately, with the development of machine learning (ML) in material science, this challenge has been well addressed \cite{Material_informatics}.
Various machine learning force fields (MLFFs) have been developed, including the Gaussian approximation potential (GAP) \cite{GAP}, TensorMol \cite{TensorMol}, the neuroevolution machine learning potential (NEP) \cite{NEP}, and the deep neural network potential (NNP) \cite{NNP}. 
Notably, neural networks possess the unique ability to theoretically represent unknown multidimensional real-valued functions with arbitrary precision by selecting appropriate network models.
Recent research demonstrates that the NNP model matches the accuracy of quantum mechanics for both finite and extended systems, showcasing its size extensive nature \cite{NNP_2, NNP_3}.
At present, MLFFs have been successfully applied in diverse materials, including TiO$_{2}$ \cite{DeepMD_TiO2}, Ga$_2$O$_3$ \cite{DeepMD_Ga2O3} and TMDs \cite{RMSE_Compare_1, RMSE_Compare_2, NNP_2}, making the NNP with the broadest range of applications and the most promising prospects in current research.

In this work, we study the friction response of MX$_2$ monolayers (M = Mo, W and X = S, Se) deposited on Au(111) and Ag(111) substrates by means of classical MD simulations at different loads and velocities of a scanning tip, employing MLFFs developed for this purpose.
Our Fourier analysis of the friction signal allows us to quantify and characterise the nanoscale contributions to the friction response.
In particular, we examine the microscopic phenomena responsible for the breakdown of Amontons's law as reflected in the mean friction force;
this results in a nonmonotonic friction--load relationship and contributes to significant uncertainty in the extracted coefficient of friction from a linear fit.
Significant contribution to such deviation originates from the tip's lateral motion, which also reduces the average friction force, except in the Au/MoSe$_2$/Si system, as the lateral motion is missing from the force profile.

The article is organized as follows:
we describe the computational methods in section Computational Details, while the results of the friction simulations of TMDs on different substrates are presented in section Results and Discussion.
%%%%%%%%%%%%%%%%%%%%%%%%%%%%%%%%%%%%%%%%%%%%%%%%%%%%%%%%%%%%%%%%%%%%%
\section{Computational Details}
\label{sec:comp_det}
\begin{figure}
  \centering
  \includegraphics[width=0.4\textwidth]{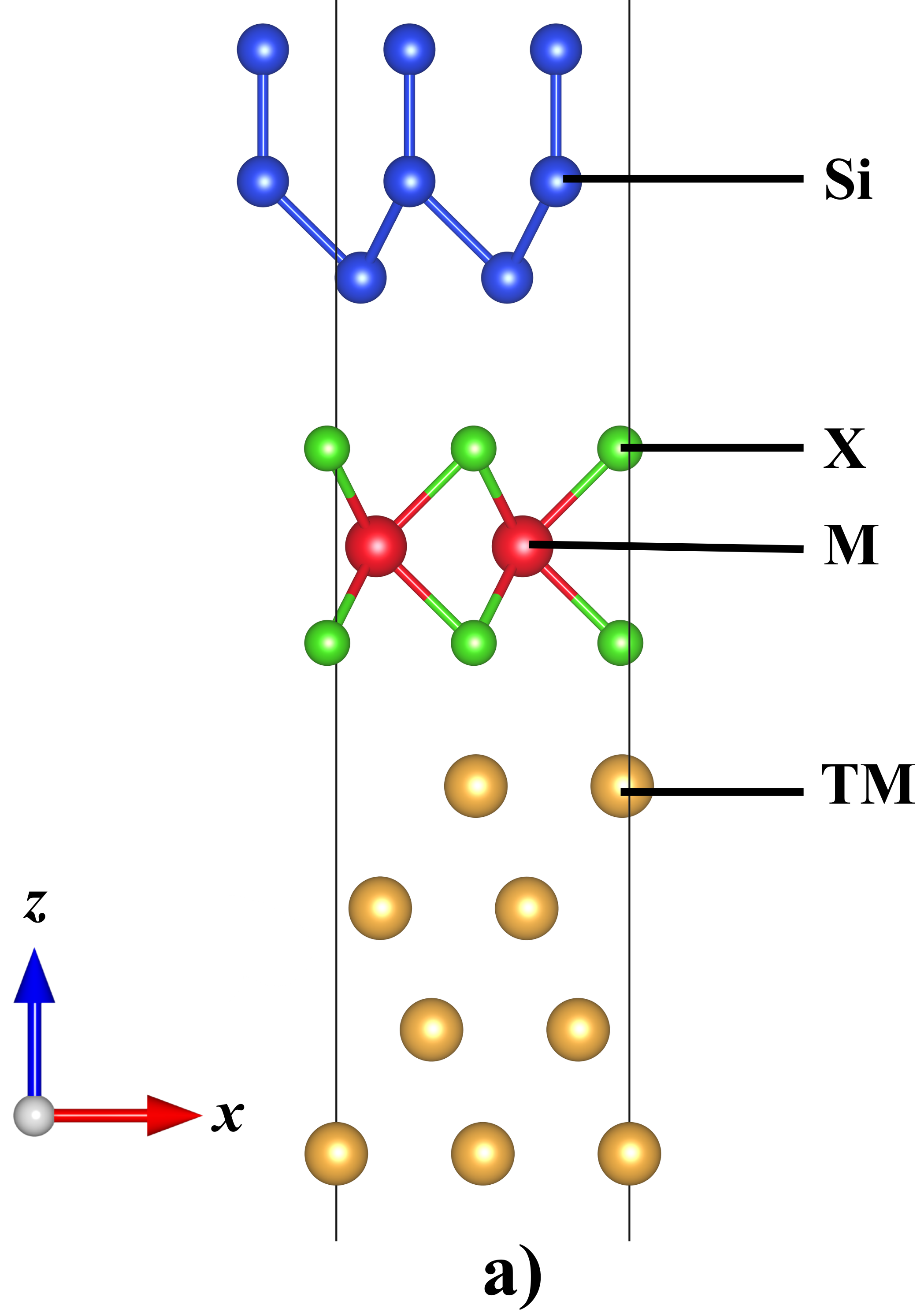}\hspace{1cm}
  \includegraphics[width=0.5\textwidth]{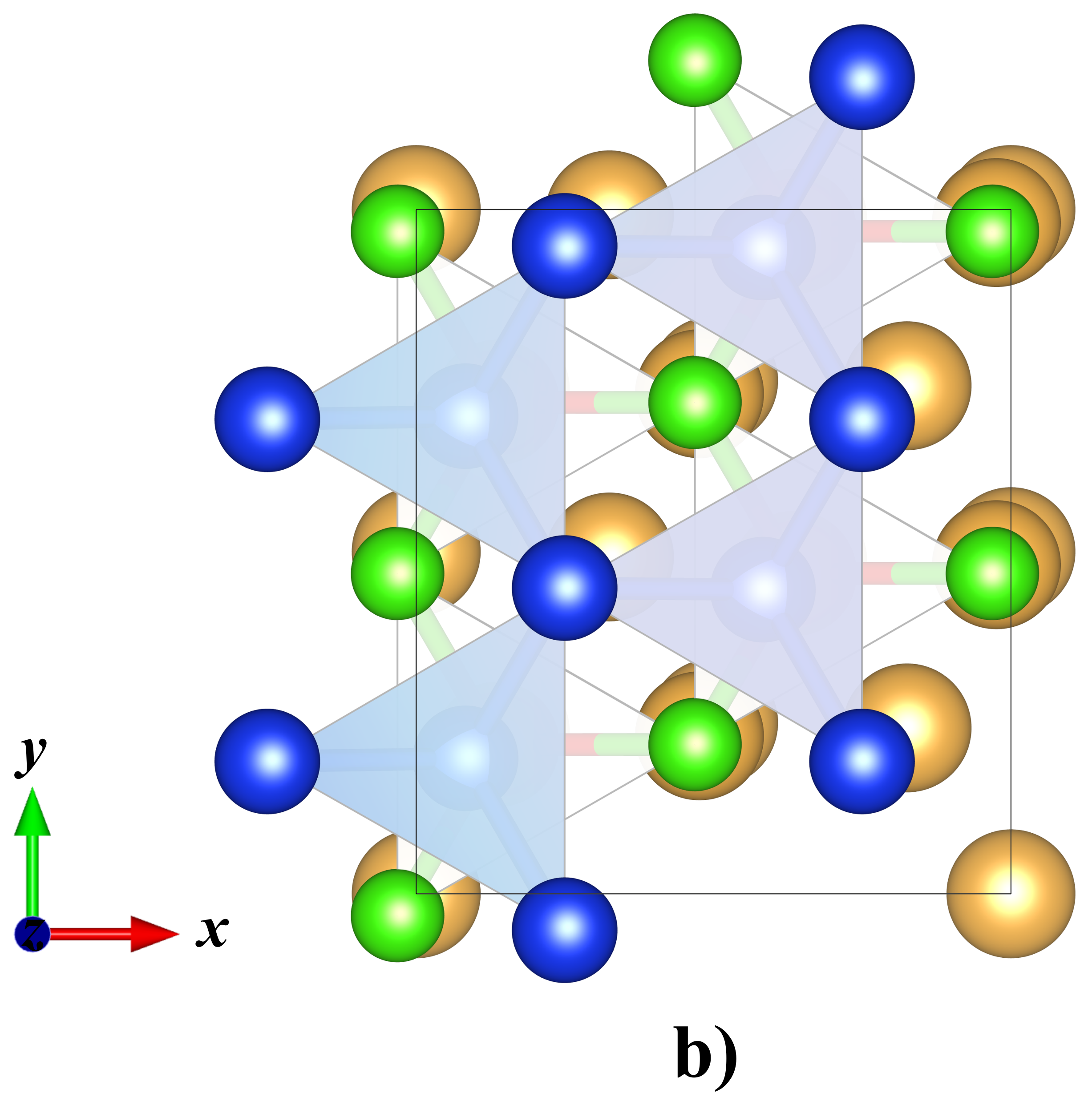}
  \caption{Initial setup to generate strucuture model of TM/MX$_2$/Si. a) Front view and b) top view. The thin black lines indicating the unit cell.}
  \label{fig:model_geometry}
\end{figure}

Density functional theory (DFT) based calculations for the structural optimisation of TM/MX$_2$/Si systems (TM = Au, Ag; M = Mo, W and X = S, Se) are performed using the Vienna ab-initio simulation package (\textsc{vasp}) \cite{vasp}.
The interactions between valence electrons and ionic cores are described using the projector augmented-wave (PAW) method \cite{PAW}. 
The generalized gradient approximation (GGA) in the Perdew--Burke--Ernzerhof (PBE) form is employed to treat the exchange-correlation energy \cite{PBE}.
The electronic self-consistent field calculations are converged to within 10$^{-8}$ eV, and the atomic positions are relaxed until the residual forces in each atom are less than 10$^{-2}$ eV/\r{A}. 
A $7\times7\times1$ Monkhorst--Pack $\bm{k}$-point mesh \cite{MP_kpoints} is used for Brillouin zone sampling, and the plane-wave basis set is truncated at a kinetic energy cutoff of 500 eV.
To accurately capture long-range dispersion interactions, the Grimme DFT-D3 correction \cite{DFT_D3} is incorporated to account for van der Waals forces between the layers.
The ML potential is trained using the DeepMD-kit package with the DeepPot–SE (Deep Potential-Smooth Edition) model \cite{DeepMD,DeepSE}.
The model includes an embedding network and a fitting network. 
The sizes of these networks are set to (25, 50, 100) and (240, 240, 240), respectively.
The cutoff radius is set to 8.0 \r{A} and the descriptors decay smoothly from 0.5 to 8.0 \r{A}.
The initial learning rate is set to 0.001 at the beginning of the training process to achieve a final value of 10$^{-8}$. 
The total number of training batches are 10$^{5}$ for the training in the initial iterations and the active learning process.
To automate the active learning process, the DP-GEN workflow is used \cite{DP-GEN}.
All molecular dynamics simulations are performed using the Large-scale Atomic/Molecular Massively Parallel Simulator (\textsc{lammps}) \cite{LAMMPS} package through the trained ML potentials.
The optimised structure of each system is equilibrated at 300 K for 6 ns with a time step of 1 fs by means of a Nos\'e-Hoover chain with three thermostats.
To construct TM/MX$_2$/Si interfaces, the TM substrate is cleaved from bulk Au and Ag as reported in Ref. \citenum{Ag_Au_Lattice} and subsequently optimized within the DFT framework.
Next, a TMD monolayer is placed on top of the TM substrate in such a way to minimize the lattice mismatch of the resulting heterostructure.
A 30 \r{A} vacuum slab orthogonal to the $\bm{c}$-direction is added on top of the geometry to prevent any interaction of the periodic replicas along the same direction;
the geometry of the heterostructure is then relaxed.
During the optimization, the bottom layer of the TM substrate is fixed to better mimic the experimental conditions;
finally, a Si (111) tip is placed above the heterostructure and the whole system is relaxed.
In this last geometric optimization, the bottom TM layer kept fixed, while the Si tip is constrained to relax only along the $\bm{c}$-direction.
An example of model geometry of the full system is reported in \autoref{fig:model_geometry}.
%%%%%%%%%%%%%%%%%%%%%%%%%%%%%%%%%%%%%%%%%%%%%%%%%%%%%%%%%%%%%%%%%%%%%%%%%%%%%%%%%%%%%%%%%%%%%%%%%%%%%%%%%%%%%%%%%%%%%%%%%%%%%%%%%%%%%%%%%%%%%%%%%%%%%%%%%%%%%%%%%%%%%%%%%%%
\subsection{Training of force field}
\begin{figure}
  \centering
  \includegraphics[width=0.99\textwidth]{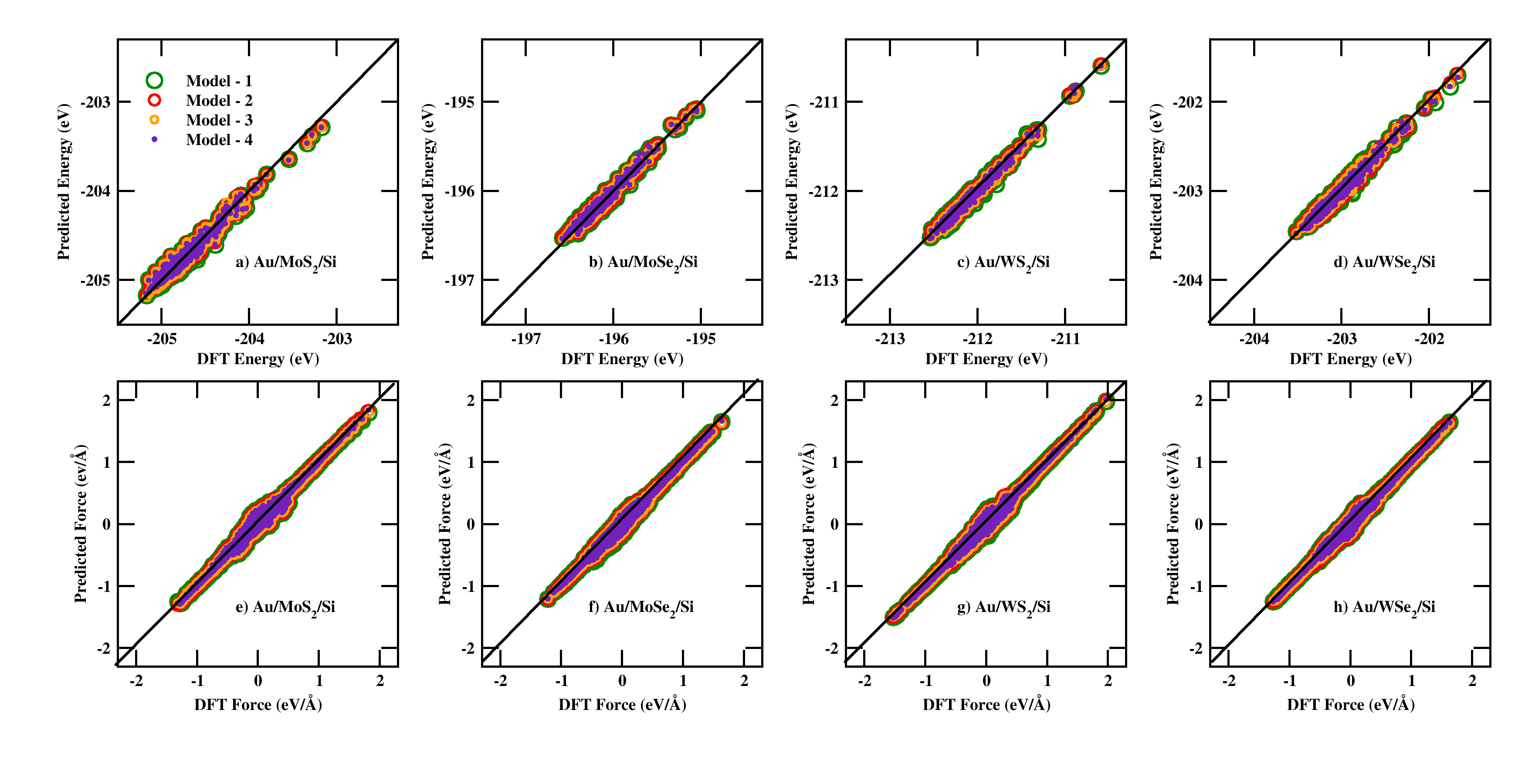}
  \caption{Energy (a–d) and force (e–h) comparisons between predicted and DFT results for Au/MX$_2$/Si systems.}
  \label{fig:MLFF_Au}
\end{figure}
\begin{figure}
  \centering
  \textbf{\includegraphics[width=0.99\textwidth]{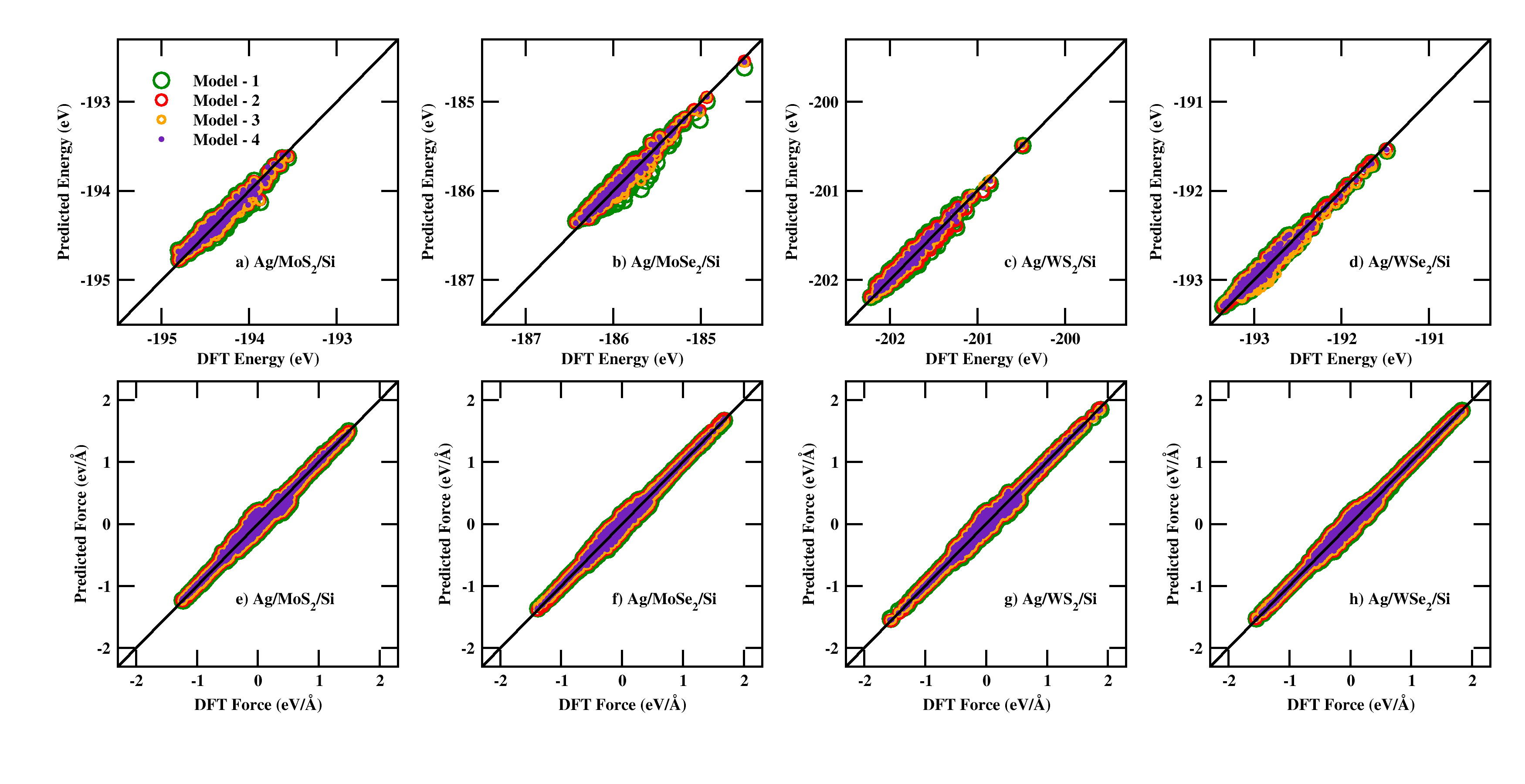}}
  \caption{Energy (a–d) and force (e–h) comparisons between predicted and DFT results for Ag/MX$_2$/Si systems.}
  \label{fig:MLFF_Ag}
\end{figure}
After optimising the full Au/MX$_2$/Si configuration, we proceed to construct the machine-learning force field (MLFF) for the system.
Once we obtain the optimised structure, the training data for the MLFF are generated by perturbing the atom positions and cell size by at least 0.01 \r{A} and 0.03 \r{A}, respectively.
A total of $\sim$400 configurations are generated to train the MLFF of all the systems and 80 configurations are taken for the validation during the active learning process.
For all configurations, energy and forces are calculated with the selected DFT setup and a single point calculation, that is at fixed atom geometry and lattice parameters.
To effectively improve the MLFF, additional data are generated using an active learning approach as proposed by Zhang et al. \cite{DP-GEN}.
At each iteration, four models are trained using the same input setup but starting with different random seeds. 
To explore new configurations to enrich the dataset, deep potential molecular dynamics (DPMD) simulations are performed with the initial models as reference, collecting the system's atomic configurations every 10 fs along the trajectories.
DPMD simulations are performed for a duration of 500 ps.
During the DPMD simulations, the force acting on each atoms in all configurations are evaluated by all four models. 
The maximum standard deviation of the atomic forces ${\sigma}_{max}$ is evaluated as a criterion for the convergence of neural network training
\begin{equation}\label{eq:S.D}
\sigma_{\max }=\max{_i} \sqrt{\left\langle\left\|\bm{f}_i-\left\langle \bm{f}_i\right\rangle\right\|^2\right\rangle}
\end{equation}
where $\bm{f}_i$ is the force acting on $i^{th}$ atom and ${ \left\langle \bm{f}_i\right\rangle }$ is the average value taken from four models.
A structure is considered a candidate if the maximum standard deviation falls within the range of 0.05 $<$ ${\sigma}_{max}$ $<$ 0.15;
while structures with ${\sigma}_{max}<0.05$ are considered correct, those with ${\sigma}_{max}$ $>$ 0.15 can be highly distorted configurations and are discarded.
The selected structures are then subjected to DFT calculations to obtain the corresponding energies and forces, and included in the dataset.
The training with four new models is repeated until the root mean square errors (RMSE) value is reduced.
The RMSE values of energy and forces for all the systems are presented in \autoref{tab:RMSE_comparision}.
All the final models have RMSE of energy and forces ranging from 0.92 to 1.16 meV/atom and 0.019 to 0.021 eV/\r{A}, respectively.
This range of RMSE values indicates that the MLFFs model is properly trained with values for the errors typical of surface and interface systems \cite{RMSE_Compare_1, RMSE_Compare_2}.  
To validate our MLFFs, we optimise the geometry of the heterostructures by using the trained MLFFs;
we then compute the RDF scalar products with the DFT optimised structure using the \textsc{maise} package \cite{MAISE}, in order to check if the two methods yield the same optimised geometry.
This analysis shows that the structures optimised with the two methods are almost identical, then confirming the reliability of the MLFFs parametrisation.
The same procedure is applied to the case of the Ag substrate.
The predicted energies and forces obtained from the machine-learning force field show very good agreement with the reference DFT results (\autoref{fig:MLFF_Au} and \autoref{fig:MLFF_Ag}).
This agreement further validates the accuracy of the trained models.

\label{subsec:train_ff}
\begin{table}
\begin{center}
\caption{The RMSE value of all the trained models.}
\begin{tabular}{| c | c | c | c | c |}
 \hline
\textbf{Monolayer}  & \textbf{Substrate} & \textbf{Energy RMSE} & \textbf{Force RMSE} & \textbf{RDF Scalar Product} \\
& & (meV/atoms) & (eV/\r{A}) & \\
\hline
 & & & &\\
        MoS$_2$ & & 1.5796 & 0.0213 & 0.904 \\
        MoSe$_2$ & Au & 1.0288 & 0.0204 & 0.892 \\
        WS$_2$ &  & 0.9198 & 0.0191 & 0.946 \\
        WSe$_2$ & & 1.1787 & 0.0193 & 0.941 \\
& & & &\\
\hline
& & & &\\
        MoS$_2$ & & 1.33 & 0.0205 & 0.879\\
        MoSe$_2$ & Ag & 0.941 & 0.0196 & 0.878  \\
        WS$_2$ &  & 0.989 & 0.0203 & 0.947 \\
        WSe$_2$ & & 1.298 & 0.0206 & 0.920 \\
& & & &\\
         \hline
      \end{tabular}
      \label{tab:RMSE_comparision}
\end{center}
\end{table}
\subsection{MD Setup for Friction Analysis}
\label{subsec:mdsetup}
\begin{figure}
  \centering
  {\includegraphics[width=0.7\textwidth]{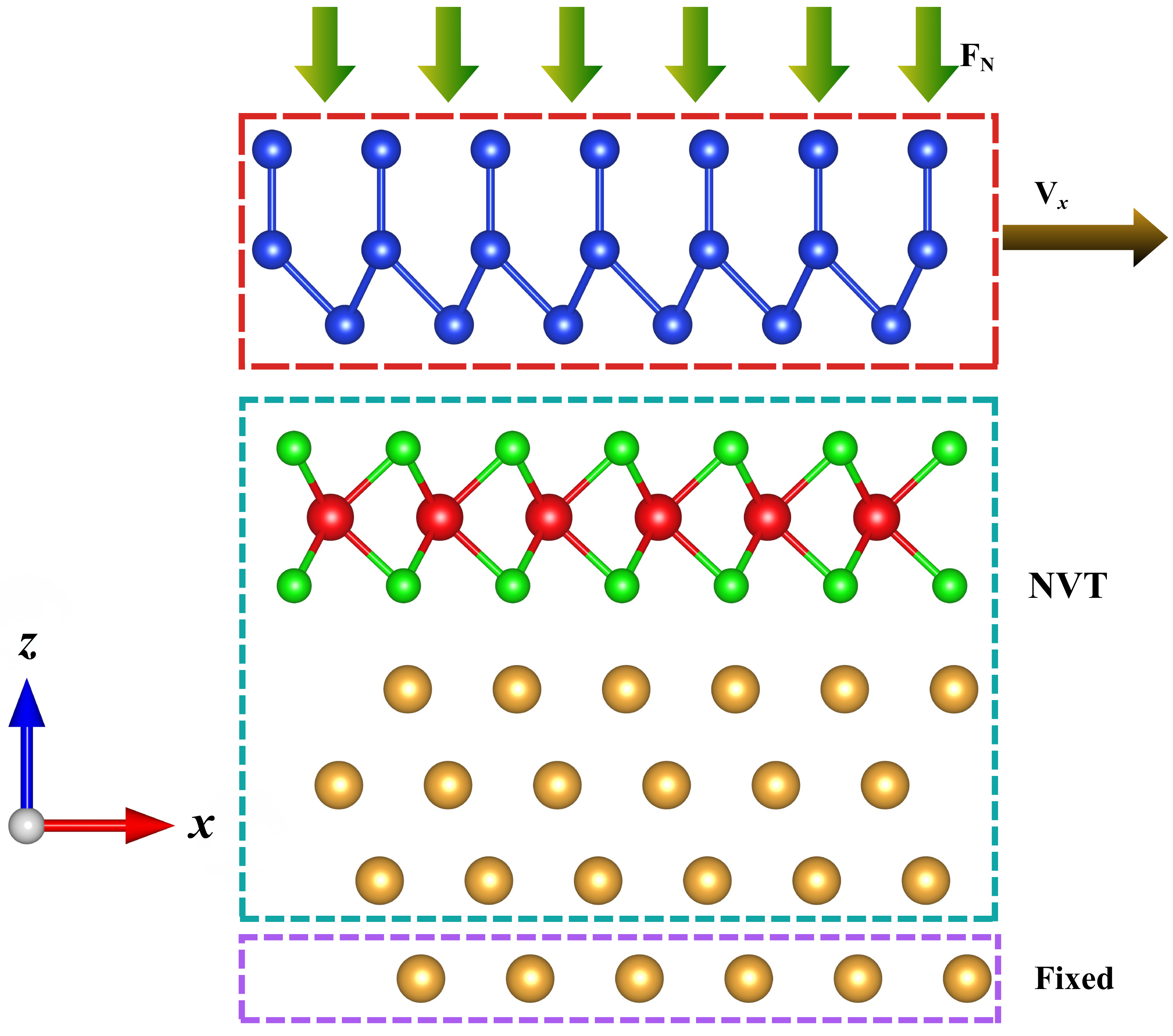}}
  \caption{MD simulations setup.}
  \label{fig:Simulation_Setup}
\end{figure}

For the MD simulations, a $3\times3\times1$ supercell of the optimised heterostructure was constructed in order to minimise the effects of the system's finite size and to provide a sufficiently large contact area for the sliding process (see \autoref{fig:Simulation_Setup}).
The resulting simulation cell contains a total of 360 atoms (See the POSCAR data in \cref{si:1},  Section I).
The bottom layer of the TM substrate is kept fixed throughout the simulation.
This constraint prevents artificial translation of the entire system and represents the bulk support typically present in experimental setups;
the remaining atoms are allowed to evolve dynamically during the simulation in the NVT ensemble at a temperature equal to 300 K.
To model the sliding process, all Si layers are treated as a rigid body to simulate a rigid scanning tip.
The tip is moved along the $\bm{x}$-direction at constant velocities of 2, 3, 4 and 5 m/s, while there is no constraint on the motion along the $\bm{y}$- and $\bm{z}$-directions.
%
% The values of the velocity are chosen to explore the velocity dependence of the frictional response within accessible MD time scales.
%
During sliding, normal loads of 0.0, 0.4, 0.6 and 0.8 nN are applied to the tip along the $\bm{z}$-direction.
Throughout the simulation, the instantaneous forces acting on the tip and the atomic trajectories are recorded at every 1 fs time step. 

%
%%%%%%%%%%%%%%%%%%%%%%%%%%%%%%%%%%%%%%%%%%%%%%%%%%%%%%%%%%%%%%%%%%%%%%%%%%%%%%%%%%%%%%%%%%%%%%%%%%%%%%%%%%%%%%%%%%%%%%%%%%%%%%%%%%%%%%%%%%%%%%%%%%%%%%%%%%%%%%%%%%%%%%%%%%%%%%%%%%%%%%%%%%%%%%%%%%%%%%%%%%%%%%%%%%%%
\section{Results and Discussion}
\label{results}
\subsection{Friction Force Analysis }
\begin{figure}
  \centering
  \includegraphics[width=0.9\textwidth]{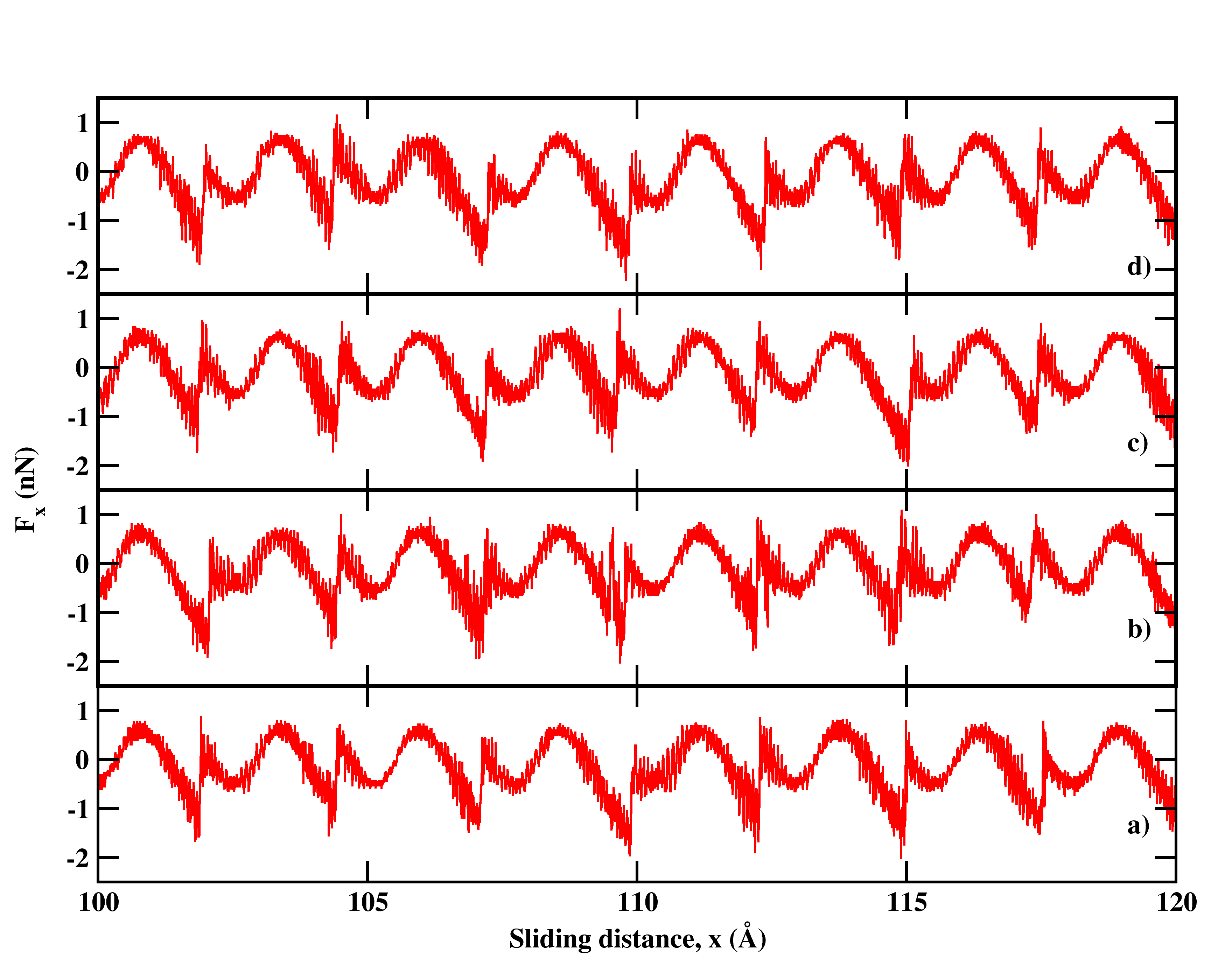}
  \caption{Friction force profile of Au/MoS$_2$/Si with tip moving at 2 m/s for under loads of a) 0 nN, b) 0.4 nN, c) 0.6 nN and d) 0.8 nN.}
  \label{fig:FAvg_Au}
\end{figure}
\begin{figure}
  \centering
  \includegraphics[width=0.9\textwidth]{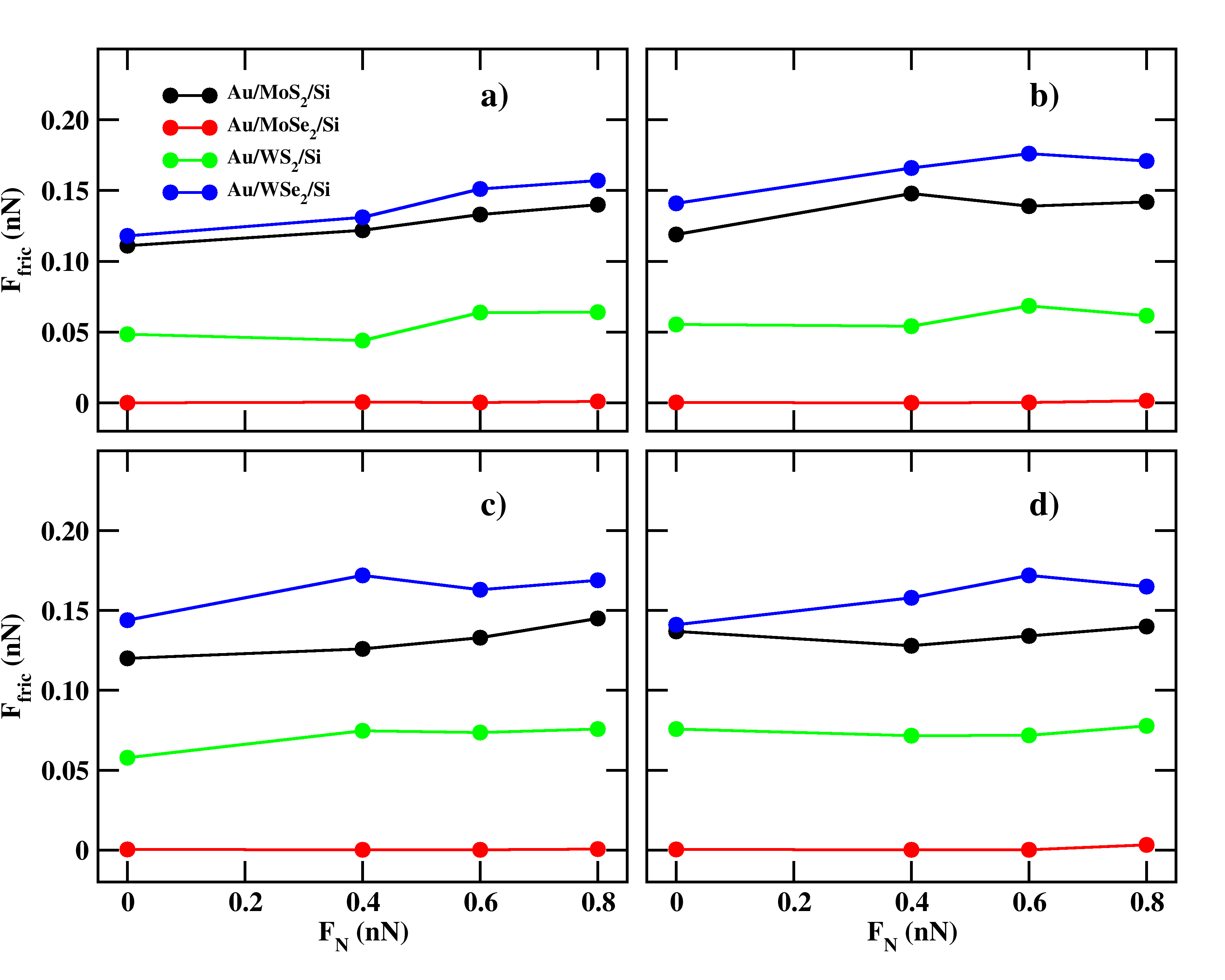}
  \caption{Average friction force as a function of applied load for all systems at tip velocities of (a) 2 m/s, (b) 3 m/s, (c) 4 m/s, and (d) 5 m/s.}
  \label{fig:COF}
\end{figure}
\begin{figure}
  \centering
  \includegraphics[width=0.9\textwidth]{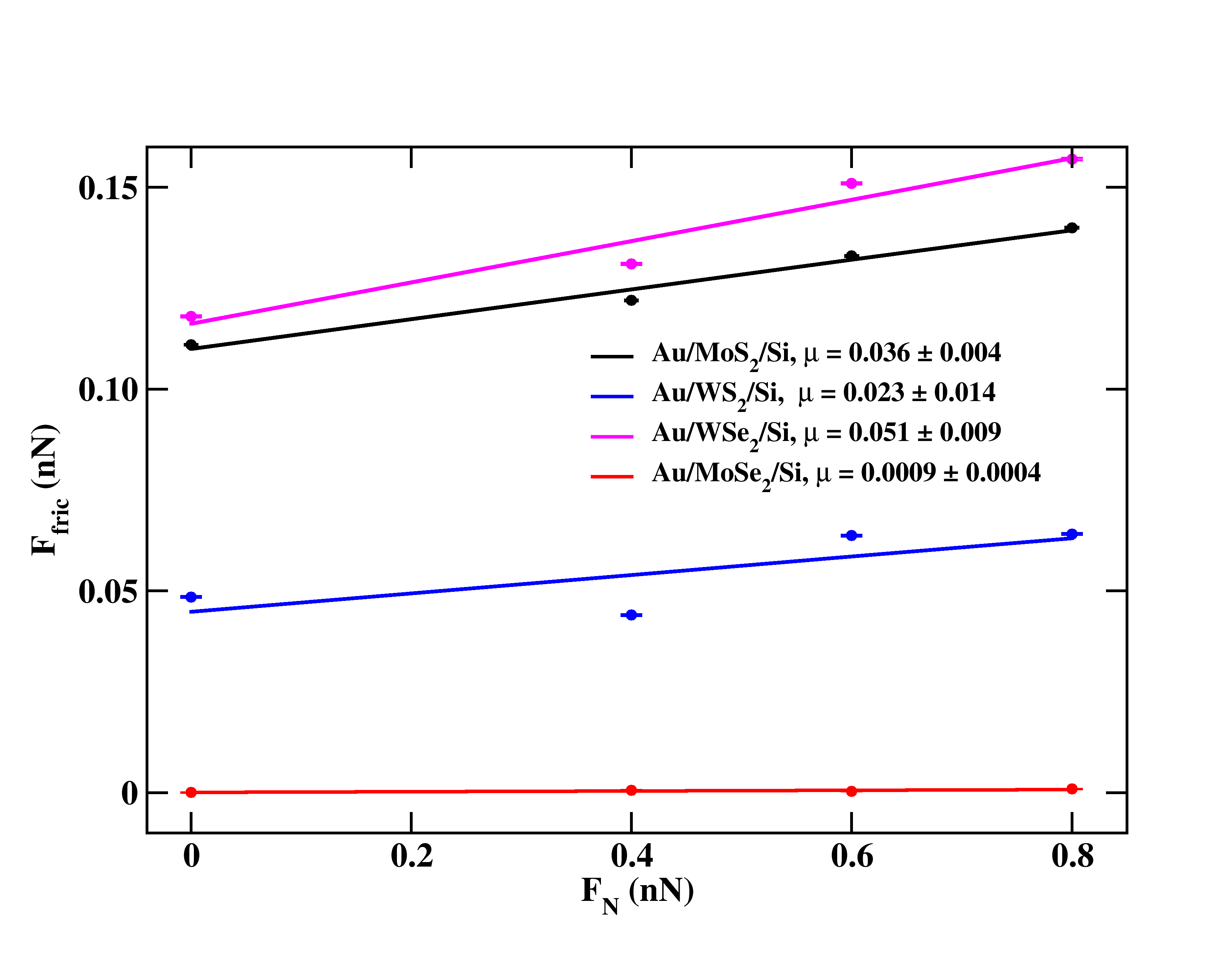}
  \caption{COF obtained from a linear fit of the average friction force as a function of loads for all systems at a tip velocity of 2 m/s.}
  \label{fig:COF_fit}
\end{figure}
\label{para:fric_forc}
We begin our study of the friction response by considering the friction force profile using 1 ns trajectories under different applied normal loads, as shown in \autoref{fig:FAvg_Au}.
Interestingly, the overall shape of the friction force profile remains nearly identical across all applied loads, indicating negligible load dependence of the profile shape.
We then calculate the average friction force acting on the Si tip along the $\bm{x}$-direction $F_{fric}$, by averaging over all the snapshots of the trajectory:
\begin{equation}
\label{eq:F_fric}
F_{fric}= -\frac{1}{N_s}\sum_{i=1}^{N_s} \sum_{j=1}^{N_{tip}} F_{xj}(t_i)
\end{equation}
where $N_s$ is the number of time steps, $N_{tip}$ is the number of atoms in the tip, $F_{xj}$ is the force acting on the $j$-th atom of the tip at the $t_i$ time step along the sliding $\bm{x}$-direction, and the negative sign accounts for the fact that friction opposes the direction of motion. 
The coefficient of friction $\mu$ is then calculated according to Amontons's law with the following linearly fitting function:
\begin{equation}\label{eq:COF}
 F_{fric} = \mu F_N + F_{fric}^0
\end{equation}
where $F_{N}$ is the applied normal load and $F_{fric}^0$ is the friction in the absence of load.
The calculated average friction force at varying loads with different velocities for all the systems are reported in \autoref{fig:COF}.
Among all systems, Au/WSe$_2$/Si system exhibits the highest friction force across all velocities, followed by Au/MoS$_2$/Si and Au/WS$_2$/Si.
Au/MoSe$_2$/Si displays nearly negligible friction force across all the applied normal loads.
This behaviour is consistent with the corresponding friction force profile, which will be discussed in detail below.
In most cases, the friction--load relationship deviates from simple linear behaviour, exhibiting nonmonotonic trends irrespective of the velocities. Looking at \autoref{fig:COF_fit}, which represents the fitted curve corresponding to \autoref{fig:COF}a, the force–fit analysis gives $\mu$ values of 0.036, 0.023, 0.051, and 0.0009, with corresponding errors of 0.003, 0.014, 0.009, and 0.0004 for Au/MoS$_2$/Si, Au/WS$_2$/Si, Au/WSe$_2$/Si, and Au/MoSe$_2$/Si, respectively.
As a result, the extracted $\mu$ shows relatively large errors, indicating that the assumption of a linear friction--load relationship is no longer strictly valid. 
A similar trend is observed at other velocities (see \cref{si:1}, Section II, Figure S1, S2 and S3).
This behaviour suggests that in our systems, a simple proportional relationship between \textit{F$_{fric}$} and  \textit{F$_{N}$} cannot fully describe the friction response.
This behaviour has already been experimentally observed for an  Au/MoS$_2$/Si monolayer grown on an Au surface \cite{PhysRevLett.133.136201}.
\begin{figure}
  \centering
  \includegraphics[width=0.99\textwidth]{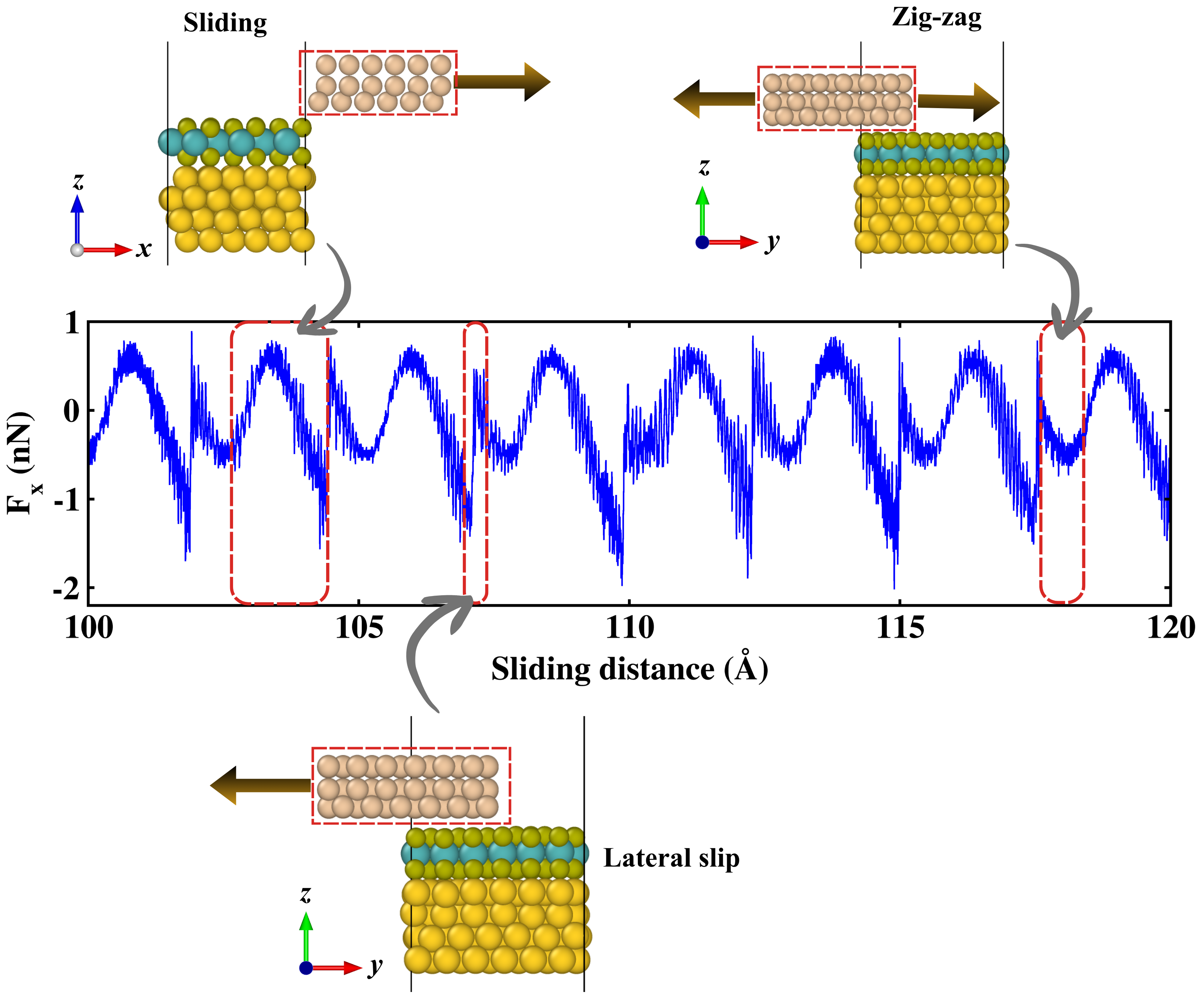}
  \caption{Friction force profile associated with the motion of the tip.}
  \label{fig:Atomic_Trajectory}
\end{figure}
\begin{figure}
  \centering
\includegraphics[width=0.8\textwidth]{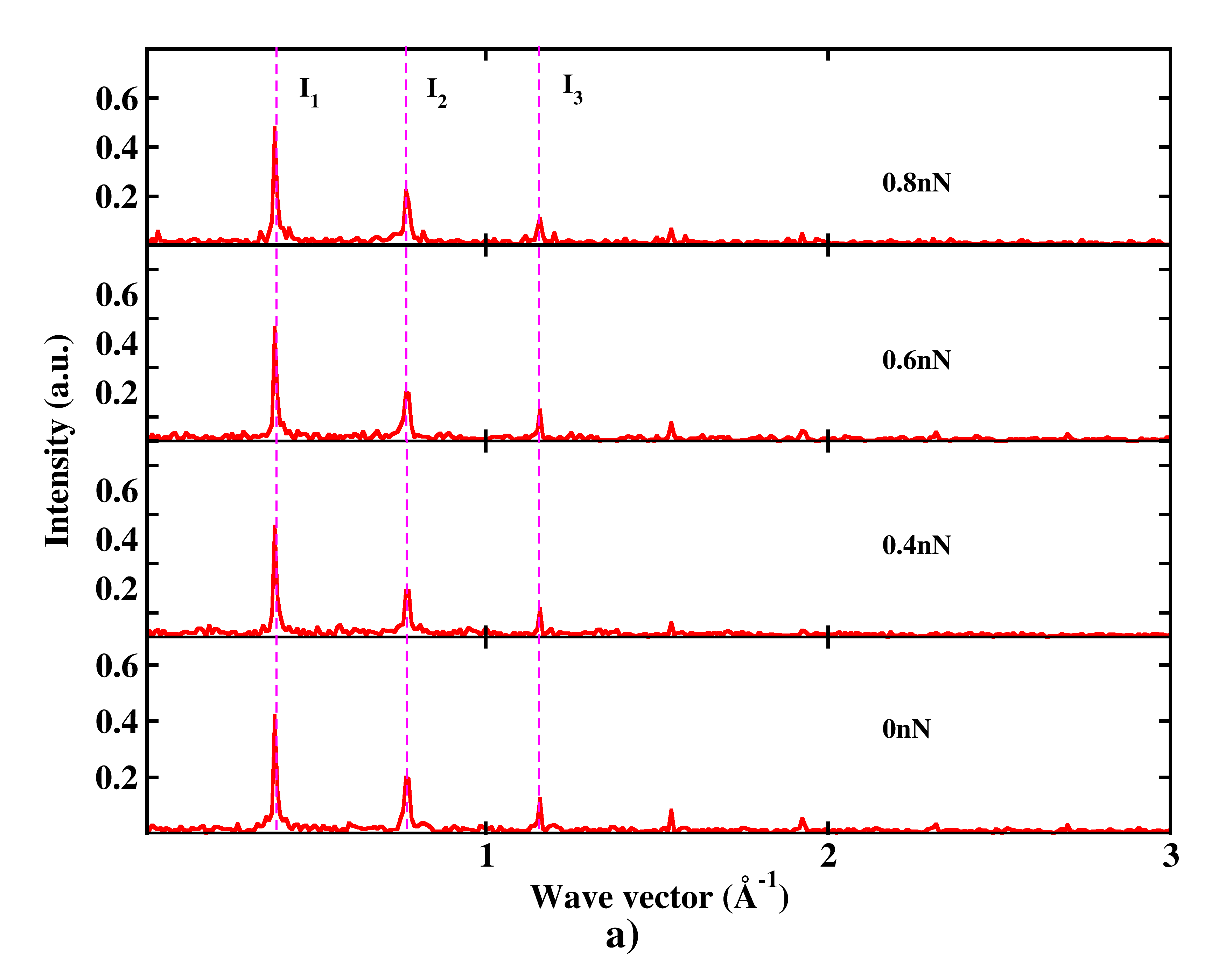}\\
\includegraphics[width=0.8\textwidth]{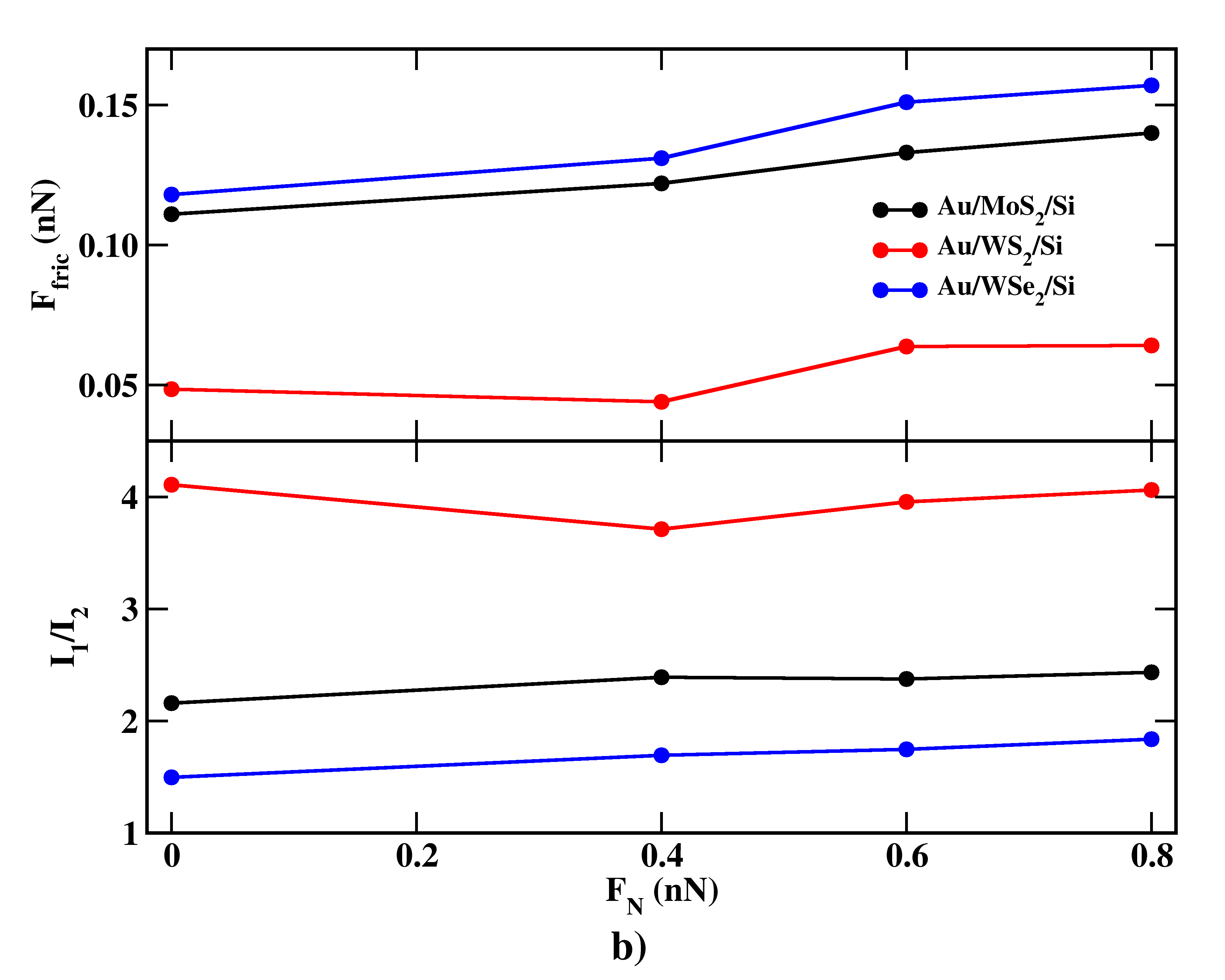}
  \caption{a) Spatial Fourier transform of Au/MoS$_2$/Si with varying loads at velocity 2 m/s, b) Peak intensity ratio and their average friction force with varying loads for all the systems except Au/MoSe$_2$/Si. } 
  \label{fig:ftransf}
\end{figure}
\phantomsection
\label{para:non_monotonic}
To understand the origin of this nonmonotonic behaviour, we look at the detail of the friction force on the tip during the sliding.
\autoref{fig:Atomic_Trajectory} shows the friction force on Si tip as a function of the sliding distance for the case of Au/MoS$_2$/Si system with a velocity of 2 m/s and normal load of 0 nN. 
Apart from the sliding events along the sliding direction, some extra features can also be observed.
Analysis of the geometric sequence along the trajectory reveals that large sudden drops in the lateral force correspond to slip events associated with the primary sliding motion along $\bm{x}$-direction, named as sliding. 
In contrast, abrupt increases in the force indicate lateral slip motion involving transverse displacement along $\bm{y}$-direction, named as lateral slip. 
Following these events, a gradual decrease in the force is observed, which corresponds to a zigzag-like motion of the tip as it relaxes while moving laterally along $\bm{y}$-direction, named as zig-zag. 
These observations confirm that the tip does not follow a strictly one-dimensional sliding pathway; 
instead, the motion involves coupled displacements along both $\bm{x}$ and $\bm{y}$ directions, leading to the different frictional behaviour along the $\bm{x}$-direction.
The zig-zag events indicate that the system explores multiple local minima on the interfacial potential energy surface during sliding;
such lateral motion effectively modifies the instantaneous contact configuration and the energy dissipation pathway, leading to fluctuations in the lateral force along $\bm{x}$-direction.
Interestingly, these features in the force profile occur in most of the cases except the system Au/MoSe$_2$/Si.
In the latter cases, the lateral slip and zig-zag part in the force profile is missing which leads to a reduction in the average friction force, as shown in \cref{si:1} Figure S4.
When two surfaces interact, the surface interactions create a corrugated potential energy landscape that governs the lateral motion during sliding.
As the tip moves across this landscape, the motion may not remain strictly confined to the imposed sliding direction;
instead, the system can explore neighbouring energy minima, resulting in transverse displacements and complex sliding pathways.

To quantitatively characterise these slip events, we perform a spatial Fourier transform of the forces experienced by the Si tip along the $\bm{x}$-direction. 
This analysis provides insight into the nature of the forces that arise from different modes of tip motion, including sliding, lateral motion, and possible nonlinearities in the tip trajectory.
\autoref{fig:ftransf}a displays the spatial Fourier transform of the average force experienced by the tip for the case of Au/MoS$_2$/Si at different normal loads with velocity of 2 m/s.
Three magenta dotted lines indicate the main peaks at wave vectors $\bm{k}_{1}$, $\bm{k}_{2}$ and $\bm{k}_{3}$ with the respective intensities $I_1$, $I_2$ and $I_{3}$. 
In the Fourier transform, the intensity of a peak is proportional to the number of times the corresponding event occurs, while the inverse of the peak position gives the width of the events on a distance scale.
In our simulations, the first peak at $\bm{k}_1$ corresponds to the sliding process along the $\bm{x}$-direction, while the second and third peaks (at $\bm{k}_2$ and $\bm{k}_3$, respectively) correspond to lateral zig-zag movement and lateral slip along the $\bm{y}$-direction, respectively.
We do not consider other peaks, as their intensity is almost as small as the background noise corresponding to thermal fluctuations of the atomic motions.
This is confirmed from the friction force profile where length of the events correlate to the peak positions.
The Fourier transform spectra reveal clear correlation between the peak intensities and the average friction force. 
In particular, the overall ratio $I_1/I_2$ of the first two peak intensities for all the systems increases as the mean friction force increases (\autoref{fig:ftransf}b).
This trend is observed for all systems across different velocities, except in the case of a Au/MoSe$_2$/Si system;
in fact, in this case only a single peak appears in the Fourier transform only at $\bm{k}_1$, that is, the peak of the sliding, making it impossible to establish a correlation based on the ratio of peak intensities (see \cref{si:1}, Section II, Figure S5).
This trend suggests that when the number of lateral slip events increases, the effective friction force decreases (\autoref{fig:ftransf}b).
Since the friction force is evaluated under different normal loads, the relative contributions of these sliding modes vary with load.
As a result, the mean friction force does not exhibit a strictly linear dependence on the applied load, which leads to the nonmonotonic behavior observed in the friction-load relationship, and contributes to the large uncertainty in the extracted coefficient of friction using Amontons's law.
%
%%%%%%%%%%%%%%%%%%%%%%%%%%%%%%%%%%%%%%%%%%%%%%%%%%%%%%%%%%%%%%%%%%%%%%%%%%%%%%%%%%%%%%%%%%%%%%%%%%%%%%%%%%%%%%%%%%%%%%%%%%%%%%%%%%%%%%%%%%%%%%%%%%%%%%%%%%%%%%%%%%%%%%%%%%%%%%%%%%%%%%%%%%%%%%%%%%%%%%%%%%%%%%%%%%%%%%%%%%%%%%%%%%%%%%%%
\subsection{Effect of substrate on friction force}
\label{subsec:effect_substrate}
\begin{figure}
  \centering
  \includegraphics[width=0.9\textwidth]{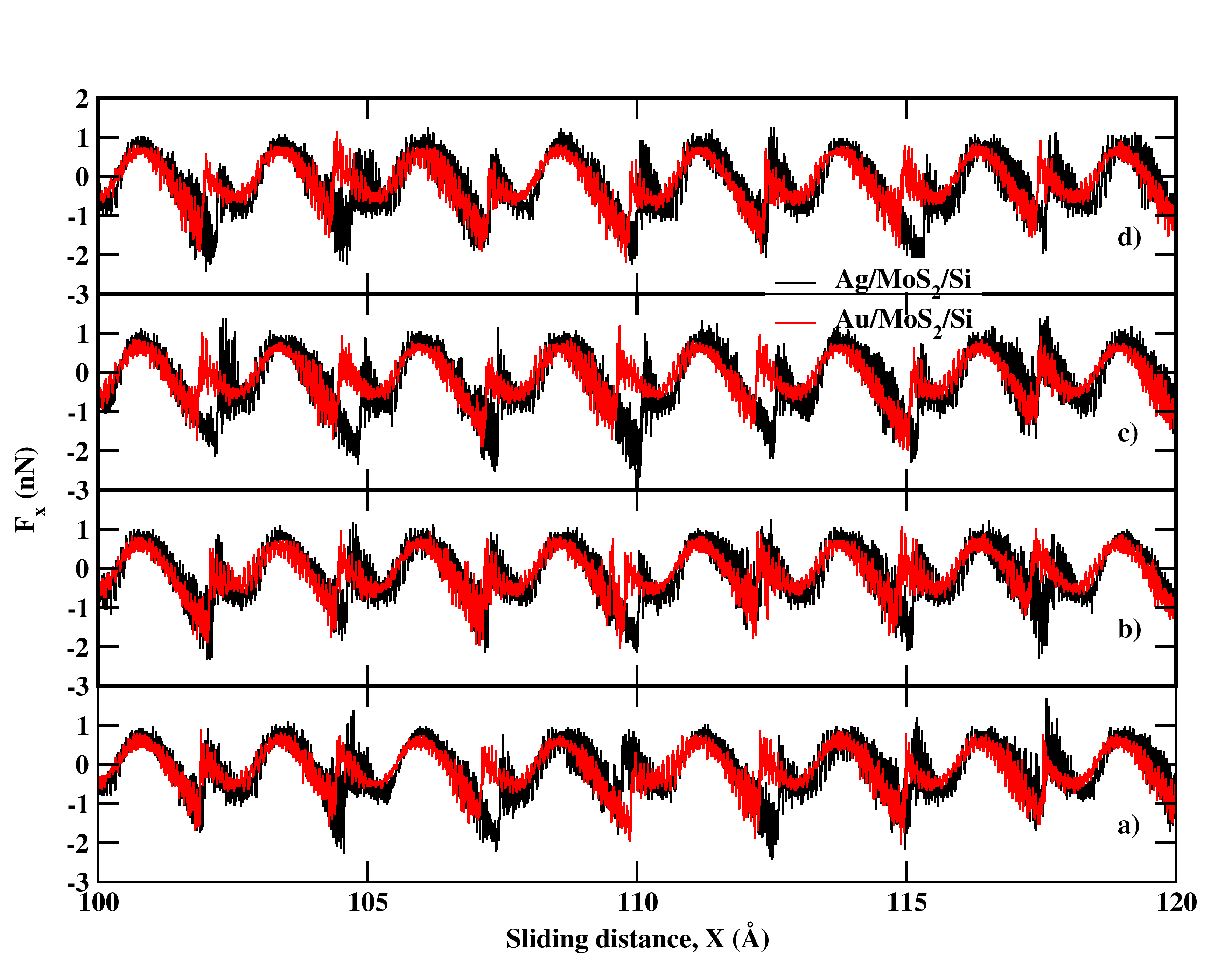}
  \caption{Comparison of friction force profile between Au/MoS$_2$/Si and Ag/MoS$_2$/Si as a function of displacement at loads a) 0 nN, b) 0.4 nN, c) 0.6 nN and d) 0.8 nN.}
   \label{fig:F_Fn}
\end{figure}
\begin{figure}
  \centering
  \includegraphics[width=0.9\textwidth]{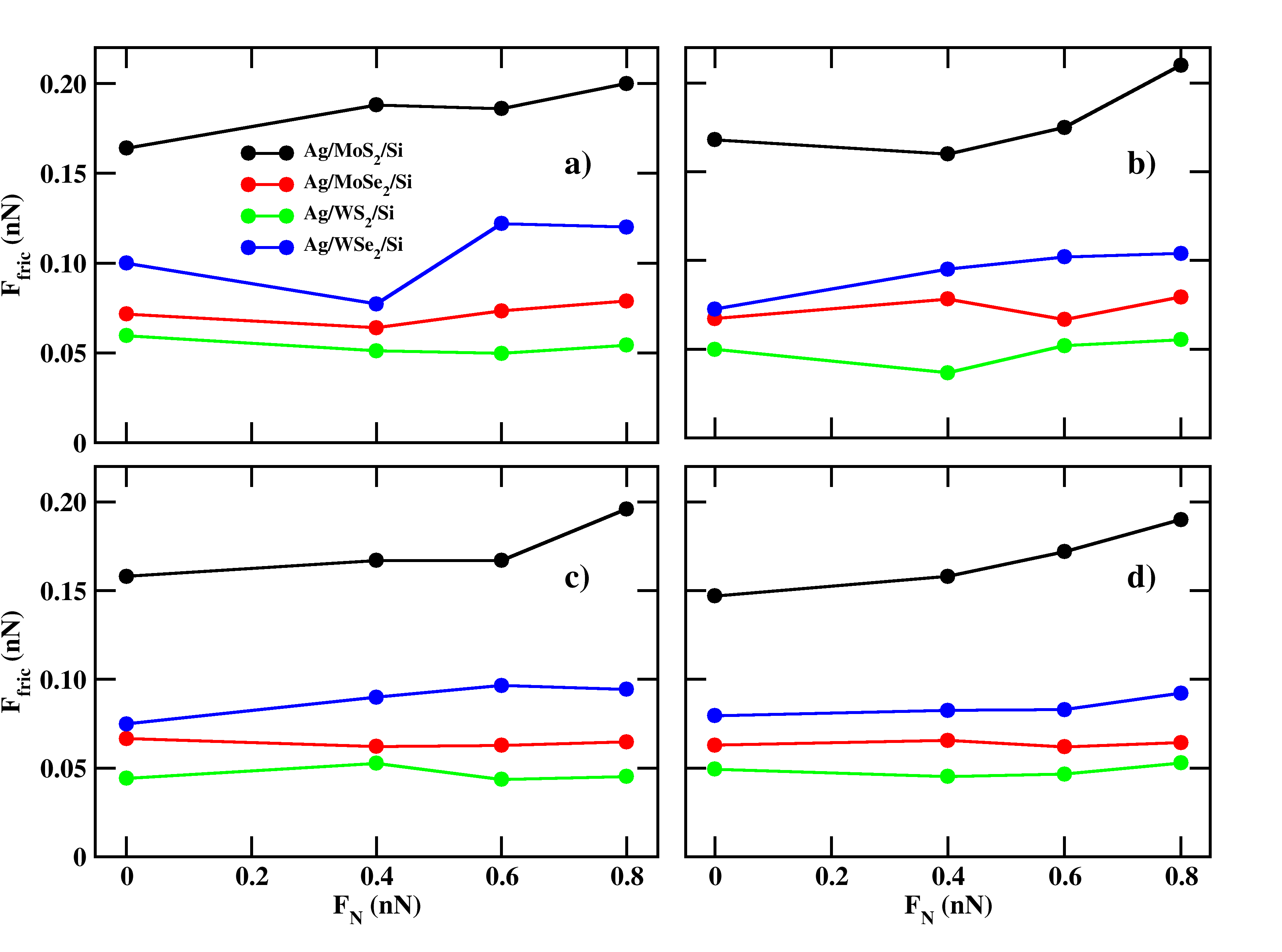}
  \caption{ Average friction force as a function of applied loads for all the systems with tip velocities  a) 2 m/s, b) 3 m/s, c) 4 m/s d) 5 m/s.} 
  \label{fig:fric}
\end{figure}
\begin{figure}
  \centering
  \includegraphics[width=0.8\textwidth]{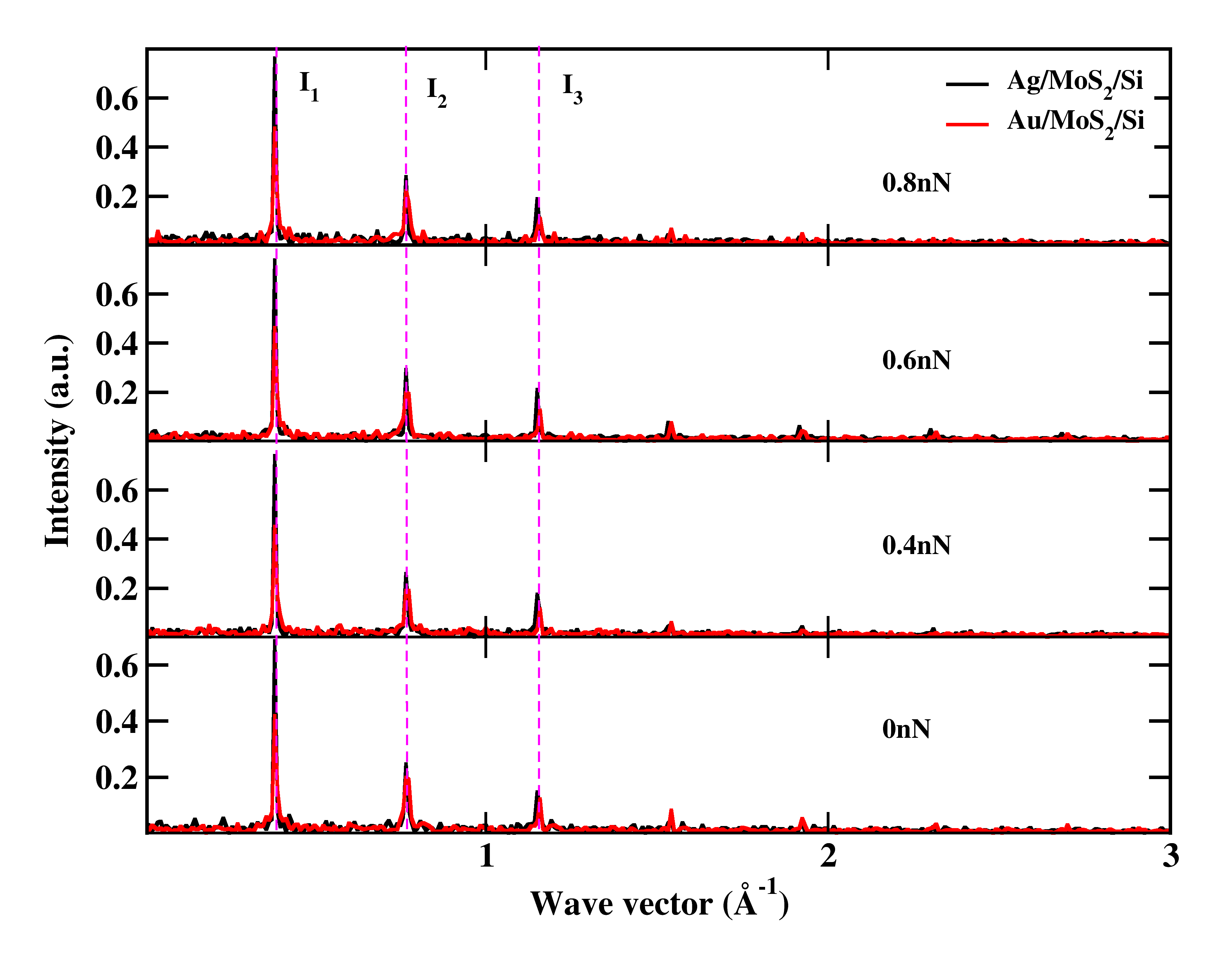}
  \caption{Fourier transform of the force on a tip varying with loads and different substrate for velocity 2 m/s. }
  \label{fig:FT_Ag_Au}
\end{figure}
\begin{figure}
  \centering
  \includegraphics[width=0.8\textwidth]{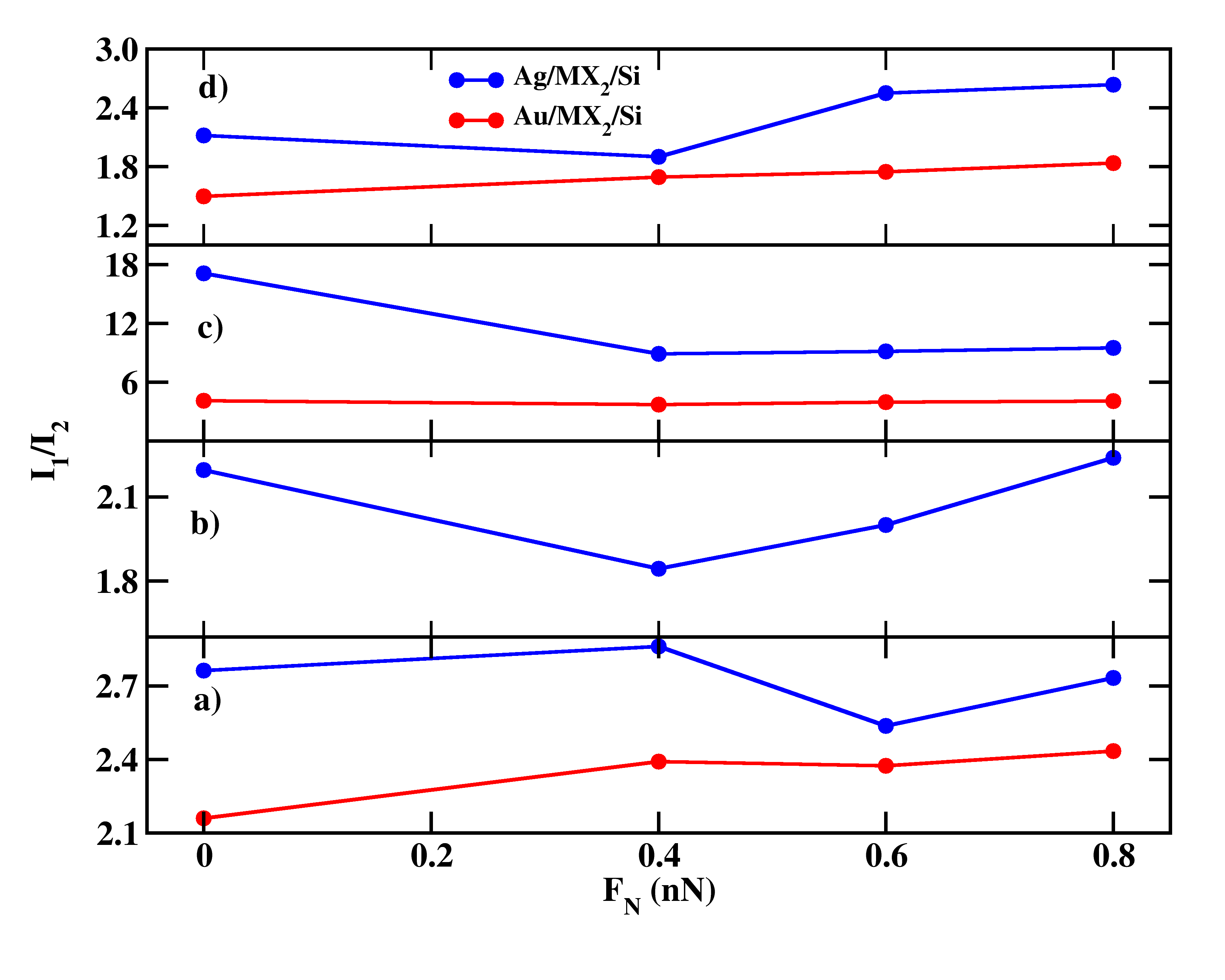}
  \caption{ Ratio of the peak intensity \textit{I$_{1}$} and \textit{I$_{2}$} varying with loads and different substrate for velocity 2 m/s.}
  \label{fig:I1_I2_Ag_Au}
\end{figure}
\begin{figure}
  \centering
  \includegraphics[width=0.8\textwidth]{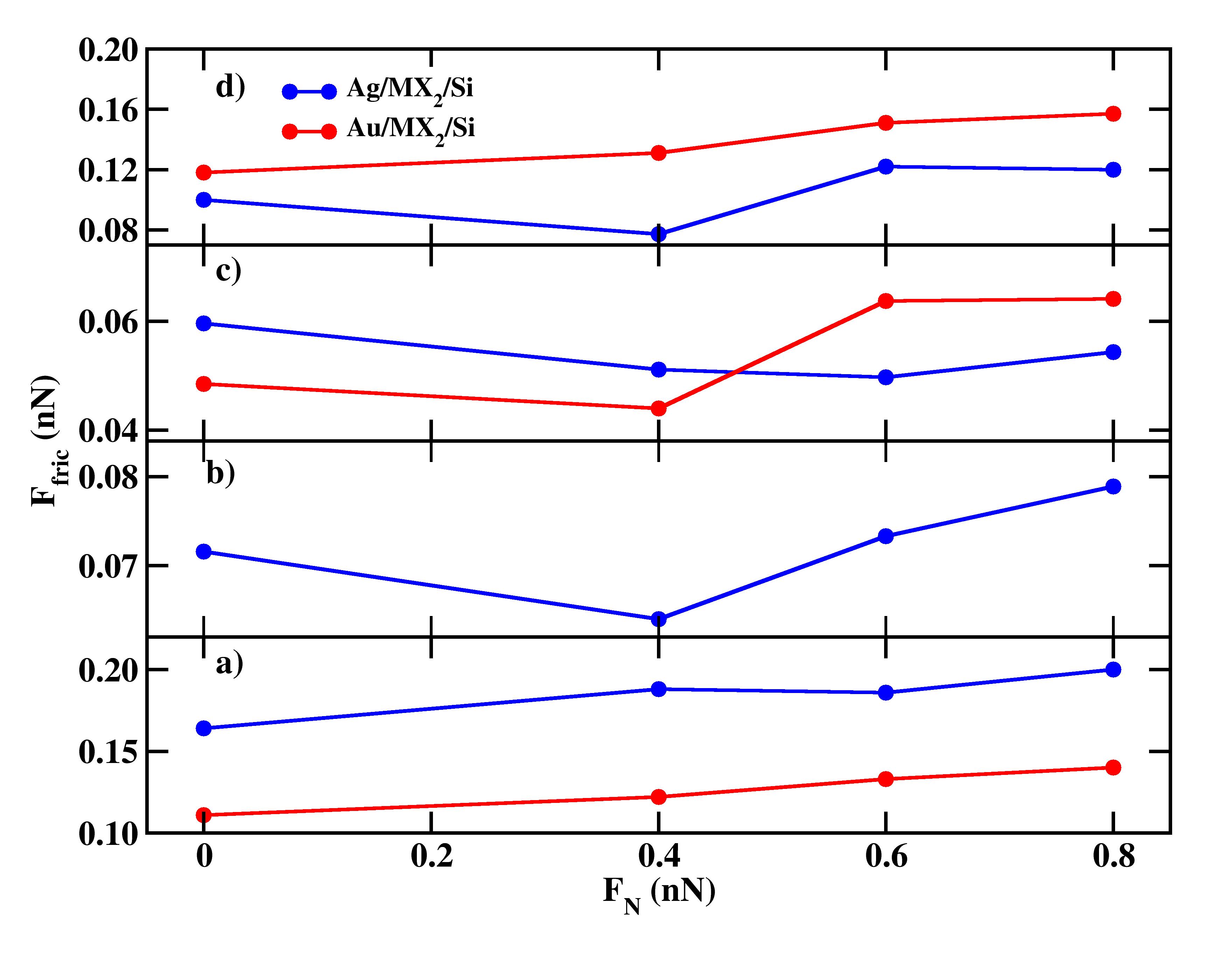}
  \caption{Average friction force, varying with loads and different substrate for velocity 2 m/s.}
  \label{fig:Favg_Ag_Au}
\end{figure}

To investigate the effect of the substrate on the lateral friction force, we analyse the trend of the force experienced by the tip as a function of sliding distance under different applied normal loads and different substrates, namely Au and Ag, as shown in \autoref{fig:F_Fn}.
We observe that the overall nature of the force signal—particularly the shape of the force profile—remains essentially unchanged as the load increases, provided the monolayer and velocity are kept constant; moreover, in the presence of the Ag substrate, the average force on the tip follows a trend very similar to that observed for the Au substrate.
Furthermore, we plot the average friction force as a function of the applied load for all systems and for all sliding velocities, in the presence of the Ag substrate, as shown in \autoref{fig:fric}. 
We found that the variation of friction force as a function of loads shows nonmonotonic trend and it is consistent with the case of Au substrate that we already have discussed before (see \autoref{fig:COF}). 
Notably, this behavior is consistently observed for all sliding velocities, indicating that the nonmonotonic features are largely independent of velocity.
Overall, among all systems, the Ag/MoS$_2$/Si system exhibits the highest friction force across all velocities, followed by Ag/WSe$_2$/Si and Ag/MoSe$_2$/Si, while the Ag/WS$_2$/Si system shows the lowest friction force over the entire velocity range. In contrast, when considering the Au substrate, the Au/WSe$_2$/Si system demonstrates the highest friction force at all velocities, followed by Au/MoS$_2$/Si and Au/WS$_2$/Si, as discussed in above paragraph.
This indicates that the friction force can change significantly depending on the substrate.
Furthermore, the calculated $\mu$ (as shown in \cref{si:1}, Section~II,  Figure~S6, S7, S8 and S9) show quite large errors, arising from the nonmonotonic dependence of the friction force on the applied load. 
We already observed a similar trend in the case of the Au substrate, suggesting that the overall nature of frictional behavior in our systems is largely independent of the substrate.
\autoref{fig:FT_Ag_Au} shows the three sharp peaks in FT at velocity 2 m/s for both Ag/MoS$_2$/Si and Au/MoS$_2$/Si systems, which have very similar profiles.  
We find that the positions of all three peaks remain nearly unchanged for different applied loads and different substrates, indicating that they are independent of loads and substrates.
We also observe that the peak positions in the Fourier transforms shift as the monolayer material is varied, confirming that the forces experienced by the tip depend on the specific monolayer (\cref{si:1} Section II, Figure S10).
In particular, the first peak intensities at varying loads are significantly higher with the Ag/MX$_2$/Si systems compared to the case with Au/MX$_2$/Si systems;
this indicates that the friction contribution along the sliding direction is greater when using the Ag substrate (\autoref{fig:Favg_Ag_Au}a).
In contrast, the opposite trend is observed with the Au/WSe$_2$/Si system, where the friction force is higher compared with the Ag/WSe$_2$/Si (\autoref{fig:Favg_Ag_Au}d).
This behaviour correlates with the Fourier transform results (which is shown in \cref{si:1}, Section II, Figure S10), where the first peak intensity is higher leads to higher in peak intensity ratio ( see \autoref{fig:I1_I2_Ag_Au}d) in systems with the Au substrate.
Overall, these findings suggest that the contribution of multiple sliding and lateral motion in the friction profile strongly influences the average friction force, depending on the specific monolayer and substrate combination.

Focusing to the intensity ratio of both Au/MX$_2$/Si and Ag/MX$_2$/Si system (see \autoref{fig:I1_I2_Ag_Au}), the variation of the peak-intensity ratio follows a trend similar to that of the friction force (see \autoref{fig:Favg_Ag_Au}). 
In particular, the peak intensity ratios are higher in the case of the Ag substrate compared with the Au substrate, irrespective of the applied loads.
Furthermore, the peak-intensity ratio exhibits a clear nonmonotonic dependence on the applied load, consistent with the nonmonotonic behaviour observed in the friction force, resulting in high errors in the estimation of the friction coefficient from a linear fit against the applied load. 
Therefore, the lateral motion of the tip plays an important role in determining the overall friction force and contributes to both the reduction of the average friction force and the emergence of its nonmonotonic behaviour.
%
%%%%%%%%%%%%%%%%%%%%%%%%%%%%%%%%%%%%%%%%%%%%%%%%%%%%%%%%%%%%%%%%%%%%%%%%%%%%%%%%%%%%%%%%%%%%%%%%%%%%%%%%%%%%%%%%%%%%%%%%%%%%%%%%%%%%%%%%%%%%%%%%%%%%%%%%%%%%%%%%%%%%%%%%
\section{Conclusions}
\label{sec:conclusion}
In this study, we investigate the nanoscale friction behaviour of TM/MX$_2$/Si systems (TM = Au, Ag; M = Mo, W; X = S, Se) using classical molecular dynamics simulations with machine-learning-based force fields trained on DFT data.
Our results show that the mean friction force does not follow a strictly linear dependence on the applied normal load.
Instead, a nonmonotonic friction–load relationship is observed, which introduces significant uncertainty in determining the coefficient of friction based on Amontons's law.
Analysis of the spatial Fourier transform of lateral force signals reveals multiple spectral peaks, with their intensities strongly correlated with the average friction force. 
The presence of these peaks indicates that the tip motion is not limited to the primary sliding direction but also includes significant lateral contributions.
Furthermore, the qualitative features of both the friction force profiles and the Fourier spectra remain similar across different substrates. 
While the magnitude of friction depends on the specific substrate material, the underlying mechanism responsible for the nonmonotonic behavior—arising from the coupling between longitudinal sliding and lateral motion of the tip—remains essentially substrate-independent.
Since our findings demonstrate that frictional forces are influenced by both the monolayer and the underlying substrate, these methods can be effectively employed to investigate nanoscale frictional behaviour in a wide range of heterostructures.
%%%%%%%%%%%%%%%%%%%%%%%%%%%%%%%%%%%%%%%%%%%%%%%%%%%%%%%%%%%%%%%%%%%%%%%%%%%%%%%%%%%%%%%%%%%%%%%%%%%%%%%%%%%%%%%%%%%%%%%%%%%%%%%%%%%%%%%%%%%%%%%%%%%%%%%%%%%%%%%%%%%%
%%%%%%%%%%%%%%%%%%%%%%%%%%%%%%%%%%%%%%%%%%%%%%%%%%%%%%%%%%%%%%%%%%%%%
\begin{suppinfo}
Supporting information is in this file after the bibliography section.
\sifile{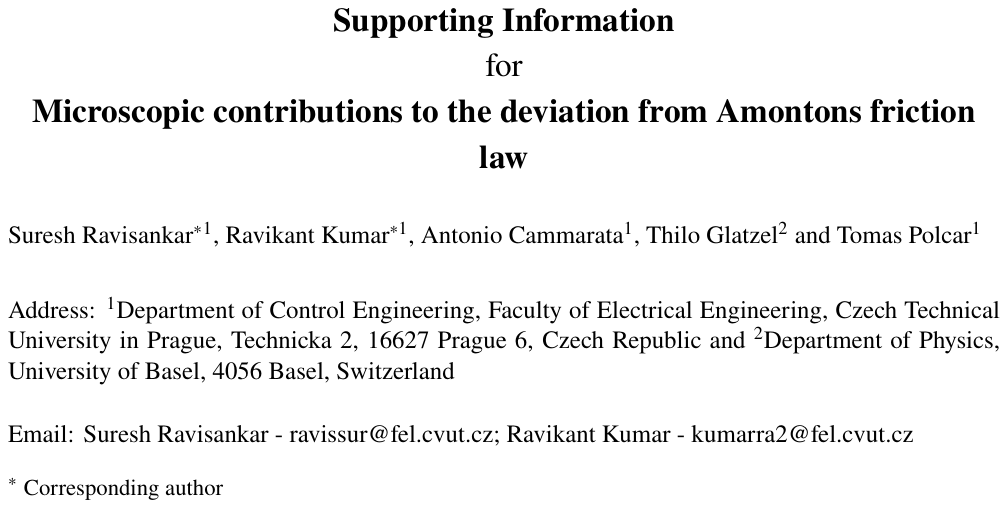}{PDF}{Additional data and Figures}\label{si:1}
\end{suppinfo}

%%%%%%%%%%%%%%%%%%%%%%%%%%%%%%%%%%%%%%%%%%%%%%%%%%%%%%%%%%%%%%%%%%%%%
\begin{acknowledgements}
The authors would like to acknowledge the financial support provided by the joint grant 'Nanocontrol' from the Czech Science Foundation (GARC) and the Swiss National Science Foundation (SNSF), with grant numbers 24-12643L and 200021L-219983, respectively.
This work was co-funded by the European Union under the project ``Robotics and advanced industrial production'' (reg. no. CZ.02.01.01/00/22\_008/0004590).
This work was supported by the Ministry of Education, Youth and Sports of the Czech Republic through the e-INFRA CZ (ID:90254).
Use of \textsc{vesta} software \cite{vesta} is also acknowledged.
\end{acknowledgements}
%%%%%%%%%%%%%%%%%%%%%%%%%%%%%%%%%%%%%%%%%%%%%%%%%%%%%%%%%%%%%%%%%%%%%
% \bibliography{paper}

\newpage



\section*{Supplementary material}

# Supporting Information

for

# Microscopic contributions to the deviation from Amontons friction law

Suresh Ravisankar[*1], Ravikant Kumar[*1], Antonio Cammarata[1], Thilo Glatzel[2] and Tomas Polcar[1]

Address: [1]Department of Control Engineering, Faculty of Electrical Engineering, Czech Technical University in Prague, Technicka 2, 16627 Prague 6, Czech Republic and [2]Department of Physics, University of Basel, 4056 Basel, Switzerland

Email: Suresh Ravisankar - ravissur@fel.cvut.cz; Ravikant Kumar - kumarra2@fel.cvut.cz

[*] Corresponding author

# Contents

# Structural Information

The geometry of each hetrostructure used in the molecular dynamics simulations with a total number of 360 atoms (144 TM, 36 M, 72 X, 108 Si) is provided in what follows in POSCAR format.

```
Au/MoS2/Si
1.0
        15.5672988892          0.0000000000          0.0000000000
         0.0000000000         17.9267997742          0.0000000000
         0.0000000000          0.0000000000         50.0000000000
    Au    Mo    S    Si
   144    36   72   108
Cartesian
        3.904829471          1.172879994         16.812799871
        3.904829471          7.148479830         16.812799871
        3.904829471         13.124079933         16.812799871
        9.093929255          1.172879994         16.812799871
        9.093929255          7.148479830         16.812799871
        9.093929255         13.124079933         16.812799871
       14.283028576          1.172879994         16.812799871
       14.283028576          7.148479830         16.812799871
       14.283028576         13.124079933         16.812799871
        3.894879778          1.172879994         23.676000535
        3.894879778          7.148479830         23.676000535
        3.894879778         13.124079933         23.676000535
        9.083979562          1.172879994         23.676000535
        9.083979562          7.148479830         23.676000535
        9.083979562         13.124079933         23.676000535
       14.273078883          1.172879994         23.676000535
       14.273078883          7.148479830         23.676000535
       14.273078883         13.124079933         23.676000535
        3.904829471          4.160680046         16.812799871
```

| | | |
|---|---|---|
| 3.904829471 | 10.136279614 | 16.812799871 |
| 3.904829471 | 16.111880252 | 16.812799871 |
| 9.093929255 | 4.160680046 | 16.812799871 |
| 9.093929255 | 10.136279614 | 16.812799871 |
| 9.093929255 | 16.111880252 | 16.812799871 |
| 14.283028576 | 4.160680046 | 16.812799871 |
| 14.283028576 | 10.136279614 | 16.812799871 |
| 14.283028576 | 16.111880252 | 16.812799871 |
| 3.894879778 | 4.160680046 | 23.676000535 |
| 3.894879778 | 10.136279614 | 23.676000535 |
| 3.894879778 | 16.111880252 | 23.676000535 |
| 9.083979562 | 4.160680046 | 23.676000535 |
| 9.083979562 | 10.136279614 | 23.676000535 |
| 9.083979562 | 16.111880252 | 23.676000535 |
| 14.273078883 | 4.160680046 | 23.676000535 |
| 14.273078883 | 10.136279614 | 23.676000535 |
| 14.273078883 | 16.111880252 | 23.676000535 |
| 0.441922960 | 1.172879994 | 19.131700695 |
| 0.441922960 | 7.148479830 | 19.131700695 |
| 0.441922960 | 13.124079933 | 19.131700695 |
| 5.631022773 | 1.172879994 | 19.131700695 |
| 5.631022773 | 7.148479830 | 19.131700695 |
| 5.631022773 | 13.124079933 | 19.131700695 |
| 10.820122558 | 1.172879994 | 19.131700695 |
| 10.820122558 | 7.148479830 | 19.131700695 |
| 10.820122558 | 13.124079933 | 19.131700695 |
| 0.441922960 | 4.160680046 | 19.131700695 |
| 0.441922960 | 10.136279614 | 19.131700695 |
| 0.441922960 | 16.111880252 | 19.131700695 |
| 5.631022773 | 4.160680046 | 19.131700695 |
| 5.631022773 | 10.136279614 | 19.131700695 |
| 5.631022773 | 16.111880252 | 19.131700695 |
| 10.820122558 | 4.160680046 | 19.131700695 |
| 10.820122558 | 10.136279614 | 19.131700695 |
| 10.820122558 | 16.111880252 | 19.131700695 |
| 2.168479790 | 1.172879994 | 21.399800479 |
| 2.168479790 | 7.148479830 | 21.399800479 |
| 2.168479790 | 13.124079933 | 21.399800479 |
| 7.357579342 | 1.172879994 | 21.399800479 |
| 7.357579342 | 7.148479830 | 21.399800479 |
| 7.357579342 | 13.124079933 | 21.399800479 |
| 12.546678663 | 1.172879994 | 21.399800479 |
| 12.546678663 | 7.148479830 | 21.399800479 |
| 12.546678663 | 13.124079933 | 21.399800479 |
| 2.168479790 | 4.160680046 | 21.399800479 |

| | | |
|---|---|---|
| 2.168479790 | 10.136279614 | 21.399800479 |
| 2.168479790 | 16.111880252 | 21.399800479 |
| 7.357579342 | 4.160680046 | 21.399800479 |
| 7.357579342 | 10.136279614 | 21.399800479 |
| 7.357579342 | 16.111880252 | 21.399800479 |
| 12.546678663 | 4.160680046 | 21.399800479 |
| 12.546678663 | 10.136279614 | 21.399800479 |
| 12.546678663 | 16.111880252 | 21.399800479 |
| 3.036469634 | 2.666779887 | 19.131700695 |
| 3.036469634 | 8.642379989 | 19.131700695 |
| 3.036469634 | 14.617979558 | 19.131700695 |
| 8.225569650 | 2.666779887 | 19.131700695 |
| 8.225569650 | 8.642379989 | 19.131700695 |
| 8.225569650 | 14.617979558 | 19.131700695 |
| 13.414668970 | 2.666779887 | 19.131700695 |
| 13.414668970 | 8.642379989 | 19.131700695 |
| 13.414668970 | 14.617979558 | 19.131700695 |
| 3.036469634 | 5.654580205 | 19.131700695 |
| 3.036469634 | 11.630180308 | 19.131700695 |
| 3.036469634 | 17.605779877 | 19.131700695 |
| 8.225569650 | 5.654580205 | 19.131700695 |
| 8.225569650 | 11.630180308 | 19.131700695 |
| 8.225569650 | 17.605779877 | 19.131700695 |
| 13.414668970 | 5.654580205 | 19.131700695 |
| 13.414668970 | 11.630180308 | 19.131700695 |
| 13.414668970 | 17.605779877 | 19.131700695 |
| 15.141231570 | 2.666779887 | 21.399800479 |
| 15.141231570 | 8.642379989 | 21.399800479 |
| 15.141231570 | 14.617979558 | 21.399800479 |
| 4.763032466 | 2.666779887 | 21.399800479 |
| 4.763032466 | 8.642379989 | 21.399800479 |
| 4.763032466 | 14.617979558 | 21.399800479 |
| 9.952132250 | 2.666779887 | 21.399800479 |
| 9.952132250 | 8.642379989 | 21.399800479 |
| 9.952132250 | 14.617979558 | 21.399800479 |
| 15.141231570 | 5.654580205 | 21.399800479 |
| 15.141231570 | 11.630180308 | 21.399800479 |
| 15.141231570 | 17.605779877 | 21.399800479 |
| 4.763032466 | 5.654580205 | 21.399800479 |
| 4.763032466 | 11.630180308 | 21.399800479 |
| 4.763032466 | 17.605779877 | 21.399800479 |
| 9.952132250 | 5.654580205 | 21.399800479 |
| 9.952132250 | 11.630180308 | 21.399800479 |
| 9.952132250 | 17.605779877 | 21.399800479 |
| 1.310279927 | 2.666779887 | 16.812799871 |

| | | |
|---|---|---|
| 1.310279927 | 8.642379989 | 16.812799871 |
| 1.310279927 | 14.617979558 | 16.812799871 |
| 6.499379595 | 2.666779887 | 16.812799871 |
| 6.499379595 | 8.642379989 | 16.812799871 |
| 6.499379595 | 14.617979558 | 16.812799871 |
| 11.688479379 | 2.666779887 | 16.812799871 |
| 11.688479379 | 8.642379989 | 16.812799871 |
| 11.688479379 | 14.617979558 | 16.812799871 |
| 1.300330002 | 2.666779887 | 23.676000535 |
| 1.300330002 | 8.642379989 | 23.676000535 |
| 1.300330002 | 14.617979558 | 23.676000535 |
| 6.489429438 | 2.666779887 | 23.676000535 |
| 6.489429438 | 8.642379989 | 23.676000535 |
| 6.489429438 | 14.617979558 | 23.676000535 |
| 11.678529687 | 2.666779887 | 23.676000535 |
| 11.678529687 | 8.642379989 | 23.676000535 |
| 11.678529687 | 14.617979558 | 23.676000535 |
| 1.310279927 | 5.654580205 | 16.812799871 |
| 1.310279927 | 11.630180308 | 16.812799871 |
| 1.310279927 | 17.605779877 | 16.812799871 |
| 6.499379595 | 5.654580205 | 16.812799871 |
| 6.499379595 | 11.630180308 | 16.812799871 |
| 6.499379595 | 17.605779877 | 16.812799871 |
| 11.688479379 | 5.654580205 | 16.812799871 |
| 11.688479379 | 11.630180308 | 16.812799871 |
| 11.688479379 | 17.605779877 | 16.812799871 |
| 1.300330002 | 5.654580205 | 23.676000535 |
| 1.300330002 | 11.630180308 | 23.676000535 |
| 1.300330002 | 17.605779877 | 23.676000535 |
| 6.489429438 | 5.654580205 | 23.676000535 |
| 6.489429438 | 11.630180308 | 23.676000535 |
| 6.489429438 | 17.605779877 | 23.676000535 |
| 11.678529687 | 5.654580205 | 23.676000535 |
| 11.678529687 | 11.630180308 | 23.676000535 |
| 11.678529687 | 17.605779877 | 23.676000535 |
| 15.139013001 | 2.666749968 | 27.767300606 |
| 15.139013001 | 8.642350071 | 27.767300606 |
| 15.139013001 | 14.617949639 | 27.767300606 |
| 4.760813897 | 2.666749968 | 27.767300606 |
| 4.760813897 | 8.642350071 | 27.767300606 |
| 4.760813897 | 14.617949639 | 27.767300606 |
| 9.949913681 | 2.666749968 | 27.767300606 |
| 9.949913681 | 8.642350071 | 27.767300606 |
| 9.949913681 | 14.617949639 | 27.767300606 |
| 15.139013001 | 5.654549752 | 27.767300606 |

| | | |
|---|---|---|
| 15.139013001 | 11.630149321 | 27.767300606 |
| 15.139013001 | 17.605749958 | 27.767300606 |
| 4.760813897 | 5.654549752 | 27.767300606 |
| 4.760813897 | 11.630149321 | 27.767300606 |
| 4.760813897 | 17.605749958 | 27.767300606 |
| 9.949913681 | 5.654549752 | 27.767300606 |
| 9.949913681 | 11.630149321 | 27.767300606 |
| 9.949913681 | 17.605749958 | 27.767300606 |
| 2.166259829 | 1.172900029 | 27.767300606 |
| 2.166259829 | 7.148500132 | 27.767300606 |
| 2.166259829 | 13.124100235 | 27.767300606 |
| 7.355359381 | 1.172900029 | 27.767300606 |
| 7.355359381 | 7.148500132 | 27.767300606 |
| 7.355359381 | 13.124100235 | 27.767300606 |
| 12.544459166 | 1.172900029 | 27.767300606 |
| 12.544459166 | 7.148500132 | 27.767300606 |
| 12.544459166 | 13.124100235 | 27.767300606 |
| 2.166259829 | 4.160700081 | 27.767300606 |
| 2.166259829 | 10.136299916 | 27.767300606 |
| 2.166259829 | 16.111899485 | 27.767300606 |
| 7.355359381 | 4.160700081 | 27.767300606 |
| 7.355359381 | 10.136299916 | 27.767300606 |
| 7.355359381 | 16.111899485 | 27.767300606 |
| 12.544459166 | 4.160700081 | 27.767300606 |
| 12.544459166 | 10.136299916 | 27.767300606 |
| 12.544459166 | 16.111899485 | 27.767300606 |
| 1.300399825 | 2.666869909 | 29.409000278 |
| 1.300399825 | 8.642469745 | 29.409000278 |
| 1.300399825 | 14.618069314 | 29.409000278 |
| 6.489499494 | 2.666869909 | 29.409000278 |
| 6.489499494 | 8.642469745 | 29.409000278 |
| 6.489499494 | 14.618069314 | 29.409000278 |
| 11.678599278 | 2.666869909 | 29.409000278 |
| 11.678599278 | 8.642469745 | 29.409000278 |
| 11.678599278 | 14.618069314 | 29.409000278 |
| 1.300399825 | 5.654669961 | 29.409000278 |
| 1.300399825 | 11.630270064 | 29.409000278 |
| 1.300399825 | 17.605869632 | 29.409000278 |
| 6.489499494 | 5.654669961 | 29.409000278 |
| 6.489499494 | 11.630270064 | 29.409000278 |
| 6.489499494 | 17.605869632 | 29.409000278 |
| 11.678599278 | 5.654669961 | 29.409000278 |
| 11.678599278 | 11.630270064 | 29.409000278 |
| 11.678599278 | 17.605869632 | 29.409000278 |
| 3.894949833 | 1.172789972 | 29.409000278 |

| | | |
|---|---|---|
| 3.894949833 | 7.148390074 | 29.409000278 |
| 3.894949833 | 13.123990177 | 29.409000278 |
| 9.084049154 | 1.172789972 | 29.409000278 |
| 9.084049154 | 7.148390074 | 29.409000278 |
| 9.084049154 | 13.123990177 | 29.409000278 |
| 14.273149402 | 1.172789972 | 29.409000278 |
| 14.273149402 | 7.148390074 | 29.409000278 |
| 14.273149402 | 13.123990177 | 29.409000278 |
| 3.894949833 | 4.160590023 | 29.409000278 |
| 3.894949833 | 10.136189859 | 29.409000278 |
| 3.894949833 | 16.111789427 | 29.409000278 |
| 9.084049154 | 4.160590023 | 29.409000278 |
| 9.084049154 | 10.136189859 | 29.409000278 |
| 9.084049154 | 16.111789427 | 29.409000278 |
| 14.273149402 | 4.160590023 | 29.409000278 |
| 14.273149402 | 10.136189859 | 29.409000278 |
| 14.273149402 | 16.111789427 | 29.409000278 |
| 1.300379876 | 2.666790038 | 26.137998700 |
| 1.300379876 | 8.642390140 | 26.137998700 |
| 1.300379876 | 14.617990243 | 26.137998700 |
| 6.489479544 | 2.666790038 | 26.137998700 |
| 6.489479544 | 8.642390140 | 26.137998700 |
| 6.489479544 | 14.617990243 | 26.137998700 |
| 11.678578864 | 2.666790038 | 26.137998700 |
| 11.678578864 | 8.642390140 | 26.137998700 |
| 11.678578864 | 14.617990243 | 26.137998700 |
| 1.300379876 | 5.654589822 | 26.137998700 |
| 1.300379876 | 11.630189925 | 26.137998700 |
| 1.300379876 | 17.605789493 | 26.137998700 |
| 6.489479544 | 5.654589822 | 26.137998700 |
| 6.489479544 | 11.630189925 | 26.137998700 |
| 6.489479544 | 17.605789493 | 26.137998700 |
| 11.678578864 | 5.654589822 | 26.137998700 |
| 11.678578864 | 11.630189925 | 26.137998700 |
| 11.678578864 | 17.605789493 | 26.137998700 |
| 3.894929884 | 1.172859960 | 26.137998700 |
| 3.894929884 | 7.148460063 | 26.137998700 |
| 3.894929884 | 13.124059631 | 26.137998700 |
| 9.084029668 | 1.172859960 | 26.137998700 |
| 9.084029668 | 7.148460063 | 26.137998700 |
| 9.084029668 | 13.124059631 | 26.137998700 |
| 14.273128989 | 1.172859960 | 26.137998700 |
| 14.273128989 | 7.148460063 | 26.137998700 |
| 14.273128989 | 13.124059631 | 26.137998700 |
| 3.894929884 | 4.160660011 | 26.137998700 |

| | | |
|---|---|---|
| 3.894929884 | 10.136260381 | 26.137998700 |
| 3.894929884 | 16.111859950 | 26.137998700 |
| 9.084029668 | 4.160660011 | 26.137998700 |
| 9.084029668 | 10.136260381 | 26.137998700 |
| 9.084029668 | 16.111859950 | 26.137998700 |
| 14.273128989 | 4.160660011 | 26.137998700 |
| 14.273128989 | 10.136260381 | 26.137998700 |
| 14.273128989 | 16.111859950 | 26.137998700 |
| 0.430496947 | 1.183179998 | 32.341098785 |
| 0.430496947 | 7.158779834 | 32.341098785 |
| 0.430496947 | 13.134379402 | 32.341098785 |
| 5.619596354 | 1.183179998 | 32.341098785 |
| 5.619596354 | 7.158779834 | 32.341098785 |
| 5.619596354 | 13.134379402 | 32.341098785 |
| 10.808696603 | 1.183179998 | 32.341098785 |
| 10.808696603 | 7.158779834 | 32.341098785 |
| 10.808696603 | 13.134379402 | 32.341098785 |
| 3.025029760 | 2.656389860 | 32.341098785 |
| 3.025029760 | 8.631989696 | 32.341098785 |
| 3.025029760 | 14.607589264 | 32.341098785 |
| 8.214129776 | 2.656389860 | 32.341098785 |
| 8.214129776 | 8.631989696 | 32.341098785 |
| 8.214129776 | 14.607589264 | 32.341098785 |
| 13.403229097 | 2.656389860 | 32.341098785 |
| 13.403229097 | 8.631989696 | 32.341098785 |
| 13.403229097 | 14.607589264 | 32.341098785 |
| 0.430496947 | 4.170969899 | 32.341098785 |
| 0.430496947 | 10.146569467 | 32.341098785 |
| 0.430496947 | 16.122170104 | 32.341098785 |
| 5.619596354 | 4.170969899 | 32.341098785 |
| 5.619596354 | 10.146569467 | 32.341098785 |
| 5.619596354 | 16.122170104 | 32.341098785 |
| 10.808696603 | 4.170969899 | 32.341098785 |
| 10.808696603 | 10.146569467 | 32.341098785 |
| 10.808696603 | 16.122170104 | 32.341098785 |
| 3.025029760 | 5.644179761 | 32.341098785 |
| 3.025029760 | 11.619779329 | 32.341098785 |
| 3.025029760 | 17.595379966 | 32.341098785 |
| 8.214129776 | 5.644179761 | 32.341098785 |
| 8.214129776 | 11.619779329 | 32.341098785 |
| 8.214129776 | 17.595379966 | 32.341098785 |
| 13.403229097 | 5.644179761 | 32.341098785 |
| 13.403229097 | 11.619779329 | 32.341098785 |
| 13.403229097 | 17.595379966 | 32.341098785 |
| 1.295339845 | 5.657970087 | 34.130498767 |

| | | |
|---|---|---|
| 1.295339845 | 11.633569655 | 34.130498767 |
| 1.295339845 | 17.609170292 | 34.130498767 |
| 6.484439282 | 5.657970087 | 34.130498767 |
| 6.484439282 | 11.633569655 | 34.130498767 |
| 6.484439282 | 17.609170292 | 34.130498767 |
| 11.673539530 | 5.657970087 | 34.130498767 |
| 11.673539530 | 11.633569655 | 34.130498767 |
| 11.673539530 | 17.609170292 | 34.130498767 |
| 3.889879647 | 1.155600014 | 34.130498767 |
| 3.889879647 | 7.131199716 | 34.130498767 |
| 3.889879647 | 13.106799819 | 34.130498767 |
| 9.078979199 | 1.155600014 | 34.130498767 |
| 9.078979199 | 7.131199716 | 34.130498767 |
| 9.078979199 | 13.106799819 | 34.130498767 |
| 14.268078520 | 1.155600014 | 34.130498767 |
| 14.268078520 | 7.131199716 | 34.130498767 |
| 14.268078520 | 13.106799819 | 34.130498767 |
| 1.295339845 | 2.670179919 | 34.130600095 |
| 1.295339845 | 8.645780022 | 34.130600095 |
| 1.295339845 | 14.621379590 | 34.130600095 |
| 6.484439282 | 2.670179919 | 34.130600095 |
| 6.484439282 | 8.645780022 | 34.130600095 |
| 6.484439282 | 14.621379590 | 34.130600095 |
| 11.673539530 | 2.670179919 | 34.130600095 |
| 11.673539530 | 8.645780022 | 34.130600095 |
| 11.673539530 | 14.621379590 | 34.130600095 |
| 3.889879647 | 4.143399932 | 34.130498767 |
| 3.889879647 | 10.118999500 | 34.130498767 |
| 3.889879647 | 16.094600137 | 34.130498767 |
| 9.078979199 | 4.143399932 | 34.130498767 |
| 9.078979199 | 10.118999500 | 34.130498767 |
| 9.078979199 | 16.094600137 | 34.130498767 |
| 14.268078520 | 4.143399932 | 34.130498767 |
| 14.268078520 | 10.118999500 | 34.130498767 |
| 14.268078520 | 16.094600137 | 34.130498767 |
| 1.295339845 | 5.657970087 | 36.510300636 |
| 1.295339845 | 11.633569655 | 36.510300636 |
| 1.295339845 | 17.609170292 | 36.510300636 |
| 6.484439282 | 5.657970087 | 36.510300636 |
| 6.484439282 | 11.633569655 | 36.510300636 |
| 6.484439282 | 17.609170292 | 36.510300636 |
| 11.673539530 | 5.657970087 | 36.510300636 |
| 11.673539530 | 11.633569655 | 36.510300636 |
| 11.673539530 | 17.609170292 | 36.510300636 |
| 3.889879647 | 1.155600014 | 36.510199308 |

```
   3.889879647          7.131199716         36.510199308
   3.889879647         13.106799819         36.510199308
   9.078979199          1.155600014         36.510199308
   9.078979199          7.131199716         36.510199308
   9.078979199         13.106799819         36.510199308
  14.268078520          1.155600014         36.510199308
  14.268078520          7.131199716         36.510199308
  14.268078520         13.106799819         36.510199308
   1.295339845          2.670179919         36.510300636
   1.295339845          8.645780022         36.510300636
   1.295339845         14.621379590         36.510300636
   6.484439282          2.670179919         36.510300636
   6.484439282          8.645780022         36.510300636
   6.484439282         14.621379590         36.510300636
  11.673539530          2.670179919         36.510300636
  11.673539530          8.645780022         36.510300636
  11.673539530         14.621379590         36.510300636
   3.889879647          4.143399932         36.510199308
   3.889879647         10.118999500         36.510199308
   3.889879647         16.094600137         36.510199308
   9.078979199          4.143399932         36.510199308
   9.078979199         10.118999500         36.510199308
   9.078979199         16.094600137         36.510199308
  14.268078520          4.143399932         36.510199308
  14.268078520         10.118999500         36.510199308
  14.268078520         16.094600137         36.510199308


Au/MoSe2/Si
1.0
      15.9442605972         0.0000000000          0.0000000000
       0.0000000000        18.3540554047          0.0000000000
       0.0000000000         0.0000000000         50.0000000000
   Au    Mo    Se    Si
  144    36    72   108
Cartesian
    0.000000000          0.000000000         20.000000298
    0.000000000          6.118018651         20.000000298
    0.000000000         12.236037301         20.000000298
    5.314753691          0.000000000         20.000000298
    5.314753691          6.118018651         20.000000298
    5.314753691         12.236037301         20.000000298
   10.629507382          0.000000000         20.000000298
   10.629507382          6.118018651         20.000000298
   10.629507382         12.236037301         20.000000298
    5.185026843          6.001260705         26.669687033
```

| | | |
|---|---|---|
| 5.185026843 | 12.119279356 | 26.669687033 |
| 5.185026843 | 18.237297459 | 26.669687033 |
| 10.499780534 | 6.001260705 | 26.669687033 |
| 10.499780534 | 12.119279356 | 26.669687033 |
| 10.499780534 | 18.237297459 | 26.669687033 |
| 15.814533749 | 6.001260705 | 26.669687033 |
| 15.814533749 | 12.119279356 | 26.669687033 |
| 15.814533749 | 18.237297459 | 26.669687033 |
| 0.000000000 | 3.059009325 | 20.000000298 |
| 0.000000000 | 9.177027702 | 20.000000298 |
| 0.000000000 | 15.295045806 | 20.000000298 |
| 5.314753691 | 3.059009325 | 20.000000298 |
| 5.314753691 | 9.177027702 | 20.000000298 |
| 5.314753691 | 15.295045806 | 20.000000298 |
| 10.629507382 | 3.059009325 | 20.000000298 |
| 10.629507382 | 9.177027702 | 20.000000298 |
| 10.629507382 | 15.295045806 | 20.000000298 |
| 5.185026843 | 2.942251380 | 26.669687033 |
| 5.185026843 | 9.060269757 | 26.669687033 |
| 5.185026843 | 15.178287860 | 26.669687033 |
| 10.499780534 | 2.942251380 | 26.669687033 |
| 10.499780534 | 9.060269757 | 26.669687033 |
| 10.499780534 | 15.178287860 | 26.669687033 |
| 15.814533749 | 2.942251380 | 26.669687033 |
| 15.814533749 | 9.060269757 | 26.669687033 |
| 15.814533749 | 15.178287860 | 26.669687033 |
| 1.726741453 | 6.070536335 | 22.249865532 |
| 1.726741453 | 12.188554985 | 22.249865532 |
| 1.726741453 | 18.306573089 | 22.249865532 |
| 7.041494787 | 6.070536335 | 22.249865532 |
| 7.041494787 | 12.188554985 | 22.249865532 |
| 7.041494787 | 18.306573089 | 22.249865532 |
| 12.356248478 | 6.070536335 | 22.249865532 |
| 12.356248478 | 12.188554985 | 22.249865532 |
| 12.356248478 | 18.306573089 | 22.249865532 |
| 1.726741453 | 3.011527557 | 22.249865532 |
| 1.726741453 | 9.129545934 | 22.249865532 |
| 1.726741453 | 15.247564584 | 22.249865532 |
| 7.041494787 | 3.011527557 | 22.249865532 |
| 7.041494787 | 9.129545934 | 22.249865532 |
| 7.041494787 | 15.247564584 | 22.249865532 |
| 12.356248478 | 3.011527557 | 22.249865532 |
| 12.356248478 | 9.129545934 | 22.249865532 |
| 12.356248478 | 15.247564584 | 22.249865532 |
| 3.452869453 | 6.037171921 | 24.451269209 |

| | | |
|---|---|---|
| 3.452869453 | 12.155190571 | 24.451269209 |
| 3.452869453 | 18.273208675 | 24.451269209 |
| 8.767623381 | 6.037171921 | 24.451269209 |
| 8.767623381 | 12.155190571 | 24.451269209 |
| 8.767623381 | 18.273208675 | 24.451269209 |
| 14.082376597 | 6.037171921 | 24.451269209 |
| 14.082376597 | 12.155190571 | 24.451269209 |
| 14.082376597 | 18.273208675 | 24.451269209 |
| 3.452869453 | 2.978162595 | 24.451269209 |
| 3.452869453 | 9.096180973 | 24.451269209 |
| 3.452869453 | 15.214199076 | 24.451269209 |
| 8.767623381 | 2.978162595 | 24.451269209 |
| 8.767623381 | 9.096180973 | 24.451269209 |
| 8.767623381 | 15.214199076 | 24.451269209 |
| 14.082376597 | 2.978162595 | 24.451269209 |
| 14.082376597 | 9.096180973 | 24.451269209 |
| 14.082376597 | 15.214199076 | 24.451269209 |
| 4.386159060 | 1.483371096 | 22.249738872 |
| 4.386159060 | 7.601389473 | 22.249738872 |
| 4.386159060 | 13.719408124 | 22.249738872 |
| 9.700912751 | 1.483371096 | 22.249738872 |
| 9.700912751 | 7.601389473 | 22.249738872 |
| 9.700912751 | 13.719408124 | 22.249738872 |
| 15.015665967 | 1.483371096 | 22.249738872 |
| 15.015665967 | 7.601389473 | 22.249738872 |
| 15.015665967 | 13.719408124 | 22.249738872 |
| 4.386159060 | 4.542380695 | 22.249738872 |
| 4.386159060 | 10.660399619 | 22.249738872 |
| 4.386159060 | 16.778417722 | 22.249738872 |
| 9.700912751 | 4.542380695 | 22.249738872 |
| 9.700912751 | 10.660399619 | 22.249738872 |
| 9.700912751 | 16.778417722 | 22.249738872 |
| 15.015665967 | 4.542380695 | 22.249738872 |
| 15.015665967 | 10.660399619 | 22.249738872 |
| 15.015665967 | 16.778417722 | 22.249738872 |
| 0.796382790 | 1.444921420 | 24.454124272 |
| 0.796382790 | 7.562939661 | 24.454124272 |
| 0.796382790 | 13.680958858 | 24.454124272 |
| 6.111136303 | 1.444921420 | 24.454124272 |
| 6.111136303 | 7.562939661 | 24.454124272 |
| 6.111136303 | 13.680958858 | 24.454124272 |
| 11.425889994 | 1.444921420 | 24.454124272 |
| 11.425889994 | 7.562939661 | 24.454124272 |
| 11.425889994 | 13.680958858 | 24.454124272 |
| 0.796382790 | 4.503931156 | 24.454124272 |

| | | |
|---|---|---|
| 0.796382790 | 10.621949259 | 24.454124272 |
| 0.796382790 | 16.739968457 | 24.454124272 |
| 6.111136303 | 4.503931156 | 24.454124272 |
| 6.111136303 | 10.621949259 | 24.454124272 |
| 6.111136303 | 16.739968457 | 24.454124272 |
| 11.425889994 | 4.503931156 | 24.454124272 |
| 11.425889994 | 10.621949259 | 24.454124272 |
| 11.425889994 | 16.739968457 | 24.454124272 |
| 2.657376845 | 1.529504663 | 20.000000298 |
| 2.657376845 | 7.647522903 | 20.000000298 |
| 2.657376845 | 13.765541553 | 20.000000298 |
| 7.972130299 | 1.529504663 | 20.000000298 |
| 7.972130299 | 7.647522903 | 20.000000298 |
| 7.972130299 | 13.765541553 | 20.000000298 |
| 13.286883514 | 1.529504663 | 20.000000298 |
| 13.286883514 | 7.647522903 | 20.000000298 |
| 13.286883514 | 13.765541553 | 20.000000298 |
| 2.531148718 | 1.412257705 | 26.669946313 |
| 2.531148718 | 7.530275945 | 26.669946313 |
| 2.531148718 | 13.648294596 | 26.669946313 |
| 7.845902172 | 1.412257705 | 26.669946313 |
| 7.845902172 | 7.530275945 | 26.669946313 |
| 7.845902172 | 13.648294596 | 26.669946313 |
| 13.160655862 | 1.412257705 | 26.669946313 |
| 13.160655862 | 7.530275945 | 26.669946313 |
| 13.160655862 | 13.648294596 | 26.669946313 |
| 2.657376845 | 4.588514398 | 20.000000298 |
| 2.657376845 | 10.706533049 | 20.000000298 |
| 2.657376845 | 16.824551152 | 20.000000298 |
| 7.972130299 | 4.588514398 | 20.000000298 |
| 7.972130299 | 10.706533049 | 20.000000298 |
| 7.972130299 | 16.824551152 | 20.000000298 |
| 13.286883514 | 4.588514398 | 20.000000298 |
| 13.286883514 | 10.706533049 | 20.000000298 |
| 13.286883514 | 16.824551152 | 20.000000298 |
| 2.531148718 | 4.471267167 | 26.669946313 |
| 2.531148718 | 10.589286091 | 26.669946313 |
| 2.531148718 | 16.707304195 | 26.669946313 |
| 7.845902172 | 4.471267167 | 26.669946313 |
| 7.845902172 | 10.589286091 | 26.669946313 |
| 7.845902172 | 16.707304195 | 26.669946313 |
| 13.160655862 | 4.471267167 | 26.669946313 |
| 13.160655862 | 10.589286091 | 26.669946313 |
| 13.160655862 | 16.707304195 | 26.669946313 |
| 0.720179648 | 1.334453490 | 31.018465757 |

| | | |
|---|---|---|
| 0.720179648 | 7.452472140 | 31.018465757 |
| 0.720179648 | 13.570490244 | 31.018465757 |
| 6.034933279 | 1.334453490 | 31.018465757 |
| 6.034933279 | 7.452472140 | 31.018465757 |
| 6.034933279 | 13.570490244 | 31.018465757 |
| 11.349686970 | 1.334453490 | 31.018465757 |
| 11.349686970 | 7.452472140 | 31.018465757 |
| 11.349686970 | 13.570490244 | 31.018465757 |
| 0.720179648 | 4.393462815 | 31.018465757 |
| 0.720179648 | 10.511481739 | 31.018465757 |
| 0.720179648 | 16.629499843 | 31.018465757 |
| 6.034933279 | 4.393462815 | 31.018465757 |
| 6.034933279 | 10.511481739 | 31.018465757 |
| 6.034933279 | 16.629499843 | 31.018465757 |
| 11.349686970 | 4.393462815 | 31.018465757 |
| 11.349686970 | 10.511481739 | 31.018465757 |
| 11.349686970 | 16.629499843 | 31.018465757 |
| 3.377824671 | 5.922980469 | 31.018307805 |
| 3.377824671 | 12.040999119 | 31.018307805 |
| 3.377824671 | 18.159017223 | 31.018307805 |
| 8.692577886 | 5.922980469 | 31.018307805 |
| 8.692577886 | 12.040999119 | 31.018307805 |
| 8.692577886 | 18.159017223 | 31.018307805 |
| 14.007332052 | 5.922980469 | 31.018307805 |
| 14.007332052 | 12.040999119 | 31.018307805 |
| 14.007332052 | 18.159017223 | 31.018307805 |
| 3.377824671 | 2.863971417 | 31.018307805 |
| 3.377824671 | 8.981990067 | 31.018307805 |
| 3.377824671 | 15.100008718 | 31.018307805 |
| 8.692577886 | 2.863971417 | 31.018307805 |
| 8.692577886 | 8.981990067 | 31.018307805 |
| 8.692577886 | 15.100008718 | 31.018307805 |
| 14.007332052 | 2.863971417 | 31.018307805 |
| 14.007332052 | 8.981990067 | 31.018307805 |
| 14.007332052 | 15.100008718 | 31.018307805 |
| 2.489290940 | 1.332780647 | 32.792428136 |
| 2.489290940 | 7.450798887 | 32.792428136 |
| 2.489290940 | 13.568817538 | 32.792428136 |
| 7.804044393 | 1.332780647 | 32.792428136 |
| 7.804044393 | 7.450798887 | 32.792428136 |
| 7.804044393 | 13.568817538 | 32.792428136 |
| 13.118797609 | 1.332780647 | 32.792428136 |
| 13.118797609 | 7.450798887 | 32.792428136 |
| 13.118797609 | 13.568817538 | 32.792428136 |
| 2.489290940 | 4.391790109 | 32.792428136 |

| | | |
|---|---|---|
| 2.489290940 | 10.509809033 | 32.792428136 |
| 2.489290940 | 16.627827136 | 32.792428136 |
| 7.804044393 | 4.391790109 | 32.792428136 |
| 7.804044393 | 10.509809033 | 32.792428136 |
| 7.804044393 | 16.627827136 | 32.792428136 |
| 13.118797609 | 4.391790109 | 32.792428136 |
| 13.118797609 | 10.509809033 | 32.792428136 |
| 13.118797609 | 16.627827136 | 32.792428136 |
| 5.146511928 | 5.921592199 | 32.792788744 |
| 5.146511928 | 12.039610850 | 32.792788744 |
| 5.146511928 | 18.157628953 | 32.792788744 |
| 10.461265619 | 5.921592199 | 32.792788744 |
| 10.461265619 | 12.039610850 | 32.792788744 |
| 10.461265619 | 18.157628953 | 32.792788744 |
| 15.776018834 | 5.921592199 | 32.792788744 |
| 15.776018834 | 12.039610850 | 32.792788744 |
| 15.776018834 | 18.157628953 | 32.792788744 |
| 5.146511928 | 2.862582874 | 32.792788744 |
| 5.146511928 | 8.980601251 | 32.792788744 |
| 5.146511928 | 15.098619355 | 32.792788744 |
| 10.461265619 | 2.862582874 | 32.792788744 |
| 10.461265619 | 8.980601251 | 32.792788744 |
| 10.461265619 | 15.098619355 | 32.792788744 |
| 15.776018834 | 2.862582874 | 32.792788744 |
| 15.776018834 | 8.980601251 | 32.792788744 |
| 15.776018834 | 15.098619355 | 32.792788744 |
| 2.493253195 | 1.336985386 | 29.265865684 |
| 2.493253195 | 7.455003626 | 29.265865684 |
| 2.493253195 | 13.573022824 | 29.265865684 |
| 7.808006886 | 1.336985386 | 29.265865684 |
| 7.808006886 | 7.455003626 | 29.265865684 |
| 7.808006886 | 13.573022824 | 29.265865684 |
| 13.122760577 | 1.336985386 | 29.265865684 |
| 13.122760577 | 7.455003626 | 29.265865684 |
| 13.122760577 | 13.573022824 | 29.265865684 |
| 2.493253195 | 4.395994848 | 29.265865684 |
| 2.493253195 | 10.514013225 | 29.265865684 |
| 2.493253195 | 16.632031328 | 29.265865684 |
| 7.808006886 | 4.395994848 | 29.265865684 |
| 7.808006886 | 10.514013225 | 29.265865684 |
| 7.808006886 | 16.632031328 | 29.265865684 |
| 13.122760577 | 4.395994848 | 29.265865684 |
| 13.122760577 | 10.514013225 | 29.265865684 |
| 13.122760577 | 16.632031328 | 29.265865684 |
| 5.150610796 | 5.925518518 | 29.265716672 |

| | | |
|---|---|---|
| 5.150610796 | 12.043537169 | 29.265716672 |
| 5.150610796 | 18.161555273 | 29.265716672 |
| 10.465364487 | 5.925518518 | 29.265716672 |
| 10.465364487 | 12.043537169 | 29.265716672 |
| 10.465364487 | 18.161555273 | 29.265716672 |
| 15.780117702 | 5.925518518 | 29.265716672 |
| 15.780117702 | 12.043537169 | 29.265716672 |
| 15.780117702 | 18.161555273 | 29.265716672 |
| 5.150610796 | 2.866509740 | 29.265716672 |
| 5.150610796 | 8.984528117 | 29.265716672 |
| 5.150610796 | 15.102546768 | 29.265716672 |
| 10.465364487 | 2.866509740 | 29.265716672 |
| 10.465364487 | 8.984528117 | 29.265716672 |
| 10.465364487 | 15.102546768 | 29.265716672 |
| 15.780117702 | 2.866509740 | 29.265716672 |
| 15.780117702 | 8.984528117 | 29.265716672 |
| 15.780117702 | 15.102546768 | 29.265716672 |
| 0.440924002 | 1.211381598 | 35.891878605 |
| 0.440924002 | 7.329400249 | 35.891878605 |
| 0.440924002 | 13.447418900 | 35.891878605 |
| 5.755677574 | 1.211381598 | 35.891878605 |
| 5.755677574 | 7.329400249 | 35.891878605 |
| 5.755677574 | 13.447418900 | 35.891878605 |
| 11.070430790 | 1.211381598 | 35.891878605 |
| 11.070430790 | 7.329400249 | 35.891878605 |
| 11.070430790 | 13.447418900 | 35.891878605 |
| 3.098300729 | 2.719712554 | 35.891494155 |
| 3.098300729 | 8.837730931 | 35.891494155 |
| 3.098300729 | 14.955749035 | 35.891494155 |
| 8.413054657 | 2.719712554 | 35.891494155 |
| 8.413054657 | 8.837730931 | 35.891494155 |
| 8.413054657 | 14.955749035 | 35.891494155 |
| 13.727807873 | 2.719712554 | 35.891494155 |
| 13.727807873 | 8.837730931 | 35.891494155 |
| 13.727807873 | 14.955749035 | 35.891494155 |
| 0.440924002 | 4.270391197 | 35.891878605 |
| 0.440924002 | 10.388409301 | 35.891878605 |
| 0.440924002 | 16.506428498 | 35.891878605 |
| 5.755677574 | 4.270391197 | 35.891878605 |
| 5.755677574 | 10.388409301 | 35.891878605 |
| 5.755677574 | 16.506428498 | 35.891878605 |
| 11.070430790 | 4.270391197 | 35.891878605 |
| 11.070430790 | 10.388409301 | 35.891878605 |
| 11.070430790 | 16.506428498 | 35.891878605 |
| 3.098300729 | 5.778721879 | 35.891494155 |

| | | |
|---|---|---|
| 3.098300729 | 11.896740530 | 35.891494155 |
| 3.098300729 | 18.014758633 | 35.891494155 |
| 8.413054657 | 5.778721879 | 35.891494155 |
| 8.413054657 | 11.896740530 | 35.891494155 |
| 8.413054657 | 18.014758633 | 35.891494155 |
| 13.727807873 | 5.778721879 | 35.891494155 |
| 13.727807873 | 11.896740530 | 35.891494155 |
| 13.727807873 | 18.014758633 | 35.891494155 |
| 1.326716324 | 5.792836499 | 37.640830874 |
| 1.326716324 | 11.910855149 | 37.640830874 |
| 1.326716324 | 18.028873253 | 37.640830874 |
| 6.641469777 | 5.792836499 | 37.640830874 |
| 6.641469777 | 11.910855149 | 37.640830874 |
| 6.641469777 | 18.028873253 | 37.640830874 |
| 11.956223468 | 5.792836499 | 37.640830874 |
| 11.956223468 | 11.910855149 | 37.640830874 |
| 11.956223468 | 18.028873253 | 37.640830874 |
| 3.984093169 | 1.183149898 | 37.640309334 |
| 3.984093169 | 7.301168275 | 37.640309334 |
| 3.984093169 | 13.419186378 | 37.640309334 |
| 9.298846385 | 1.183149898 | 37.640309334 |
| 9.298846385 | 7.301168275 | 37.640309334 |
| 9.298846385 | 13.419186378 | 37.640309334 |
| 14.613600551 | 1.183149898 | 37.640309334 |
| 14.613600551 | 7.301168275 | 37.640309334 |
| 14.613600551 | 13.419186378 | 37.640309334 |
| 1.326716324 | 2.733827994 | 37.640830874 |
| 1.326716324 | 8.851846645 | 37.640830874 |
| 1.326716324 | 14.969864748 | 37.640830874 |
| 6.641469777 | 2.733827994 | 37.640830874 |
| 6.641469777 | 8.851846645 | 37.640830874 |
| 6.641469777 | 14.969864748 | 37.640830874 |
| 11.956223468 | 2.733827994 | 37.640830874 |
| 11.956223468 | 8.851846645 | 37.640830874 |
| 11.956223468 | 14.969864748 | 37.640830874 |
| 3.984093169 | 4.242158949 | 37.640309334 |
| 3.984093169 | 10.360177874 | 37.640309334 |
| 3.984093169 | 16.478195977 | 37.640309334 |
| 9.298846385 | 4.242158949 | 37.640309334 |
| 9.298846385 | 10.360177874 | 37.640309334 |
| 9.298846385 | 16.478195977 | 37.640309334 |
| 14.613600551 | 4.242158949 | 37.640309334 |
| 14.613600551 | 10.360177874 | 37.640309334 |
| 14.613600551 | 16.478195977 | 37.640309334 |
| 1.326716324 | 5.792836499 | 40.030124784 |

```
1.326716324              11.910855149              40.030124784
1.326716324              18.028873253              40.030124784
6.641469777               5.792836499              40.030124784
6.641469777              11.910855149              40.030124784
6.641469777              18.028873253              40.030124784
11.956223468              5.792836499              40.030124784
11.956223468             11.910855149              40.030124784
11.956223468             18.028873253              40.030124784
3.984093169               1.183149898              40.029674768
3.984093169               7.301168275              40.029674768
3.984093169              13.419186378              40.029674768
9.298846385               1.183149898              40.029674768
9.298846385               7.301168275              40.029674768
9.298846385              13.419186378              40.029674768
14.613600551              1.183149898              40.029674768
14.613600551              7.301168275              40.029674768
14.613600551             13.419186378              40.029674768
1.326716324               2.733827994              40.030124784
1.326716324               8.851846645              40.030124784
1.326716324              14.969864748              40.030124784
6.641469777               2.733827994              40.030124784
6.641469777               8.851846645              40.030124784
6.641469777              14.969864748              40.030124784
11.956223468              2.733827994              40.030124784
11.956223468              8.851846645              40.030124784
11.956223468             14.969864748              40.030124784
3.984093169               4.242158949              40.029674768
3.984093169              10.360177874              40.029674768
3.984093169              16.478195977              40.029674768
9.298846385               4.242158949              40.029674768
9.298846385              10.360177874              40.029674768
9.298846385              16.478195977              40.029674768
14.613600551              4.242158949              40.029674768
14.613600551             10.360177874              40.029674768
14.613600551             16.478195977              40.029674768


Au/WS2/Si
1.0
        15.6610240936           0.0000000000            0.0000000000
         0.0000000000          17.9489860535            0.0000000000
         0.0000000000           0.0000000000           50.0000000000
     Au    W    S    Si
    144   36   72   108
Cartesian
       0.000000000            0.000000000           20.000000298
```

| | | |
|---|---|---|
| 0.000000000 | 5.982999809 | 20.000000298 |
| 0.000000000 | 11.965999618 | 20.000000298 |
| 5.220340120 | 0.000000000 | 20.000000298 |
| 5.220340120 | 5.982999809 | 20.000000298 |
| 5.220340120 | 11.965999618 | 20.000000298 |
| 10.440699843 | 0.000000000 | 20.000000298 |
| 10.440699843 | 5.982999809 | 20.000000298 |
| 10.440699843 | 11.965999618 | 20.000000298 |
| 5.123150047 | 5.875819734 | 26.829200983 |
| 5.123150047 | 11.858800285 | 26.829200983 |
| 5.123150047 | 17.841799559 | 26.829200983 |
| 10.343500435 | 5.875819734 | 26.829200983 |
| 10.343500435 | 11.858800285 | 26.829200983 |
| 10.343500435 | 17.841799559 | 26.829200983 |
| 15.563800416 | 5.875819734 | 26.829200983 |
| 15.563800416 | 11.858800285 | 26.829200983 |
| 15.563800416 | 17.841799559 | 26.829200983 |
| 0.000000000 | 2.991499904 | 20.000000298 |
| 0.000000000 | 8.974489817 | 20.000000298 |
| 0.000000000 | 14.957499790 | 20.000000298 |
| 5.220340120 | 2.991499904 | 20.000000298 |
| 5.220340120 | 8.974489817 | 20.000000298 |
| 5.220340120 | 14.957499790 | 20.000000298 |
| 10.440699843 | 2.991499904 | 20.000000298 |
| 10.440699843 | 8.974489817 | 20.000000298 |
| 10.440699843 | 14.957499790 | 20.000000298 |
| 5.123150047 | 2.884329993 | 26.829200983 |
| 5.123150047 | 8.867319906 | 26.829200983 |
| 5.123150047 | 14.850300457 | 26.829200983 |
| 10.343500435 | 2.884329993 | 26.829200983 |
| 10.343500435 | 8.867319906 | 26.829200983 |
| 10.343500435 | 14.850300457 | 26.829200983 |
| 15.563800416 | 2.884329993 | 26.829200983 |
| 15.563800416 | 8.867319906 | 26.829200983 |
| 15.563800416 | 14.850300457 | 26.829200983 |
| 1.702759958 | 5.940040266 | 22.307200730 |
| 1.702759958 | 11.923000490 | 22.307200730 |
| 1.702759958 | 17.905999764 | 22.307200730 |
| 6.923109879 | 5.940040266 | 22.307200730 |
| 6.923109879 | 11.923000490 | 22.307200730 |
| 6.923109879 | 17.905999764 | 22.307200730 |
| 12.143400058 | 5.940040266 | 22.307200730 |
| 12.143400058 | 11.923000490 | 22.307200730 |
| 12.143400058 | 17.905999764 | 22.307200730 |
| 1.702759958 | 2.948540094 | 22.307200730 |

| | | |
|---|---|---|
| 1.702759958 | 8.931539903 | 22.307200730 |
| 1.702759958 | 14.914499592 | 22.307200730 |
| 6.923109879 | 2.948540094 | 22.307200730 |
| 6.923109879 | 8.931539903 | 22.307200730 |
| 6.923109879 | 14.914499592 | 22.307200730 |
| 12.143400058 | 2.948540094 | 22.307200730 |
| 12.143400058 | 8.931539903 | 22.307200730 |
| 12.143400058 | 14.914499592 | 22.307200730 |
| 3.412740071 | 5.907780222 | 24.564200640 |
| 3.412740071 | 11.890800358 | 24.564200640 |
| 3.412740071 | 17.873799631 | 24.564200640 |
| 8.633080424 | 5.907780222 | 24.564200640 |
| 8.633080424 | 11.890800358 | 24.564200640 |
| 8.633080424 | 17.873799631 | 24.564200640 |
| 13.853400008 | 5.907780222 | 24.564200640 |
| 13.853400008 | 11.890800358 | 24.564200640 |
| 13.853400008 | 17.873799631 | 24.564200640 |
| 3.412740071 | 2.916289946 | 24.564200640 |
| 3.412740071 | 8.899279859 | 24.564200640 |
| 3.412740071 | 14.882300529 | 24.564200640 |
| 8.633080424 | 2.916289946 | 24.564200640 |
| 8.633080424 | 8.899279859 | 24.564200640 |
| 8.633080424 | 14.882300529 | 24.564200640 |
| 13.853400008 | 2.916289946 | 24.564200640 |
| 13.853400008 | 8.899279859 | 24.564200640 |
| 13.853400008 | 14.882300529 | 24.564200640 |
| 4.312930018 | 1.452790008 | 22.307200730 |
| 4.312930018 | 7.435789817 | 22.307200730 |
| 4.312930018 | 13.418800324 | 22.307200730 |
| 9.533280406 | 1.452790008 | 22.307200730 |
| 9.533280406 | 7.435789817 | 22.307200730 |
| 9.533280406 | 13.418800324 | 22.307200730 |
| 14.753599989 | 1.452790008 | 22.307200730 |
| 14.753599989 | 7.435789817 | 22.307200730 |
| 14.753599989 | 13.418800324 | 22.307200730 |
| 4.312930018 | 4.444289912 | 22.307200730 |
| 4.312930018 | 10.427300152 | 22.307200730 |
| 4.312930018 | 16.410300496 | 22.307200730 |
| 9.533280406 | 4.444289912 | 22.307200730 |
| 9.533280406 | 10.427300152 | 22.307200730 |
| 9.533280406 | 16.410300496 | 22.307200730 |
| 14.753599989 | 4.444289912 | 22.307200730 |
| 14.753599989 | 10.427300152 | 22.307200730 |
| 14.753599989 | 16.410300496 | 22.307200730 |
| 0.802572986 | 1.420539994 | 24.564200640 |

| | | |
|---|---|---|
| 0.802572986 | 7.403529773 | 24.564200640 |
| 0.802572986 | 13.386499626 | 24.564200640 |
| 6.022909898 | 1.420539994 | 24.564200640 |
| 6.022909898 | 7.403529773 | 24.564200640 |
| 6.022909898 | 13.386499626 | 24.564200640 |
| 11.243299958 | 1.420539994 | 24.564200640 |
| 11.243299958 | 7.403529773 | 24.564200640 |
| 11.243299958 | 13.386499626 | 24.564200640 |
| 0.802572986 | 4.412040032 | 24.564200640 |
| 0.802572986 | 10.395000524 | 24.564200640 |
| 0.802572986 | 16.377999798 | 24.564200640 |
| 6.022909898 | 4.412040032 | 24.564200640 |
| 6.022909898 | 10.395000524 | 24.564200640 |
| 6.022909898 | 16.377999798 | 24.564200640 |
| 11.243299958 | 4.412040032 | 24.564200640 |
| 11.243299958 | 10.395000524 | 24.564200640 |
| 11.243299958 | 16.377999798 | 24.564200640 |
| 2.610170060 | 1.495749952 | 20.000000298 |
| 2.610170060 | 7.478740266 | 20.000000298 |
| 2.610170060 | 13.461699956 | 20.000000298 |
| 7.830510180 | 1.495749952 | 20.000000298 |
| 7.830510180 | 7.478740266 | 20.000000298 |
| 7.830510180 | 13.461699956 | 20.000000298 |
| 13.050899774 | 1.495749952 | 20.000000298 |
| 13.050899774 | 7.478740266 | 20.000000298 |
| 13.050899774 | 13.461699956 | 20.000000298 |
| 2.512979987 | 1.388570011 | 26.829200983 |
| 2.512979987 | 7.371569820 | 26.829200983 |
| 2.512979987 | 13.354600119 | 26.829200983 |
| 7.733320107 | 1.388570011 | 26.829200983 |
| 7.733320107 | 7.371569820 | 26.829200983 |
| 7.733320107 | 13.354600119 | 26.829200983 |
| 12.953700366 | 1.388570011 | 26.829200983 |
| 12.953700366 | 7.371569820 | 26.829200983 |
| 12.953700366 | 13.354600119 | 26.829200983 |
| 2.610170060 | 4.487250258 | 20.000000298 |
| 2.610170060 | 10.470199784 | 20.000000298 |
| 2.610170060 | 16.453200128 | 20.000000298 |
| 7.830510180 | 4.487250258 | 20.000000298 |
| 7.830510180 | 10.470199784 | 20.000000298 |
| 7.830510180 | 16.453200128 | 20.000000298 |
| 13.050899774 | 4.487250258 | 20.000000298 |
| 13.050899774 | 10.470199784 | 20.000000298 |
| 13.050899774 | 16.453200128 | 20.000000298 |
| 2.512979987 | 4.380069915 | 26.829200983 |

| | | |
|---|---|---|
| 2.512979987 | 10.363099947 | 26.829200983 |
| 2.512979987 | 16.346100291 | 26.829200983 |
| 7.733320107 | 4.380069915 | 26.829200983 |
| 7.733320107 | 10.363099947 | 26.829200983 |
| 7.733320107 | 16.346100291 | 26.829200983 |
| 12.953700366 | 4.380069915 | 26.829200983 |
| 12.953700366 | 10.363099947 | 26.829200983 |
| 12.953700366 | 16.346100291 | 26.829200983 |
| 3.362009956 | 2.848309986 | 30.914399028 |
| 3.362009956 | 8.831310062 | 30.914399028 |
| 3.362009956 | 14.814300242 | 30.914399028 |
| 8.582360344 | 2.848309986 | 30.914399028 |
| 8.582360344 | 8.831310062 | 30.914399028 |
| 8.582360344 | 14.814300242 | 30.914399028 |
| 13.802700464 | 2.848309986 | 30.914399028 |
| 13.802700464 | 8.831310062 | 30.914399028 |
| 13.802700464 | 14.814300242 | 30.914399028 |
| 3.362009956 | 5.839809890 | 30.914399028 |
| 3.362009956 | 11.822800071 | 30.914399028 |
| 3.362009956 | 17.805800414 | 30.914399028 |
| 8.582360344 | 5.839809890 | 30.914399028 |
| 8.582360344 | 11.822800071 | 30.914399028 |
| 8.582360344 | 17.805800414 | 30.914399028 |
| 13.802700464 | 5.839809890 | 30.914399028 |
| 13.802700464 | 11.822800071 | 30.914399028 |
| 13.802700464 | 17.805800414 | 30.914399028 |
| 0.751833012 | 1.352579960 | 30.914399028 |
| 0.751833012 | 7.335570140 | 30.914399028 |
| 0.751833012 | 13.318599904 | 30.914399028 |
| 5.972170214 | 1.352579960 | 30.914399028 |
| 5.972170214 | 7.335570140 | 30.914399028 |
| 5.972170214 | 13.318599904 | 30.914399028 |
| 11.192499599 | 1.352579960 | 30.914399028 |
| 11.192499599 | 7.335570140 | 30.914399028 |
| 11.192499599 | 13.318599904 | 30.914399028 |
| 0.751833012 | 4.344080131 | 30.914399028 |
| 0.751833012 | 10.327099732 | 30.914399028 |
| 0.751833012 | 16.310100076 | 30.914399028 |
| 5.972170214 | 4.344080131 | 30.914399028 |
| 5.972170214 | 10.327099732 | 30.914399028 |
| 5.972170214 | 16.310100076 | 30.914399028 |
| 11.192499599 | 4.344080131 | 30.914399028 |
| 11.192499599 | 10.327099732 | 30.914399028 |
| 11.192499599 | 16.310100076 | 30.914399028 |
| 5.097550105 | 2.846580050 | 32.559499145 |

| | | |
|---|---|---|
| 5.097550105 | 8.829569963 | 32.559499145 |
| 5.097550105 | 14.812600262 | 32.559499145 |
| 10.317900026 | 2.846580050 | 32.559499145 |
| 10.317900026 | 8.829569963 | 32.559499145 |
| 10.317900026 | 14.812600262 | 32.559499145 |
| 15.538200007 | 2.846580050 | 32.559499145 |
| 15.538200007 | 8.829569963 | 32.559499145 |
| 15.538200007 | 14.812600262 | 32.559499145 |
| 5.097550105 | 5.838069791 | 32.559499145 |
| 5.097550105 | 11.821100090 | 32.559499145 |
| 5.097550105 | 17.804100434 | 32.559499145 |
| 10.317900026 | 5.838069791 | 32.559499145 |
| 10.317900026 | 11.821100090 | 32.559499145 |
| 10.317900026 | 17.804100434 | 32.559499145 |
| 15.538200007 | 5.838069791 | 32.559499145 |
| 15.538200007 | 11.821100090 | 32.559499145 |
| 15.538200007 | 17.804100434 | 32.559499145 |
| 2.487400115 | 1.350749993 | 32.559400797 |
| 2.487400115 | 7.333740173 | 32.559400797 |
| 2.487400115 | 13.316699863 | 32.559400797 |
| 7.707739768 | 1.350749993 | 32.559400797 |
| 7.707739768 | 7.333740173 | 32.559400797 |
| 7.707739768 | 13.316699863 | 32.559400797 |
| 12.928099957 | 1.350749993 | 32.559400797 |
| 12.928099957 | 7.333740173 | 32.559400797 |
| 12.928099957 | 13.316699863 | 32.559400797 |
| 2.487400115 | 4.342240002 | 32.559400797 |
| 2.487400115 | 10.325199691 | 32.559400797 |
| 2.487400115 | 16.308200035 | 32.559400797 |
| 7.707739768 | 4.342240002 | 32.559400797 |
| 7.707739768 | 10.325199691 | 32.559400797 |
| 7.707739768 | 16.308200035 | 32.559400797 |
| 12.928099957 | 4.342240002 | 32.559400797 |
| 12.928099957 | 10.325199691 | 32.559400797 |
| 12.928099957 | 16.308200035 | 32.559400797 |
| 5.099859976 | 2.850210027 | 29.288899899 |
| 5.099859976 | 8.833210103 | 29.288899899 |
| 5.099859976 | 14.816200284 | 29.288899899 |
| 10.320200096 | 2.850210027 | 29.288899899 |
| 10.320200096 | 8.833210103 | 29.288899899 |
| 10.320200096 | 14.816200284 | 29.288899899 |
| 15.540500077 | 2.850210027 | 29.288899899 |
| 15.540500077 | 8.833210103 | 29.288899899 |
| 15.540500077 | 14.816200284 | 29.288899899 |
| 5.099859976 | 5.841709931 | 29.288899899 |

| | | |
|---|---|---|
| 5.099859976 | 11.824700112 | 29.288899899 |
| 5.099859976 | 17.807700455 | 29.288899899 |
| 10.320200096 | 5.841709931 | 29.288899899 |
| 10.320200096 | 11.824700112 | 29.288899899 |
| 10.320200096 | 17.807700455 | 29.288899899 |
| 15.540500077 | 5.841709931 | 29.288899899 |
| 15.540500077 | 11.824700112 | 29.288899899 |
| 15.540500077 | 17.807700455 | 29.288899899 |
| 2.489680115 | 1.354450045 | 29.288899899 |
| 2.489680115 | 7.337449854 | 29.288899899 |
| 2.489680115 | 13.320400450 | 29.288899899 |
| 7.710019768 | 1.354450045 | 29.288899899 |
| 7.710019768 | 7.337449854 | 29.288899899 |
| 7.710019768 | 13.320400450 | 29.288899899 |
| 12.930400027 | 1.354450045 | 29.288899899 |
| 12.930400027 | 7.337449854 | 29.288899899 |
| 12.930400027 | 13.320400450 | 29.288899899 |
| 2.489680115 | 4.345949949 | 29.288899899 |
| 2.489680115 | 10.328900278 | 29.288899899 |
| 2.489680115 | 16.311899552 | 29.288899899 |
| 7.710019768 | 4.345949949 | 29.288899899 |
| 7.710019768 | 10.328900278 | 29.288899899 |
| 7.710019768 | 16.311899552 | 29.288899899 |
| 12.930400027 | 4.345949949 | 29.288899899 |
| 12.930400027 | 10.328900278 | 29.288899899 |
| 12.930400027 | 16.311899552 | 29.288899899 |
| 0.433091005 | 1.184650057 | 35.513699055 |
| 0.433091005 | 7.167640103 | 35.513699055 |
| 0.433091005 | 13.150600328 | 35.513699055 |
| 5.653429958 | 1.184650057 | 35.513699055 |
| 5.653429958 | 7.167640103 | 35.513699055 |
| 5.653429958 | 13.150600328 | 35.513699055 |
| 10.873799949 | 1.184650057 | 35.513699055 |
| 10.873799949 | 7.167640103 | 35.513699055 |
| 10.873799949 | 13.150600328 | 35.513699055 |
| 3.043259898 | 2.659689989 | 35.513699055 |
| 3.043259898 | 8.642680170 | 35.513699055 |
| 3.043259898 | 14.625699771 | 35.513699055 |
| 8.263600018 | 2.659689989 | 35.513699055 |
| 8.263600018 | 8.642680170 | 35.513699055 |
| 8.263600018 | 14.625699771 | 35.513699055 |
| 13.483899998 | 2.659689989 | 35.513699055 |
| 13.483899998 | 8.642680170 | 35.513699055 |
| 13.483899998 | 14.625699771 | 35.513699055 |
| 0.433091005 | 4.176139932 | 35.513699055 |

| | | |
|---|---|---|
| 0.433091005 | 10.159100156 | 35.513699055 |
| 0.433091005 | 16.142100500 | 35.513699055 |
| 5.653429958 | 4.176139932 | 35.513699055 |
| 5.653429958 | 10.159100156 | 35.513699055 |
| 5.653429958 | 16.142100500 | 35.513699055 |
| 10.873799949 | 4.176139932 | 35.513699055 |
| 10.873799949 | 10.159100156 | 35.513699055 |
| 10.873799949 | 16.142100500 | 35.513699055 |
| 3.043259898 | 5.651190161 | 35.513699055 |
| 3.043259898 | 11.634199599 | 35.513699055 |
| 3.043259898 | 17.617199943 | 35.513699055 |
| 8.263600018 | 5.651190161 | 35.513699055 |
| 8.263600018 | 11.634199599 | 35.513699055 |
| 8.263600018 | 17.617199943 | 35.513699055 |
| 13.483899998 | 5.651190161 | 35.513699055 |
| 13.483899998 | 11.634199599 | 35.513699055 |
| 13.483899998 | 17.617199943 | 35.513699055 |
| 1.303149947 | 5.664990065 | 37.314298749 |
| 1.303149947 | 11.647999503 | 37.314298749 |
| 1.303149947 | 17.630999847 | 37.314298749 |
| 6.523490067 | 5.664990065 | 37.314298749 |
| 6.523490067 | 11.647999503 | 37.314298749 |
| 6.523490067 | 17.630999847 | 37.314298749 |
| 11.743800316 | 5.664990065 | 37.314298749 |
| 11.743800316 | 11.647999503 | 37.314298749 |
| 11.743800316 | 17.630999847 | 37.314298749 |
| 3.913320007 | 1.157039952 | 37.314298749 |
| 3.913320007 | 7.140030132 | 37.314298749 |
| 3.913320007 | 13.123000520 | 37.314298749 |
| 9.133660127 | 1.157039952 | 37.314298749 |
| 9.133660127 | 7.140030132 | 37.314298749 |
| 9.133660127 | 13.123000520 | 37.314298749 |
| 14.354000247 | 1.157039952 | 37.314298749 |
| 14.354000247 | 7.140030132 | 37.314298749 |
| 14.354000247 | 13.123000520 | 37.314298749 |
| 1.303149947 | 2.673489894 | 37.314298749 |
| 1.303149947 | 8.656490237 | 37.314298749 |
| 1.303149947 | 14.639499675 | 37.314298749 |
| 6.523490067 | 2.673489894 | 37.314298749 |
| 6.523490067 | 8.656490237 | 37.314298749 |
| 6.523490067 | 14.639499675 | 37.314298749 |
| 11.743800316 | 2.673489894 | 37.314298749 |
| 11.743800316 | 8.656490237 | 37.314298749 |
| 11.743800316 | 14.639499675 | 37.314298749 |
| 3.913320007 | 4.148540123 | 37.314298749 |

| | | |
|---|---|---|
| 3.913320007 | 10.131500348 | 37.314298749 |
| 3.913320007 | 16.114499622 | 37.314298749 |
| 9.133660127 | 4.148540123 | 37.314298749 |
| 9.133660127 | 10.131500348 | 37.314298749 |
| 9.133660127 | 16.114499622 | 37.314298749 |
| 14.354000247 | 4.148540123 | 37.314298749 |
| 14.354000247 | 10.131500348 | 37.314298749 |
| 14.354000247 | 16.114499622 | 37.314298749 |
| 1.303149947 | 5.664990065 | 39.695098996 |
| 1.303149947 | 11.647999503 | 39.695098996 |
| 1.303149947 | 17.630999847 | 39.695098996 |
| 6.523490067 | 5.664990065 | 39.695098996 |
| 6.523490067 | 11.647999503 | 39.695098996 |
| 6.523490067 | 17.630999847 | 39.695098996 |
| 11.743800316 | 5.664990065 | 39.695098996 |
| 11.743800316 | 11.647999503 | 39.695098996 |
| 11.743800316 | 17.630999847 | 39.695098996 |
| 3.913320007 | 1.157039952 | 39.695000648 |
| 3.913320007 | 7.140030132 | 39.695000648 |
| 3.913320007 | 13.123000520 | 39.695000648 |
| 9.133660127 | 1.157039952 | 39.695000648 |
| 9.133660127 | 7.140030132 | 39.695000648 |
| 9.133660127 | 13.123000520 | 39.695000648 |
| 14.354000247 | 1.157039952 | 39.695000648 |
| 14.354000247 | 7.140030132 | 39.695000648 |
| 14.354000247 | 13.123000520 | 39.695000648 |
| 1.303149947 | 2.673489894 | 39.695098996 |
| 1.303149947 | 8.656490237 | 39.695098996 |
| 1.303149947 | 14.639499675 | 39.695098996 |
| 6.523490067 | 2.673489894 | 39.695098996 |
| 6.523490067 | 8.656490237 | 39.695098996 |
| 6.523490067 | 14.639499675 | 39.695098996 |
| 11.743800316 | 2.673489894 | 39.695098996 |
| 11.743800316 | 8.656490237 | 39.695098996 |
| 11.743800316 | 14.639499675 | 39.695098996 |
| 3.913320007 | 4.148540123 | 39.695000648 |
| 3.913320007 | 10.131500348 | 39.695000648 |
| 3.913320007 | 16.114499622 | 39.695000648 |
| 9.133660127 | 4.148540123 | 39.695000648 |
| 9.133660127 | 10.131500348 | 39.695000648 |
| 9.133660127 | 16.114499622 | 39.695000648 |
| 14.354000247 | 4.148540123 | 39.695000648 |
| 14.354000247 | 10.131500348 | 39.695000648 |
| 14.354000247 | 16.114499622 | 39.695000648 |

```
Au/WSe2/Si
1.0
      15.9505681992        0.0000000000        0.0000000000
       0.0000000000       18.4212131500        0.0000000000
       0.0000000000        0.0000000000       50.0000000000
   Au    W    Se    Si
  144   36   72   108
Cartesian
     0.000000000        0.000000000       20.000000298
     0.000000000        6.140400174       20.000000298
     0.000000000       12.280800349       20.000000298
     5.316860028        0.000000000       20.000000298
     5.316860028        6.140400174       20.000000298
     5.316860028       12.280800349       20.000000298
    10.633700090        0.000000000       20.000000298
    10.633700090        6.140400174       20.000000298
    10.633700090       12.280800349       20.000000298
     5.334040157        6.160960035       26.642000675
     5.334070105       12.301499654       26.641801000
     5.334070105        0.020486295       26.641899347
    10.650899710        6.161130223       26.641801000
    10.650799883       12.301399737       26.641899347
    10.650899710        0.020587310       26.641899347
     0.017230993        6.161180182       26.641899347
     0.017230993       12.301299820       26.641899347
     0.017230993        0.020587310       26.641899347
     0.000000000        3.070200087       20.000000298
     0.000000000        9.210609869       20.000000298
     0.000000000       15.350999612       20.000000298
     5.316860028        3.070200087       20.000000298
     5.316860028        9.210609869       20.000000298
     5.316860028       15.350999612       20.000000298
    10.633700090        3.070200087       20.000000298
    10.633700090        9.210609869       20.000000298
    10.633700090       15.350999612       20.000000298
     5.334050140        3.090770104       26.641899347
     5.334040157        9.231230119       26.641899347
     5.334040157       15.371600098       26.641801000
    10.650899710        3.090930136       26.641899347
    10.650999536        9.231220237       26.641899347
    10.650799883       15.371500181       26.641899347
     0.017230993        3.090890059       26.641899347
     0.017230993        9.231220237       26.641899347
     0.017230993       15.371500181       26.641899347
     1.778879974        6.148199745       22.235900164
```

| | | |
|---|---|---|
| 1.778860009 | 12.288700386 | 22.235900164 |
| 1.778840044 | 0.007885763 | 22.236000001 |
| 7.095670124 | 6.148220058 | 22.236000001 |
| 7.095710054 | 12.288600469 | 22.235900164 |
| 7.095730019 | 0.007885763 | 22.235900164 |
| 12.412599554 | 6.148279899 | 22.235900164 |
| 12.412499728 | 12.288700386 | 22.235900164 |
| 12.412499728 | 0.007885763 | 22.236000001 |
| 1.778900058 | 3.078080086 | 22.236000001 |
| 1.778879974 | 9.218400107 | 22.235900164 |
| 1.778819960 | 15.358899650 | 22.235900164 |
| 7.095670124 | 3.078049891 | 22.235900164 |
| 7.095710054 | 9.218400107 | 22.236000001 |
| 7.095720037 | 15.358799732 | 22.235900164 |
| 12.412599554 | 3.078040009 | 22.235900164 |
| 12.412599554 | 9.218509906 | 22.235900164 |
| 12.412599554 | 15.358899650 | 22.236000001 |
| 3.556050064 | 6.153650167 | 24.433200061 |
| 3.556009896 | 12.294000383 | 24.433100224 |
| 3.556080012 | 0.013186858 | 24.433100224 |
| 8.872919599 | 6.153589778 | 24.433100224 |
| 8.872850196 | 12.294100300 | 24.433100224 |
| 8.872919599 | 0.013287873 | 24.433100224 |
| 14.189800068 | 6.153600209 | 24.433100224 |
| 14.189800068 | 12.294000383 | 24.433100224 |
| 14.189700241 | 0.013085843 | 24.433100224 |
| 3.556080012 | 3.083480000 | 24.433200061 |
| 3.556030099 | 9.223809903 | 24.433100224 |
| 3.556070030 | 15.364199647 | 24.433000386 |
| 8.872919599 | 3.083419885 | 24.433100224 |
| 8.872930057 | 9.223850529 | 24.433100224 |
| 8.872839738 | 15.364299564 | 24.433100224 |
| 14.189800068 | 3.083299930 | 24.433100224 |
| 14.189800068 | 9.223850529 | 24.433100224 |
| 14.189800068 | 15.364199647 | 24.433200061 |
| 4.437289785 | 1.542830030 | 22.235900164 |
| 4.437310226 | 7.683209891 | 22.236000001 |
| 4.437289785 | 13.823600184 | 22.235900164 |
| 9.754179761 | 1.542799972 | 22.235900164 |
| 9.754110358 | 7.683230204 | 22.235900164 |
| 9.754139830 | 13.823700101 | 22.235900164 |
| 15.071000333 | 1.542779934 | 22.235900164 |
| 15.071000333 | 7.683180245 | 22.235900164 |
| 15.070899556 | 13.823700101 | 22.235900164 |
| 4.437289785 | 4.613060037 | 22.236000001 |

| | | |
|---|---|---|
| 4.437289785 | 10.753399822 | 22.235900164 |
| 4.437320208 | 16.893800545 | 22.235900164 |
| 9.754210184 | 4.612979884 | 22.235900164 |
| 9.754110358 | 10.753399822 | 22.235900164 |
| 9.754110358 | 16.893900462 | 22.235900164 |
| 15.071000333 | 4.612929925 | 22.235900164 |
| 15.071000333 | 10.753499739 | 22.235900164 |
| 15.071000333 | 16.893800545 | 22.235900164 |
| 0.897605017 | 1.548640043 | 24.433200061 |
| 0.897688027 | 7.688929869 | 24.433100224 |
| 0.897595985 | 13.829299849 | 24.433100224 |
| 6.214479840 | 1.548490030 | 24.433100224 |
| 6.214449893 | 7.688870029 | 24.433200061 |
| 6.214499806 | 13.829299849 | 24.433000386 |
| 11.531300413 | 1.548580065 | 24.433200061 |
| 11.531400239 | 7.689000141 | 24.433100224 |
| 11.531300413 | 13.829399766 | 24.433100224 |
| 0.897559026 | 4.618839856 | 24.433100224 |
| 0.897705972 | 10.759099487 | 24.433100224 |
| 0.897648988 | 16.899600128 | 24.433100224 |
| 6.214489823 | 4.618709744 | 24.433200061 |
| 6.214439910 | 10.759099487 | 24.433100224 |
| 6.214580142 | 16.899500211 | 24.433100224 |
| 11.531400239 | 4.618799779 | 24.433100224 |
| 11.531300413 | 10.759200502 | 24.433100224 |
| 11.531199636 | 16.899500211 | 24.433100224 |
| 2.658430014 | 1.535100044 | 20.000000298 |
| 2.658430014 | 7.675510237 | 20.000000298 |
| 2.658430014 | 13.815899981 | 20.000000298 |
| 7.975279821 | 1.535100044 | 20.000000298 |
| 7.975279821 | 7.675510237 | 20.000000298 |
| 7.975279821 | 13.815899981 | 20.000000298 |
| 13.292099918 | 1.535100044 | 20.000000298 |
| 13.292099918 | 7.675510237 | 20.000000298 |
| 13.292099918 | 13.815899981 | 20.000000298 |
| 2.675680022 | 1.555370035 | 26.641899347 |
| 2.675640091 | 7.695909791 | 26.642000675 |
| 2.675619888 | 13.836199617 | 26.641801000 |
| 7.992499881 | 1.555549968 | 26.641899347 |
| 7.992530305 | 7.695830187 | 26.641899347 |
| 7.992450443 | 13.836299534 | 26.641899347 |
| 13.309400315 | 1.555339977 | 26.641899347 |
| 13.309400315 | 7.695870263 | 26.641801000 |
| 13.309299538 | 13.836199617 | 26.641899347 |
| 2.658430014 | 4.605299994 | 20.000000298 |

| | | |
|---|---|---|
| 2.658430014 | 10.745699619 | 20.000000298 |
| 2.658430014 | 16.886100342 | 20.000000298 |
| 7.975279821 | 4.605299994 | 20.000000298 |
| 7.975279821 | 10.745699619 | 20.000000298 |
| 7.975279821 | 16.886100342 | 20.000000298 |
| 13.292099918 | 4.605299994 | 20.000000298 |
| 13.292099918 | 10.745699619 | 20.000000298 |
| 13.292099918 | 16.886100342 | 20.000000298 |
| 2.675530044 | 4.625650138 | 26.641899347 |
| 2.675709970 | 10.766100271 | 26.641899347 |
| 2.675690004 | 16.906399979 | 26.641801000 |
| 7.992559777 | 4.625660020 | 26.641899347 |
| 7.992510339 | 10.766100271 | 26.641899347 |
| 7.992539812 | 16.906599813 | 26.641899347 |
| 13.309500141 | 4.625619943 | 26.641899347 |
| 13.309400315 | 10.766100271 | 26.641899347 |
| 13.309199711 | 16.906300062 | 26.641899347 |
| 0.917297980 | 1.574730066 | 31.008499861 |
| 0.917239985 | 7.715189944 | 31.008499861 |
| 0.917275994 | 13.855600000 | 31.008398533 |
| 6.234159909 | 1.574770006 | 31.008499861 |
| 6.234100013 | 7.715189944 | 31.008499861 |
| 6.234109996 | 13.855600000 | 31.008398533 |
| 11.551000447 | 1.574780025 | 31.008499861 |
| 11.551000447 | 7.715149867 | 31.008398533 |
| 11.551000447 | 13.855600000 | 31.008499861 |
| 0.917249017 | 4.644969818 | 31.008499861 |
| 0.917267973 | 10.785399639 | 31.008499861 |
| 0.917275994 | 16.925800362 | 31.008499861 |
| 6.234129961 | 4.644980249 | 31.008499861 |
| 6.234090031 | 10.785399639 | 31.008499861 |
| 6.234139944 | 16.925800362 | 31.008499861 |
| 11.551000447 | 4.644959936 | 31.008398533 |
| 11.551000447 | 10.785399639 | 31.008499861 |
| 11.551000447 | 16.925800362 | 31.008499861 |
| 3.575719913 | 6.180089763 | 31.008499861 |
| 3.575740116 | 12.320500368 | 31.008398533 |
| 3.575750098 | 0.039687942 | 31.008398533 |
| 8.892599668 | 6.180050235 | 31.008499861 |
| 8.892550230 | 12.320500368 | 31.008499861 |
| 8.892619633 | 0.039687942 | 31.008499861 |
| 14.209400275 | 6.180050235 | 31.008499861 |
| 14.209400275 | 12.320500368 | 31.008499861 |
| 14.209500102 | 0.039586927 | 31.008499861 |
| 3.575750098 | 3.109860030 | 31.008499861 |

| | | |
|---|---|---|
| 3.575709930 | 9.250280243 | 31.008499861 |
| 3.575770063 | 15.390699632 | 31.008398533 |
| 8.892610126 | 3.109860030 | 31.008499861 |
| 8.892579703 | 9.250260479 | 31.008499861 |
| 8.892610126 | 15.390699632 | 31.008499861 |
| 14.209500102 | 3.109849874 | 31.008499861 |
| 14.209400275 | 9.250260479 | 31.008499861 |
| 14.209400275 | 15.390699632 | 31.008499861 |
| 2.690219980 | 1.575549990 | 32.784798741 |
| 2.690190032 | 7.716020024 | 32.784900069 |
| 2.690239945 | 13.856400435 | 32.784700394 |
| 8.007049822 | 1.575599949 | 32.784798741 |
| 8.007080245 | 7.715960184 | 32.784798741 |
| 8.007029857 | 13.856400435 | 32.784798741 |
| 13.323899867 | 1.575570029 | 32.784798741 |
| 13.323899867 | 7.715989830 | 32.784798741 |
| 13.323899867 | 13.856400435 | 32.784798741 |
| 2.690200014 | 4.645780135 | 32.784798741 |
| 2.690200014 | 10.786200073 | 32.784798741 |
| 2.690249928 | 16.926599699 | 32.784700394 |
| 8.007100210 | 4.645770253 | 32.784798741 |
| 8.007049822 | 10.786200073 | 32.784798741 |
| 8.007080245 | 16.926599699 | 32.784798741 |
| 13.323899867 | 4.645759822 | 32.784798741 |
| 13.323899867 | 10.786200073 | 32.784798741 |
| 13.323899867 | 16.926599699 | 32.784798741 |
| 5.348640011 | 6.180829808 | 32.784798741 |
| 5.348640011 | 12.321299705 | 32.784798741 |
| 5.348649994 | 0.040487278 | 32.784700394 |
| 10.665500039 | 6.180819926 | 32.784700394 |
| 10.665500039 | 12.321299705 | 32.784798741 |
| 10.665500039 | 0.040487278 | 32.784798741 |
| 0.031731496 | 6.180829808 | 32.784798741 |
| 0.031832273 | 12.321199788 | 32.784798741 |
| 0.031832273 | 0.040386263 | 32.784798741 |
| 5.348649994 | 3.110650034 | 32.784798741 |
| 5.348630028 | 9.251059816 | 32.784798741 |
| 5.348630028 | 15.391500067 | 32.784700394 |
| 10.665500039 | 3.110620114 | 32.784798741 |
| 10.665500039 | 9.251030170 | 32.784798741 |
| 10.665500039 | 15.391500067 | 32.784798741 |
| 0.031832273 | 3.110589919 | 32.784798741 |
| 0.031731496 | 9.251069698 | 32.784798741 |
| 0.031832273 | 15.391400150 | 32.784798741 |
| 2.689790013 | 1.574210031 | 29.243999720 |

| | | |
|---|---|---|
| 2.689719897 | 7.714629969 | 29.243999720 |
| 2.689790013 | 13.855000498 | 29.243901372 |
| 8.006649565 | 1.574200012 | 29.243999720 |
| 8.006620093 | 7.714610205 | 29.243999720 |
| 8.006679989 | 13.855000498 | 29.243999720 |
| 13.323600388 | 1.574150053 | 29.243999720 |
| 13.323499611 | 7.714580011 | 29.243999720 |
| 13.323499611 | 13.855000498 | 29.243999720 |
| 2.689790013 | 4.644420274 | 29.243999720 |
| 2.689790013 | 10.784800136 | 29.243999720 |
| 2.689770048 | 16.925199761 | 29.243901372 |
| 8.006640058 | 4.644399962 | 29.243999720 |
| 8.006649565 | 10.784800136 | 29.243999720 |
| 8.006690447 | 16.925199761 | 29.243999720 |
| 13.323499611 | 4.644350003 | 29.243999720 |
| 13.323499611 | 10.784800136 | 29.243999720 |
| 13.323499611 | 16.925199761 | 29.243999720 |
| 5.348169876 | 6.179519906 | 29.243999720 |
| 5.348199824 | 12.319899768 | 29.243901372 |
| 5.348250213 | 0.039086243 | 29.243999720 |
| 10.664999956 | 6.179500143 | 29.243901372 |
| 10.664999956 | 12.319899768 | 29.243999720 |
| 10.665099782 | 0.039086243 | 29.243999720 |
| 0.031332190 | 6.179500143 | 29.243999720 |
| 0.031332190 | 12.319899768 | 29.243999720 |
| 0.031332190 | 0.039086243 | 29.243999720 |
| 5.348189841 | 3.109350014 | 29.243999720 |
| 5.348139928 | 9.249689524 | 29.243999720 |
| 5.348229772 | 15.390200047 | 29.243901372 |
| 10.665099782 | 3.109280017 | 29.243901372 |
| 10.664999956 | 9.249710386 | 29.243999720 |
| 10.664999956 | 15.390100130 | 29.243999720 |
| 0.031332190 | 3.109309937 | 29.243999720 |
| 0.031332190 | 9.249700504 | 29.243999720 |
| 0.031332190 | 15.390100130 | 29.243999720 |
| 0.441097987 | 1.215809943 | 36.240199208 |
| 0.441097987 | 7.356220274 | 36.240300536 |
| 0.441097987 | 13.496600136 | 36.240300536 |
| 5.757949785 | 1.215809943 | 36.240300536 |
| 5.757949785 | 7.356220274 | 36.240199208 |
| 5.757949785 | 13.496600136 | 36.240199208 |
| 11.074799831 | 1.215809943 | 36.240199208 |
| 11.074799831 | 7.356220274 | 36.240300536 |
| 11.074799831 | 13.496600136 | 36.240199208 |
| 3.099529992 | 2.729659903 | 36.240300536 |

| | | |
|---|---|---|
| 3.099529992 | 8.870069959 | 36.240300536 |
| 3.099529992 | 15.010499779 | 36.240300536 |
| 8.416380037 | 2.729659903 | 36.240300536 |
| 8.416380037 | 8.870069959 | 36.240300536 |
| 8.416380037 | 15.010499779 | 36.240300536 |
| 13.733199659 | 2.729659903 | 36.240300536 |
| 13.733199659 | 8.870069959 | 36.240300536 |
| 13.733199659 | 15.010499779 | 36.240300536 |
| 0.441097987 | 4.286019912 | 36.240199208 |
| 0.441097987 | 10.426399774 | 36.240300536 |
| 0.441097987 | 16.566800497 | 36.240300536 |
| 5.757949785 | 4.286019912 | 36.240300536 |
| 5.757949785 | 10.426399774 | 36.240300536 |
| 5.757949785 | 16.566800497 | 36.240199208 |
| 11.074799831 | 4.286019912 | 36.240300536 |
| 11.074799831 | 10.426399774 | 36.240300536 |
| 11.074799831 | 16.566800497 | 36.240199208 |
| 3.099529992 | 5.799870146 | 36.240300536 |
| 3.099529992 | 11.940300515 | 36.240300536 |
| 3.099529992 | 18.080700141 | 36.240300536 |
| 8.416380037 | 5.799870146 | 36.240300536 |
| 8.416380037 | 11.940300515 | 36.240300536 |
| 8.416380037 | 18.080700141 | 36.240300536 |
| 13.733199659 | 5.799870146 | 36.240300536 |
| 13.733199659 | 11.940300515 | 36.240300536 |
| 13.733199659 | 18.080700141 | 36.240300536 |
| 1.327239988 | 5.814029824 | 37.962201238 |
| 1.327239988 | 11.954399803 | 37.962201238 |
| 1.327239988 | 18.094800527 | 37.962201238 |
| 6.644100016 | 5.814029824 | 37.962201238 |
| 6.644100016 | 11.954399803 | 37.962201238 |
| 6.644100016 | 18.094800527 | 37.962201238 |
| 11.960999975 | 5.814029824 | 37.962201238 |
| 11.960999975 | 11.954399803 | 37.962201238 |
| 11.960999975 | 18.094800527 | 37.962201238 |
| 3.985670002 | 1.187480020 | 37.962201238 |
| 3.985670002 | 7.327880058 | 37.962201238 |
| 3.985670002 | 13.468300545 | 37.962201238 |
| 9.302529792 | 1.187480020 | 37.962099910 |
| 9.302529792 | 7.327880058 | 37.962201238 |
| 9.302529792 | 13.468300545 | 37.962099910 |
| 14.619399803 | 1.187480020 | 37.962099910 |
| 14.619399803 | 7.327880058 | 37.962201238 |
| 14.619399803 | 13.468300545 | 37.962099910 |
| 1.327239988 | 2.743830011 | 37.962201238 |

| | | |
|---|---|---|
| 1.327239988 | 8.884240067 | 37.962201238 |
| 1.327239988 | 15.024600165 | 37.962201238 |
| 6.644100016 | 2.743830011 | 37.962201238 |
| 6.644100016 | 8.884240067 | 37.962201238 |
| 6.644100016 | 15.024600165 | 37.962201238 |
| 11.960999975 | 2.743830011 | 37.962201238 |
| 11.960999975 | 8.884240067 | 37.962201238 |
| 11.960999975 | 15.024600165 | 37.962201238 |
| 3.985670002 | 4.257679970 | 37.962201238 |
| 3.985670002 | 10.398100183 | 37.962201238 |
| 3.985670002 | 16.538499808 | 37.962201238 |
| 9.302529792 | 4.257679970 | 37.962201238 |
| 9.302529792 | 10.398100183 | 37.962201238 |
| 9.302529792 | 16.538499808 | 37.962099910 |
| 14.619399803 | 4.257679970 | 37.962099910 |
| 14.619399803 | 10.398100183 | 37.962201238 |
| 14.619399803 | 16.538499808 | 37.962099910 |
| 1.327239988 | 5.814029824 | 40.353500843 |
| 1.327239988 | 11.954399803 | 40.353500843 |
| 1.327239988 | 18.094800527 | 40.353500843 |
| 6.644100016 | 5.814029824 | 40.353500843 |
| 6.644100016 | 11.954399803 | 40.353500843 |
| 6.644100016 | 18.094800527 | 40.353500843 |
| 11.960999975 | 5.814029824 | 40.353500843 |
| 11.960999975 | 11.954399803 | 40.353500843 |
| 11.960999975 | 18.094800527 | 40.353500843 |
| 3.985670002 | 1.187480020 | 40.353399515 |
| 3.985670002 | 7.327880058 | 40.353399515 |
| 3.985670002 | 13.468300545 | 40.353399515 |
| 9.302529792 | 1.187480020 | 40.353399515 |
| 9.302529792 | 7.327880058 | 40.353399515 |
| 9.302529792 | 13.468300545 | 40.353399515 |
| 14.619399803 | 1.187480020 | 40.353399515 |
| 14.619399803 | 7.327880058 | 40.353399515 |
| 14.619399803 | 13.468300545 | 40.353399515 |
| 1.327239988 | 2.743830011 | 40.353500843 |
| 1.327239988 | 8.884240067 | 40.353500843 |
| 1.327239988 | 15.024600165 | 40.353500843 |
| 6.644100016 | 2.743830011 | 40.353500843 |
| 6.644100016 | 8.884240067 | 40.353500843 |
| 6.644100016 | 15.024600165 | 40.353500843 |
| 11.960999975 | 2.743830011 | 40.353500843 |
| 11.960999975 | 8.884240067 | 40.353500843 |
| 11.960999975 | 15.024600165 | 40.353500843 |
| 3.985670002 | 4.257679970 | 40.353399515 |

```
      3.985670002         10.398100183        40.353399515
      3.985670002         16.538499808        40.353399515
      9.302529792          4.257679970        40.353399515
      9.302529792         10.398100183        40.353399515
      9.302529792         16.538499808        40.353399515
     14.619399803          4.257679970        40.353399515
     14.619399803         10.398100183        40.353399515
     14.619399803         16.538499808        40.353399515


Ag/MoS2/Si
1.0
       15.6201000214          0.0000000000           0.0000000000
        0.0000000000         17.9876518250           0.0000000000
        0.0000000000          0.0000000000          50.0000000000
   Ag    Mo    S    Si
  144    36   72   108
Cartesian
      0.000000000          0.000000000        20.000000298
      0.000000000          5.995884120        20.000000298
      0.000000000         11.991768241        20.000000298
      5.206700162          0.000000000        20.000000298
      5.206700162          5.995884120        20.000000298
      5.206700162         11.991768241        20.000000298
     10.413400325          0.000000000        20.000000298
     10.413400325          5.995884120        20.000000298
     10.413400325         11.991768241        20.000000298
      0.128135009          0.071177994        24.562700093
      0.128135009          6.067061855        24.562700093
      0.128135009         12.062945975        24.562700093
      5.334835099          0.071177994        24.562700093
      5.334835099          6.067061855        24.562700093
      5.334835099         12.062945975        24.562700093
     10.541535261          0.071177994        24.562700093
     10.541535261          6.067061855        24.562700093
     10.541535261         12.062945975        24.562700093
      0.000000000          2.997939916        20.000000298
      0.000000000          8.993823768        20.000000298
      0.000000000         14.989707352        20.000000298
      5.206700162          2.997939916        20.000000298
      5.206700162          8.993823768        20.000000298
      5.206700162         14.989707352        20.000000298
     10.413400325          2.997939916        20.000000298
     10.413400325          8.993823768        20.000000298
     10.413400325         14.989707352        20.000000298
      0.128135009          3.069120063        24.562700093
```

| | | |
|---|---|---|
| 0.128135009 | 9.065003647 | 24.562700093 |
| 0.128135009 | 15.060888304 | 24.562700093 |
| 5.334835099 | 3.069120063 | 24.562700093 |
| 5.334835099 | 9.065003647 | 24.562700093 |
| 5.334835099 | 15.060888304 | 24.562700093 |
| 10.541535261 | 3.069120063 | 24.562700093 |
| 10.541535261 | 9.065003647 | 24.562700093 |
| 10.541535261 | 15.060888304 | 24.562700093 |
| 1.799930102 | 0.029904598 | 22.291800380 |
| 1.799930102 | 6.025788461 | 22.291800380 |
| 1.799930102 | 12.021672581 | 22.291800380 |
| 7.006630264 | 0.029904598 | 22.291800380 |
| 7.006630264 | 6.025788461 | 22.291800380 |
| 7.006630264 | 12.021672581 | 22.291800380 |
| 12.213329961 | 0.029904598 | 22.291800380 |
| 12.213329961 | 6.025788461 | 22.291800380 |
| 12.213329961 | 12.021672581 | 22.291800380 |
| 1.799930102 | 3.027849885 | 22.291800380 |
| 1.799930102 | 9.023733470 | 22.291800380 |
| 1.799930102 | 15.019618126 | 22.291800380 |
| 7.006630264 | 3.027849885 | 22.291800380 |
| 7.006630264 | 9.023733470 | 22.291800380 |
| 7.006630264 | 15.019618126 | 22.291800380 |
| 12.213329961 | 3.027849885 | 22.291800380 |
| 12.213329961 | 9.023733470 | 22.291800380 |
| 12.213329961 | 15.019618126 | 22.291800380 |
| 3.658500229 | 0.109276994 | 26.818400621 |
| 3.658500229 | 6.105161156 | 26.818400621 |
| 3.658500229 | 12.101044740 | 26.818400621 |
| 8.865199926 | 0.109276994 | 26.818400621 |
| 8.865199926 | 6.105161156 | 26.818400621 |
| 8.865199926 | 12.101044740 | 26.818400621 |
| 14.071900553 | 0.109276994 | 26.818400621 |
| 14.071900553 | 6.105161156 | 26.818400621 |
| 14.071900553 | 12.101044740 | 26.818400621 |
| 3.658500229 | 3.107219900 | 26.818400621 |
| 3.658500229 | 9.103103484 | 26.818400621 |
| 3.658500229 | 15.098988141 | 26.818400621 |
| 8.865199926 | 3.107219900 | 26.818400621 |
| 8.865199926 | 9.103103484 | 26.818400621 |
| 8.865199926 | 15.098988141 | 26.818400621 |
| 14.071900553 | 3.107219900 | 26.818400621 |
| 14.071900553 | 9.103103484 | 26.818400621 |
| 14.071900553 | 15.098988141 | 26.818400621 |
| 4.403280415 | 1.528879927 | 22.291800380 |

| | | |
|---|---|---|
| 4.403280415 | 7.524763914 | 22.291800380 |
| 4.403280415 | 13.520647498 | 22.291800380 |
| 9.609980578 | 1.528879927 | 22.291800380 |
| 9.609980578 | 7.524763914 | 22.291800380 |
| 9.609980578 | 13.520647498 | 22.291800380 |
| 14.816680274 | 1.528879927 | 22.291800380 |
| 14.816680274 | 7.524763914 | 22.291800380 |
| 14.816680274 | 13.520647498 | 22.291800380 |
| 4.403280415 | 4.526819977 | 22.291800380 |
| 4.403280415 | 10.522704098 | 22.291800380 |
| 4.403280415 | 16.518587682 | 22.291800380 |
| 9.609980578 | 4.526819977 | 22.291800380 |
| 9.609980578 | 10.522704098 | 22.291800380 |
| 9.609980578 | 16.518587682 | 22.291800380 |
| 14.816680274 | 4.526819977 | 22.291800380 |
| 14.816680274 | 10.522704098 | 22.291800380 |
| 14.816680274 | 16.518587682 | 22.291800380 |
| 1.055150031 | 1.608249942 | 26.818400621 |
| 1.055150031 | 7.604133928 | 26.818400621 |
| 1.055150031 | 13.600017513 | 26.818400621 |
| 6.261850077 | 1.608249942 | 26.818400621 |
| 6.261850077 | 7.604133928 | 26.818400621 |
| 6.261850077 | 13.600017513 | 26.818400621 |
| 11.468550240 | 1.608249942 | 26.818400621 |
| 11.468550240 | 7.604133928 | 26.818400621 |
| 11.468550240 | 13.600017513 | 26.818400621 |
| 1.055150031 | 4.606189992 | 26.818400621 |
| 1.055150031 | 10.602074112 | 26.818400621 |
| 1.055150031 | 16.597957696 | 26.818400621 |
| 6.261850077 | 4.606189992 | 26.818400621 |
| 6.261850077 | 10.602074112 | 26.818400621 |
| 6.261850077 | 16.597957696 | 26.818400621 |
| 11.468550240 | 4.606189992 | 26.818400621 |
| 11.468550240 | 10.602074112 | 26.818400621 |
| 11.468550240 | 16.597957696 | 26.818400621 |
| 2.603350314 | 1.498969958 | 20.000000298 |
| 2.603350314 | 7.494853676 | 20.000000298 |
| 2.603350314 | 13.490737797 | 20.000000298 |
| 7.810050011 | 1.498969958 | 20.000000298 |
| 7.810050011 | 7.494853676 | 20.000000298 |
| 7.810050011 | 13.490737797 | 20.000000298 |
| 13.016750638 | 1.498969958 | 20.000000298 |
| 13.016750638 | 7.494853676 | 20.000000298 |
| 13.016750638 | 13.490737797 | 20.000000298 |
| 2.731480362 | 1.570149971 | 24.562700093 |

| | | |
|---|---|---|
| 2.731480362 | 7.566034091 | 24.562700093 |
| 2.731480362 | 13.561917676 | 24.562700093 |
| 7.938180292 | 1.570149971 | 24.562700093 |
| 7.938180292 | 7.566034091 | 24.562700093 |
| 7.938180292 | 13.561917676 | 24.562700093 |
| 13.144879989 | 1.570149971 | 24.562700093 |
| 13.144879989 | 7.566034091 | 24.562700093 |
| 13.144879989 | 13.561917676 | 24.562700093 |
| 2.603350314 | 4.496909740 | 20.000000298 |
| 2.603350314 | 10.492793324 | 20.000000298 |
| 2.603350314 | 16.488677980 | 20.000000298 |
| 7.810050011 | 4.496909740 | 20.000000298 |
| 7.810050011 | 10.492793324 | 20.000000298 |
| 7.810050011 | 16.488677980 | 20.000000298 |
| 13.016750638 | 4.496909740 | 20.000000298 |
| 13.016750638 | 10.492793324 | 20.000000298 |
| 13.016750638 | 16.488677980 | 20.000000298 |
| 2.731480362 | 4.568090155 | 24.562700093 |
| 2.731480362 | 10.563974275 | 24.562700093 |
| 2.731480362 | 16.559857859 | 24.562700093 |
| 7.938180292 | 4.568090155 | 24.562700093 |
| 7.938180292 | 10.563974275 | 24.562700093 |
| 7.938180292 | 16.559857859 | 24.562700093 |
| 13.144879989 | 4.568090155 | 24.562700093 |
| 13.144879989 | 10.563974275 | 24.562700093 |
| 13.144879989 | 16.559857859 | 24.562700093 |
| 4.768780185 | 1.745140005 | 30.968898535 |
| 4.768780185 | 7.741024125 | 30.968898535 |
| 4.768780185 | 13.736908246 | 30.968898535 |
| 9.975479881 | 1.745140005 | 30.968898535 |
| 9.975479881 | 7.741024125 | 30.968898535 |
| 9.975479881 | 13.736908246 | 30.968898535 |
| 15.182180509 | 1.745140005 | 30.968898535 |
| 15.182180509 | 7.741024125 | 30.968898535 |
| 15.182180509 | 13.736908246 | 30.968898535 |
| 4.768780185 | 4.743079653 | 30.968898535 |
| 4.768780185 | 10.738963773 | 30.968898535 |
| 4.768780185 | 16.734847357 | 30.968898535 |
| 9.975479881 | 4.743079653 | 30.968898535 |
| 9.975479881 | 10.738963773 | 30.968898535 |
| 9.975479881 | 16.734847357 | 30.968898535 |
| 15.182180509 | 4.743079653 | 30.968898535 |
| 15.182180509 | 10.738963773 | 30.968898535 |
| 15.182180509 | 16.734847357 | 30.968898535 |
| 2.165490155 | 0.246140982 | 30.968898535 |

| | | |
|---|---|---|
| 2.165490155 | 6.242025085 | 30.968898535 |
| 2.165490155 | 12.237908670 | 30.968898535 |
| 7.372190084 | 0.246140982 | 30.968898535 |
| 7.372190084 | 6.242025085 | 30.968898535 |
| 7.372190084 | 12.237908670 | 30.968898535 |
| 12.578889781 | 0.246140982 | 30.968898535 |
| 12.578889781 | 6.242025085 | 30.968898535 |
| 12.578889781 | 12.237908670 | 30.968898535 |
| 2.165490155 | 3.244079809 | 30.968898535 |
| 2.165490155 | 9.239964197 | 30.968898535 |
| 2.165490155 | 15.235847781 | 30.968898535 |
| 7.372190084 | 3.244079809 | 30.968898535 |
| 7.372190084 | 9.239964197 | 30.968898535 |
| 7.372190084 | 15.235847781 | 30.968898535 |
| 12.578889781 | 3.244079809 | 30.968898535 |
| 12.578889781 | 9.239964197 | 30.968898535 |
| 12.578889781 | 15.235847781 | 30.968898535 |
| 1.291740014 | 1.742209960 | 29.339799285 |
| 1.291740014 | 7.738093946 | 29.339799285 |
| 1.291740014 | 13.733978066 | 29.339799285 |
| 6.498439827 | 1.742209960 | 29.339799285 |
| 6.498439827 | 7.738093946 | 29.339799285 |
| 6.498439827 | 13.733978066 | 29.339799285 |
| 11.705139989 | 1.742209960 | 29.339799285 |
| 11.705139989 | 7.738093946 | 29.339799285 |
| 11.705139989 | 13.733978066 | 29.339799285 |
| 1.291740014 | 4.740160195 | 29.339799285 |
| 1.291740014 | 10.736044315 | 29.339799285 |
| 1.291740014 | 16.731927900 | 29.339799285 |
| 6.498439827 | 4.740160195 | 29.339799285 |
| 6.498439827 | 10.736044315 | 29.339799285 |
| 6.498439827 | 16.731927900 | 29.339799285 |
| 11.705139989 | 4.740160195 | 29.339799285 |
| 11.705139989 | 10.736044315 | 29.339799285 |
| 11.705139989 | 16.731927900 | 29.339799285 |
| 3.895110229 | 0.243256000 | 29.339700937 |
| 3.895110229 | 6.239139936 | 29.339700937 |
| 3.895110229 | 12.235023520 | 29.339700937 |
| 9.101810158 | 0.243256000 | 29.339700937 |
| 9.101810158 | 6.239139936 | 29.339700937 |
| 9.101810158 | 12.235023520 | 29.339700937 |
| 14.308509855 | 0.243256000 | 29.339700937 |
| 14.308509855 | 6.239139936 | 29.339700937 |
| 14.308509855 | 12.235023520 | 29.339700937 |
| 3.895110229 | 3.241199752 | 29.339700937 |

| | | |
|---|---|---|
| 3.895110229 | 9.237083337 | 29.339700937 |
| 3.895110229 | 15.232967993 | 29.339700937 |
| 9.101810158 | 3.241199752 | 29.339700937 |
| 9.101810158 | 9.237083337 | 29.339700937 |
| 9.101810158 | 15.232967993 | 29.339700937 |
| 14.308509855 | 3.241199752 | 29.339700937 |
| 14.308509855 | 9.237083337 | 29.339700937 |
| 14.308509855 | 15.232967993 | 29.339700937 |
| 1.300430020 | 1.747209920 | 32.605299354 |
| 1.300430020 | 7.743093906 | 32.605299354 |
| 1.300430020 | 13.738977491 | 32.605299354 |
| 6.507130066 | 1.747209920 | 32.605299354 |
| 6.507130066 | 7.743093906 | 32.605299354 |
| 6.507130066 | 13.738977491 | 32.605299354 |
| 11.713830228 | 1.747209920 | 32.605299354 |
| 11.713830228 | 7.743093906 | 32.605299354 |
| 11.713830228 | 13.738977491 | 32.605299354 |
| 1.300430020 | 4.745149970 | 32.605299354 |
| 1.300430020 | 10.741034090 | 32.605299354 |
| 1.300430020 | 16.736917674 | 32.605299354 |
| 6.507130066 | 4.745149970 | 32.605299354 |
| 6.507130066 | 10.741034090 | 32.605299354 |
| 6.507130066 | 16.736917674 | 32.605299354 |
| 11.713830228 | 4.745149970 | 32.605299354 |
| 11.713830228 | 10.741034090 | 32.605299354 |
| 11.713830228 | 16.736917674 | 32.605299354 |
| 3.903760433 | 0.248339002 | 32.605499029 |
| 3.903760433 | 6.244222988 | 32.605499029 |
| 3.903760433 | 12.240106572 | 32.605499029 |
| 9.110460363 | 0.248339002 | 32.605499029 |
| 9.110460363 | 6.244222988 | 32.605499029 |
| 9.110460363 | 12.240106572 | 32.605499029 |
| 14.317160059 | 0.248339002 | 32.605499029 |
| 14.317160059 | 6.244222988 | 32.605499029 |
| 14.317160059 | 12.240106572 | 32.605499029 |
| 3.903760433 | 3.246279856 | 32.605499029 |
| 3.903760433 | 9.242164244 | 32.605499029 |
| 3.903760433 | 15.238047828 | 32.605499029 |
| 9.110460363 | 3.246279856 | 32.605499029 |
| 9.110460363 | 9.242164244 | 32.605499029 |
| 9.110460363 | 15.238047828 | 32.605499029 |
| 14.317160059 | 3.246279856 | 32.605499029 |
| 14.317160059 | 9.242164244 | 32.605499029 |
| 14.317160059 | 15.238047828 | 32.605499029 |
| 0.431960015 | 1.187199890 | 35.576900840 |

| | | |
|---|---|---|
| 0.431960015 | 7.183083876 | 35.576900840 |
| 0.431960015 | 13.178967997 | 35.576900840 |
| 5.638660148 | 1.187199890 | 35.576900840 |
| 5.638660148 | 7.183083876 | 35.576900840 |
| 5.638660148 | 13.178967997 | 35.576900840 |
| 10.845360310 | 1.187199890 | 35.576900840 |
| 10.845360310 | 7.183083876 | 35.576900840 |
| 10.845360310 | 13.178967997 | 35.576900840 |
| 3.035310300 | 2.665419771 | 35.576701164 |
| 3.035310300 | 8.661303624 | 35.576701164 |
| 3.035310300 | 14.657187208 | 35.576701164 |
| 8.242009997 | 2.665419771 | 35.576701164 |
| 8.242009997 | 8.661303624 | 35.576701164 |
| 8.242009997 | 14.657187208 | 35.576701164 |
| 13.448710624 | 2.665419771 | 35.576701164 |
| 13.448710624 | 8.661303624 | 35.576701164 |
| 13.448710624 | 14.657187208 | 35.576701164 |
| 0.431960015 | 4.185139940 | 35.576900840 |
| 0.431960015 | 10.181023524 | 35.576900840 |
| 0.431960015 | 16.176908181 | 35.576900840 |
| 5.638660148 | 4.185139940 | 35.576900840 |
| 5.638660148 | 10.181023524 | 35.576900840 |
| 5.638660148 | 16.176908181 | 35.576900840 |
| 10.845360310 | 4.185139940 | 35.576900840 |
| 10.845360310 | 10.181023524 | 35.576900840 |
| 10.845360310 | 16.176908181 | 35.576900840 |
| 3.035310300 | 5.663359687 | 35.576701164 |
| 3.035310300 | 11.659243808 | 35.576701164 |
| 3.035310300 | 17.655127392 | 35.576701164 |
| 8.242009997 | 5.663359687 | 35.576701164 |
| 8.242009997 | 11.659243808 | 35.576701164 |
| 8.242009997 | 17.655127392 | 35.576701164 |
| 13.448710624 | 5.663359687 | 35.576701164 |
| 13.448710624 | 11.659243808 | 35.576701164 |
| 13.448710624 | 17.655127392 | 35.576701164 |
| 1.299740126 | 5.677189855 | 37.355598807 |
| 1.299740126 | 11.673073440 | 37.355598807 |
| 1.299740126 | 17.668958096 | 37.355598807 |
| 6.506440172 | 5.677189855 | 37.355598807 |
| 6.506440172 | 11.673073440 | 37.355598807 |
| 6.506440172 | 17.668958096 | 37.355598807 |
| 11.713140335 | 5.677189855 | 37.355598807 |
| 11.713140335 | 11.673073440 | 37.355598807 |
| 11.713140335 | 17.668958096 | 37.355598807 |
| 3.903090324 | 1.159529905 | 37.355500460 |

| | | |
|---|---|---|
| 3.903090324 | 7.155413891 | 37.355500460 |
| 3.903090324 | 13.151298012 | 37.355500460 |
| 9.109790021 | 1.159529905 | 37.355500460 |
| 9.109790021 | 7.155413891 | 37.355500460 |
| 9.109790021 | 13.151298012 | 37.355500460 |
| 14.316490649 | 1.159529905 | 37.355500460 |
| 14.316490649 | 7.155413891 | 37.355500460 |
| 14.316490649 | 13.151298012 | 37.355500460 |
| 1.299740126 | 2.679249939 | 37.355598807 |
| 1.299740126 | 8.675133792 | 37.355598807 |
| 1.299740126 | 14.671017912 | 37.355598807 |
| 6.506440172 | 2.679249939 | 37.355598807 |
| 6.506440172 | 8.675133792 | 37.355598807 |
| 6.506440172 | 14.671017912 | 37.355598807 |
| 11.713140335 | 2.679249939 | 37.355598807 |
| 11.713140335 | 8.675133792 | 37.355598807 |
| 11.713140335 | 14.671017912 | 37.355598807 |
| 3.903090324 | 4.157469955 | 37.355500460 |
| 3.903090324 | 10.153353539 | 37.355500460 |
| 3.903090324 | 16.149238195 | 37.355500460 |
| 9.109790021 | 4.157469955 | 37.355500460 |
| 9.109790021 | 10.153353539 | 37.355500460 |
| 9.109790021 | 16.149238195 | 37.355500460 |
| 14.316490649 | 4.157469955 | 37.355500460 |
| 14.316490649 | 10.153353539 | 37.355500460 |
| 14.316490649 | 16.149238195 | 37.355500460 |
| 1.299740126 | 5.677189855 | 39.734199643 |
| 1.299740126 | 11.673073440 | 39.734199643 |
| 1.299740126 | 17.668958096 | 39.734199643 |
| 6.506440172 | 5.677189855 | 39.734199643 |
| 6.506440172 | 11.673073440 | 39.734199643 |
| 6.506440172 | 17.668958096 | 39.734199643 |
| 11.713140335 | 5.677189855 | 39.734199643 |
| 11.713140335 | 11.673073440 | 39.734199643 |
| 11.713140335 | 17.668958096 | 39.734199643 |
| 3.903090324 | 1.159529905 | 39.734199643 |
| 3.903090324 | 7.155413891 | 39.734199643 |
| 3.903090324 | 13.151298012 | 39.734199643 |
| 9.109790021 | 1.159529905 | 39.734199643 |
| 9.109790021 | 7.155413891 | 39.734199643 |
| 9.109790021 | 13.151298012 | 39.734199643 |
| 14.316490649 | 1.159529905 | 39.734199643 |
| 14.316490649 | 7.155413891 | 39.734199643 |
| 14.316490649 | 13.151298012 | 39.734199643 |
| 1.299740126 | 2.679249939 | 39.734199643 |

```
    1.299740126          8.675133792          39.734199643
    1.299740126         14.671017912          39.734199643
    6.506440172          2.679249939          39.734199643
    6.506440172          8.675133792          39.734199643
    6.506440172         14.671017912          39.734199643
   11.713140335          2.679249939          39.734199643
   11.713140335          8.675133792          39.734199643
   11.713140335         14.671017912          39.734199643
    3.903090324          4.157469955          39.734199643
    3.903090324         10.153353539          39.734199643
    3.903090324         16.149238195          39.734199643
    9.109790021          4.157469955          39.734199643
    9.109790021         10.153353539          39.734199643
    9.109790021         16.149238195          39.734199643
   14.316490649          4.157469955          39.734199643
   14.316490649         10.153353539          39.734199643
   14.316490649         16.149238195          39.734199643


Ag/MoSe2/Si
1.0
     15.9722995758          0.0000000000          0.0000000000
      0.0000000000         18.3864002228          0.0000000000
      0.0000000000          0.0000000000         50.0000000000
  Ag    Mo    Se    Si
 144    36    72   108
Cartesian
    5.324110014          0.000000000          20.000000298
    5.324110014          6.128800257          20.000000298
    5.324110014         12.257600514          20.000000298
   10.648209555          0.000000000          20.000000298
   10.648209555          6.128800257          20.000000298
   10.648209555         12.257600514          20.000000298
    0.000010155          0.000000000          20.000000298
    0.000010155          6.128800257          20.000000298
    0.000010155         12.257600514          20.000000298
    5.272609840          6.086380133          26.672500372
    5.272609840         12.215179842          26.672500372
    5.272609840         18.343980647          26.672500372
   10.596709857          6.086380133          26.672500372
   10.596709857         12.215179842          26.672500372
   10.596709857         18.343980647          26.672500372
   15.920809399          6.086380133          26.672500372
   15.920809399         12.215179842          26.672500372
   15.920809399         18.343980647          26.672500372
    5.324110014          3.064400128          20.000000298
```

| | | |
|---|---|---|
| 5.324110014 | 9.193200111 | 20.000000298 |
| 5.324110014 | 15.321999820 | 20.000000298 |
| 10.648209555 | 3.064400128 | 20.000000298 |
| 10.648209555 | 9.193200111 | 20.000000298 |
| 10.648209555 | 15.321999820 | 20.000000298 |
| 0.000010155 | 3.064400128 | 20.000000298 |
| 0.000010155 | 9.193200111 | 20.000000298 |
| 0.000010155 | 15.321999820 | 20.000000298 |
| 5.272609840 | 3.021980004 | 26.672500372 |
| 5.272609840 | 9.150779987 | 26.672500372 |
| 5.272609840 | 15.279580244 | 26.672500372 |
| 10.596709857 | 3.021980004 | 26.672500372 |
| 10.596709857 | 9.150779987 | 26.672500372 |
| 10.596709857 | 15.279580244 | 26.672500372 |
| 15.920809399 | 3.021980004 | 26.672500372 |
| 15.920809399 | 9.150779987 | 26.672500372 |
| 15.920809399 | 15.279580244 | 26.672500372 |
| 1.756749925 | 6.113820197 | 22.237199545 |
| 1.756749925 | 12.242620454 | 22.237199545 |
| 1.756749925 | 18.371420163 | 22.237199545 |
| 7.080849942 | 6.113820197 | 22.237199545 |
| 7.080849942 | 12.242620454 | 22.237199545 |
| 7.080849942 | 18.371420163 | 22.237199545 |
| 12.404949483 | 6.113820197 | 22.237199545 |
| 12.404949483 | 12.242620454 | 22.237199545 |
| 12.404949483 | 18.371420163 | 22.237199545 |
| 1.756749925 | 3.049420068 | 22.237199545 |
| 1.756749925 | 9.178220051 | 22.237199545 |
| 1.756749925 | 15.307019760 | 22.237199545 |
| 7.080849942 | 3.049420068 | 22.237199545 |
| 7.080849942 | 9.178220051 | 22.237199545 |
| 7.080849942 | 15.307019760 | 22.237199545 |
| 12.404949483 | 3.049420068 | 22.237199545 |
| 12.404949483 | 9.178220051 | 22.237199545 |
| 12.404949483 | 15.307019760 | 22.237199545 |
| 3.513999900 | 6.099400149 | 24.461200833 |
| 3.513999900 | 12.228200406 | 24.461200833 |
| 3.513999900 | 18.357000115 | 24.461200833 |
| 8.838100155 | 6.099400149 | 24.461200833 |
| 8.838100155 | 12.228200406 | 24.461200833 |
| 8.838100155 | 18.357000115 | 24.461200833 |
| 14.162199696 | 6.099400149 | 24.461200833 |
| 14.162199696 | 12.228200406 | 24.461200833 |
| 14.162199696 | 18.357000115 | 24.461200833 |
| 3.514009896 | 3.035000021 | 24.461200833 |

| | | |
|---|---|---|
| 3.514009896 | 9.163800004 | 24.461200833 |
| 3.514009896 | 15.292599713 | 24.461200833 |
| 8.838109675 | 3.035000021 | 24.461200833 |
| 8.838109675 | 9.163800004 | 24.461200833 |
| 8.838109675 | 15.292599713 | 24.461200833 |
| 14.162209217 | 3.035000021 | 24.461200833 |
| 14.162209217 | 9.163800004 | 24.461200833 |
| 14.162209217 | 15.292599713 | 24.461200833 |
| 4.418799695 | 1.517220004 | 22.237199545 |
| 4.418799695 | 7.646019850 | 22.237199545 |
| 4.418799695 | 13.774820107 | 22.237199545 |
| 9.742899237 | 1.517220004 | 22.237199545 |
| 9.742899237 | 7.646019850 | 22.237199545 |
| 9.742899237 | 13.774820107 | 22.237199545 |
| 15.066999730 | 1.517220004 | 22.237199545 |
| 15.066999730 | 7.646019850 | 22.237199545 |
| 15.066999730 | 13.774820107 | 22.237199545 |
| 4.418799695 | 4.581620270 | 22.237199545 |
| 4.418799695 | 10.710420800 | 22.237199545 |
| 4.418799695 | 16.839220509 | 22.237199545 |
| 9.742899237 | 4.581620270 | 22.237199545 |
| 9.742899237 | 10.710420800 | 22.237199545 |
| 9.742899237 | 16.839220509 | 22.237199545 |
| 15.066999730 | 4.581620270 | 22.237199545 |
| 15.066999730 | 10.710420800 | 22.237199545 |
| 15.066999730 | 16.839220509 | 22.237199545 |
| 0.851953937 | 1.502799956 | 24.461200833 |
| 0.851953937 | 7.631599802 | 24.461200833 |
| 0.851953937 | 13.760400059 | 24.461200833 |
| 6.176053717 | 1.502799956 | 24.461200833 |
| 6.176053717 | 7.631599802 | 24.461200833 |
| 6.176053717 | 13.760400059 | 24.461200833 |
| 11.500153258 | 1.502799956 | 24.461200833 |
| 11.500153258 | 7.631599802 | 24.461200833 |
| 11.500153258 | 13.760400059 | 24.461200833 |
| 0.851953937 | 4.567200222 | 24.461200833 |
| 0.851953937 | 10.696000753 | 24.461200833 |
| 0.851953937 | 16.824800462 | 24.461200833 |
| 6.176053717 | 4.567200222 | 24.461200833 |
| 6.176053717 | 10.696000753 | 24.461200833 |
| 6.176053717 | 16.824800462 | 24.461200833 |
| 11.500153258 | 4.567200222 | 24.461200833 |
| 11.500153258 | 10.696000753 | 24.461200833 |
| 11.500153258 | 16.824800462 | 24.461200833 |
| 2.662060005 | 1.532200064 | 20.000000298 |

| | | |
|---|---|---|
| 2.662060005 | 7.660999910 | 20.000000298 |
| 2.662060005 | 13.789800167 | 20.000000298 |
| 7.986160260 | 1.532200064 | 20.000000298 |
| 7.986160260 | 7.660999910 | 20.000000298 |
| 7.986160260 | 13.789800167 | 20.000000298 |
| 13.310259801 | 1.532200064 | 20.000000298 |
| 13.310259801 | 7.660999910 | 20.000000298 |
| 13.310259801 | 13.789800167 | 20.000000298 |
| 2.610559832 | 1.489779940 | 26.672500372 |
| 2.610559832 | 7.618579786 | 26.672500372 |
| 2.610559832 | 13.747380591 | 26.672500372 |
| 7.934659611 | 1.489779940 | 26.672500372 |
| 7.934659611 | 7.618579786 | 26.672500372 |
| 7.934659611 | 13.747380591 | 26.672500372 |
| 13.258759152 | 1.489779940 | 26.672500372 |
| 13.258759152 | 7.618579786 | 26.672500372 |
| 13.258759152 | 13.747380591 | 26.672500372 |
| 2.662060005 | 4.596590192 | 20.000000298 |
| 2.662060005 | 10.725389901 | 20.000000298 |
| 2.662060005 | 16.854190706 | 20.000000298 |
| 7.986160260 | 4.596590192 | 20.000000298 |
| 7.986160260 | 10.725389901 | 20.000000298 |
| 7.986160260 | 16.854190706 | 20.000000298 |
| 13.310259801 | 4.596590192 | 20.000000298 |
| 13.310259801 | 10.725389901 | 20.000000298 |
| 13.310259801 | 16.854190706 | 20.000000298 |
| 2.610559832 | 4.554180205 | 26.672500372 |
| 2.610559832 | 10.682980188 | 26.672500372 |
| 2.610559832 | 16.811779897 | 26.672500372 |
| 7.934659611 | 4.554180205 | 26.672500372 |
| 7.934659611 | 10.682980188 | 26.672500372 |
| 7.934659611 | 16.811779897 | 26.672500372 |
| 13.258759152 | 4.554180205 | 26.672500372 |
| 13.258759152 | 10.682980188 | 26.672500372 |
| 13.258759152 | 16.811779897 | 26.672500372 |
| 0.758539037 | 1.413870028 | 31.071400642 |
| 0.758539037 | 7.542670148 | 31.071400642 |
| 0.758539037 | 13.671469857 | 31.071400642 |
| 6.082638816 | 1.413870028 | 31.071400642 |
| 6.082638816 | 7.542670148 | 31.071400642 |
| 6.082638816 | 13.671469857 | 31.071400642 |
| 11.406738833 | 1.413870028 | 31.071400642 |
| 11.406738833 | 7.542670148 | 31.071400642 |
| 11.406738833 | 13.671469857 | 31.071400642 |
| 0.758539037 | 4.478270019 | 31.071400642 |

| | | |
|---|---|---|
| 0.758539037 | 10.607070550 | 31.071400642 |
| 0.758539037 | 16.735870259 | 31.071400642 |
| 6.082638816 | 4.478270019 | 31.071400642 |
| 6.082638816 | 10.607070550 | 31.071400642 |
| 6.082638816 | 16.735870259 | 31.071400642 |
| 11.406738833 | 4.478270019 | 31.071400642 |
| 11.406738833 | 10.607070550 | 31.071400642 |
| 11.406738833 | 16.735870259 | 31.071400642 |
| 3.420599756 | 6.010439809 | 31.071400642 |
| 3.420599756 | 12.139239518 | 31.071400642 |
| 3.420599756 | 18.268040323 | 31.071400642 |
| 8.744700011 | 6.010439809 | 31.071400642 |
| 8.744700011 | 12.139239518 | 31.071400642 |
| 8.744700011 | 18.268040323 | 31.071400642 |
| 14.068799552 | 6.010439809 | 31.071400642 |
| 14.068799552 | 12.139239518 | 31.071400642 |
| 14.068799552 | 18.268040323 | 31.071400642 |
| 3.420599756 | 2.946039954 | 31.071400642 |
| 3.420599756 | 9.074840211 | 31.071400642 |
| 3.420599756 | 15.203639920 | 31.071400642 |
| 8.744700011 | 2.946039954 | 31.071400642 |
| 8.744700011 | 9.074840211 | 31.071400642 |
| 8.744700011 | 15.203639920 | 31.071400642 |
| 14.068799552 | 2.946039954 | 31.071400642 |
| 14.068799552 | 9.074840211 | 31.071400642 |
| 14.068799552 | 15.203639920 | 31.071400642 |
| 2.530689842 | 1.412520135 | 32.844799757 |
| 2.530689842 | 7.541319981 | 32.844799757 |
| 2.530689842 | 13.670120785 | 32.844799757 |
| 7.854789621 | 1.412520135 | 32.844799757 |
| 7.854789621 | 7.541319981 | 32.844799757 |
| 7.854789621 | 13.670120785 | 32.844799757 |
| 13.178889162 | 1.412520135 | 32.844799757 |
| 13.178889162 | 7.541319981 | 32.844799757 |
| 13.178889162 | 13.670120785 | 32.844799757 |
| 2.530689842 | 4.476920400 | 32.844799757 |
| 2.530689842 | 10.605720383 | 32.844799757 |
| 2.530689842 | 16.734520092 | 32.844799757 |
| 7.854789621 | 4.476920400 | 32.844799757 |
| 7.854789621 | 10.605720383 | 32.844799757 |
| 7.854789621 | 16.734520092 | 32.844799757 |
| 13.178889162 | 4.476920400 | 32.844799757 |
| 13.178889162 | 10.605720383 | 32.844799757 |
| 13.178889162 | 16.734520092 | 32.844799757 |
| 5.192739850 | 6.009240330 | 32.844799757 |

| | | |
|---|---|---|
| 5.192739850 | 12.138040587 | 32.844799757 |
| 5.192739850 | 18.266840296 | 32.844799757 |
| 10.516839868 | 6.009240330 | 32.844799757 |
| 10.516839868 | 12.138040587 | 32.844799757 |
| 10.516839868 | 18.266840296 | 32.844799757 |
| 15.840939409 | 6.009240330 | 32.844799757 |
| 15.840939409 | 12.138040587 | 32.844799757 |
| 15.840939409 | 18.266840296 | 32.844799757 |
| 5.192739850 | 2.944839928 | 32.844799757 |
| 5.192739850 | 9.073640184 | 32.844799757 |
| 5.192739850 | 15.202439893 | 32.844799757 |
| 10.516839868 | 2.944839928 | 32.844799757 |
| 10.516839868 | 9.073640184 | 32.844799757 |
| 10.516839868 | 15.202439893 | 32.844799757 |
| 15.840939409 | 2.944839928 | 32.844799757 |
| 15.840939409 | 9.073640184 | 32.844799757 |
| 15.840939409 | 15.202439893 | 32.844799757 |
| 2.532909960 | 1.414860050 | 29.316899180 |
| 2.532909960 | 7.543660307 | 29.316899180 |
| 2.532909960 | 13.672460564 | 29.316899180 |
| 7.857009739 | 1.414860050 | 29.316899180 |
| 7.857009739 | 7.543660307 | 29.316899180 |
| 7.857009739 | 13.672460564 | 29.316899180 |
| 13.181109280 | 1.414860050 | 29.316899180 |
| 13.181109280 | 7.543660307 | 29.316899180 |
| 13.181109280 | 13.672460564 | 29.316899180 |
| 2.532909960 | 4.479260178 | 29.316899180 |
| 2.532909960 | 10.608060161 | 29.316899180 |
| 2.532909960 | 16.736859870 | 29.316899180 |
| 7.857009739 | 4.479260178 | 29.316899180 |
| 7.857009739 | 10.608060161 | 29.316899180 |
| 7.857009739 | 16.736859870 | 29.316899180 |
| 13.181109280 | 4.479260178 | 29.316899180 |
| 13.181109280 | 10.608060161 | 29.316899180 |
| 13.181109280 | 16.736859870 | 29.316899180 |
| 5.194959968 | 6.011479832 | 29.316899180 |
| 5.194959968 | 12.140279541 | 29.316899180 |
| 5.194959968 | 18.269080346 | 29.316899180 |
| 10.519059986 | 6.011479832 | 29.316899180 |
| 10.519059986 | 12.140279541 | 29.316899180 |
| 10.519059986 | 18.269080346 | 29.316899180 |
| 15.843159527 | 6.011479832 | 29.316899180 |
| 15.843159527 | 12.140279541 | 29.316899180 |
| 15.843159527 | 18.269080346 | 29.316899180 |
| 5.194959968 | 2.947079978 | 29.316899180 |

| | | |
|---|---|---|
| 5.194959968 | 9.075880234 | 29.316899180 |
| 5.194959968 | 15.204679943 | 29.316899180 |
| 10.519059986 | 2.947079978 | 29.316899180 |
| 10.519059986 | 9.075880234 | 29.316899180 |
| 10.519059986 | 15.204679943 | 29.316899180 |
| 15.843159527 | 2.947079978 | 29.316899180 |
| 15.843159527 | 9.075880234 | 29.316899180 |
| 15.843159527 | 15.204679943 | 29.316899180 |
| 0.441700971 | 1.213520080 | 35.948398709 |
| 0.441700971 | 7.342319926 | 35.948398709 |
| 0.441700971 | 13.471120183 | 35.948398709 |
| 5.765800721 | 1.213520080 | 35.948398709 |
| 5.765800721 | 7.342319926 | 35.948398709 |
| 5.765800721 | 13.471120183 | 35.948398709 |
| 11.089900738 | 1.213520080 | 35.948398709 |
| 11.089900738 | 7.342319926 | 35.948398709 |
| 11.089900738 | 13.471120183 | 35.948398709 |
| 3.103759756 | 2.724499945 | 35.948398709 |
| 3.103759756 | 8.853299928 | 35.948398709 |
| 3.103759756 | 14.982099637 | 35.948398709 |
| 8.427860012 | 2.724499945 | 35.948398709 |
| 8.427860012 | 8.853299928 | 35.948398709 |
| 8.427860012 | 14.982099637 | 35.948398709 |
| 13.751959553 | 2.724499945 | 35.948398709 |
| 13.751959553 | 8.853299928 | 35.948398709 |
| 13.751959553 | 14.982099637 | 35.948398709 |
| 0.441700971 | 4.277910208 | 35.948398709 |
| 0.441700971 | 10.406709917 | 35.948398709 |
| 0.441700971 | 16.535510722 | 35.948398709 |
| 5.765800721 | 4.277910208 | 35.948398709 |
| 5.765800721 | 10.406709917 | 35.948398709 |
| 5.765800721 | 16.535510722 | 35.948398709 |
| 11.089900738 | 4.277910208 | 35.948398709 |
| 11.089900738 | 10.406709917 | 35.948398709 |
| 11.089900738 | 16.535510722 | 35.948398709 |
| 3.103759756 | 5.788900073 | 35.948398709 |
| 3.103759756 | 11.917700330 | 35.948398709 |
| 3.103759756 | 18.046500039 | 35.948398709 |
| 8.427860012 | 5.788900073 | 35.948398709 |
| 8.427860012 | 11.917700330 | 35.948398709 |
| 8.427860012 | 18.046500039 | 35.948398709 |
| 13.751959553 | 5.788900073 | 35.948398709 |
| 13.751959553 | 11.917700330 | 35.948398709 |
| 13.751959553 | 18.046500039 | 35.948398709 |
| 1.329049913 | 5.803040115 | 37.684699893 |

| | | |
|---|---|---|
| 1.329049913 | 11.931839823 | 37.684699893 |
| 1.329049913 | 18.060640628 | 37.684699893 |
| 6.653149692 | 5.803040115 | 37.684699893 |
| 6.653149692 | 11.931839823 | 37.684699893 |
| 6.653149692 | 18.060640628 | 37.684699893 |
| 11.977249234 | 5.803040115 | 37.684699893 |
| 11.977249234 | 11.931839823 | 37.684699893 |
| 11.977249234 | 18.060640628 | 37.684699893 |
| 3.991109918 | 1.185229997 | 37.684601545 |
| 3.991109918 | 7.314029980 | 37.684601545 |
| 3.991109918 | 13.442830237 | 37.684601545 |
| 9.315209459 | 1.185229997 | 37.684601545 |
| 9.315209459 | 7.314029980 | 37.684601545 |
| 9.315209459 | 13.442830237 | 37.684601545 |
| 14.639309953 | 1.185229997 | 37.684601545 |
| 14.639309953 | 7.314029980 | 37.684601545 |
| 14.639309953 | 13.442830237 | 37.684601545 |
| 1.329049913 | 2.738639986 | 37.684699893 |
| 1.329049913 | 8.867439969 | 37.684699893 |
| 1.329049913 | 14.996240226 | 37.684699893 |
| 6.653149692 | 2.738639986 | 37.684699893 |
| 6.653149692 | 8.867439969 | 37.684699893 |
| 6.653149692 | 14.996240226 | 37.684699893 |
| 11.977249234 | 2.738639986 | 37.684699893 |
| 11.977249234 | 8.867439969 | 37.684699893 |
| 11.977249234 | 14.996240226 | 37.684699893 |
| 3.991109918 | 4.249630125 | 37.684601545 |
| 3.991109918 | 10.378429834 | 37.684601545 |
| 3.991109918 | 16.507230639 | 37.684601545 |
| 9.315209459 | 4.249630125 | 37.684601545 |
| 9.315209459 | 10.378429834 | 37.684601545 |
| 9.315209459 | 16.507230639 | 37.684601545 |
| 14.639309953 | 4.249630125 | 37.684601545 |
| 14.639309953 | 10.378429834 | 37.684601545 |
| 14.639309953 | 16.507230639 | 37.684601545 |
| 1.329049913 | 5.803040115 | 40.072602034 |
| 1.329049913 | 11.931839823 | 40.072602034 |
| 1.329049913 | 18.060640628 | 40.072602034 |
| 6.653149692 | 5.803040115 | 40.072602034 |
| 6.653149692 | 11.931839823 | 40.072602034 |
| 6.653149692 | 18.060640628 | 40.072602034 |
| 11.977249234 | 5.803040115 | 40.072602034 |
| 11.977249234 | 11.931839823 | 40.072602034 |
| 11.977249234 | 18.060640628 | 40.072602034 |
| 3.991109918 | 1.185229997 | 40.072602034 |

```
     3.991109918          7.314029980          40.072602034
     3.991109918         13.442830237          40.072602034
     9.315209459          1.185229997          40.072602034
     9.315209459          7.314029980          40.072602034
     9.315209459         13.442830237          40.072602034
    14.639309953          1.185229997          40.072602034
    14.639309953          7.314029980          40.072602034
    14.639309953         13.442830237          40.072602034
     1.329049913          2.738639986          40.072602034
     1.329049913          8.867439969          40.072602034
     1.329049913         14.996240226          40.072602034
     6.653149692          2.738639986          40.072602034
     6.653149692          8.867439969          40.072602034
     6.653149692         14.996240226          40.072602034
    11.977249234          2.738639986          40.072602034
    11.977249234          8.867439969          40.072602034
    11.977249234         14.996240226          40.072602034
     3.991109918          4.249630125          40.072602034
     3.991109918         10.378429834          40.072602034
     3.991109918         16.507230639          40.072602034
     9.315209459          4.249630125          40.072602034
     9.315209459         10.378429834          40.072602034
     9.315209459         16.507230639          40.072602034
    14.639309953          4.249630125          40.072602034
    14.639309953         10.378429834          40.072602034
    14.639309953         16.507230639          40.072602034


Ag/WS2/Si
1.0
      15.6965999603          0.0000000000          0.0000000000
       0.0000000000         17.9897994995          0.0000000000
       0.0000000000          0.0000000000         50.0000000000
   Ag    W    S    Si
  144   36   72   108
Cartesian
      5.232180028          0.000000000         20.000000298
      5.232180028          5.996600012         20.000000298
      5.232180028         11.993200024         20.000000298
     10.464379702          0.000000000         20.000000298
     10.464379702          5.996600012         20.000000298
     10.464379702         11.993200024         20.000000298
     15.696580313          0.000000000         20.000000298
     15.696580313          5.996600012         20.000000298
     15.696580313         11.993200024         20.000000298
      5.160210214          5.948310080         26.780399680
```

| | | |
|---|---|---|
| 5.160210214 | 11.944910092 | 26.780399680 |
| 5.160210214 | 17.941509567 | 26.780399680 |
| 10.392410357 | 5.948310080 | 26.780399680 |
| 10.392410357 | 11.944910092 | 26.780399680 |
| 10.392410357 | 17.941509567 | 26.780399680 |
| 15.624610032 | 5.948310080 | 26.780399680 |
| 15.624610032 | 11.944910092 | 26.780399680 |
| 15.624610032 | 17.941509567 | 26.780399680 |
| 5.232180028 | 2.998279901 | 20.000000298 |
| 5.232180028 | 8.994879913 | 20.000000298 |
| 5.232180028 | 14.991479925 | 20.000000298 |
| 10.464379702 | 2.998279901 | 20.000000298 |
| 10.464379702 | 8.994879913 | 20.000000298 |
| 10.464379702 | 14.991479925 | 20.000000298 |
| 15.696580313 | 2.998279901 | 20.000000298 |
| 15.696580313 | 8.994879913 | 20.000000298 |
| 15.696580313 | 14.991479925 | 20.000000298 |
| 5.160210214 | 2.950010074 | 26.780399680 |
| 5.160210214 | 8.946609818 | 26.780399680 |
| 5.160210214 | 14.943209293 | 26.780399680 |
| 10.392410357 | 2.950010074 | 26.780399680 |
| 10.392410357 | 8.946609818 | 26.780399680 |
| 10.392410357 | 14.943209293 | 26.780399680 |
| 15.624610032 | 2.950010074 | 26.780399680 |
| 15.624610032 | 8.946609818 | 26.780399680 |
| 15.624610032 | 14.943209293 | 26.780399680 |
| 1.719030046 | 5.979379802 | 22.276699543 |
| 1.719030046 | 11.975979277 | 22.276699543 |
| 1.719030046 | 17.972579826 | 22.276699543 |
| 6.951229838 | 5.979379802 | 22.276699543 |
| 6.951229838 | 11.975979277 | 22.276699543 |
| 6.951229838 | 17.972579826 | 22.276699543 |
| 12.183430448 | 5.979379802 | 22.276699543 |
| 12.183430448 | 11.975979277 | 22.276699543 |
| 12.183430448 | 17.972579826 | 22.276699543 |
| 1.719030046 | 2.981079796 | 22.276699543 |
| 1.719030046 | 8.977679540 | 22.276699543 |
| 1.719030046 | 14.974279552 | 22.276699543 |
| 6.951229838 | 2.981079796 | 22.276699543 |
| 6.951229838 | 8.977679540 | 22.276699543 |
| 6.951229838 | 14.974279552 | 22.276699543 |
| 12.183430448 | 2.981079796 | 22.276699543 |
| 12.183430448 | 8.977679540 | 22.276699543 |
| 12.183430448 | 14.974279552 | 22.276699543 |
| 3.439080119 | 5.963379841 | 24.534900486 |

| | | |
|---|---|---|
| 3.439080119 | 11.959979853 | 24.534900486 |
| 3.439080119 | 17.956579329 | 24.534900486 |
| 8.671279794 | 5.963379841 | 24.534900486 |
| 8.671279794 | 11.959979853 | 24.534900486 |
| 8.671279794 | 17.956579329 | 24.534900486 |
| 13.903480404 | 5.963379841 | 24.534900486 |
| 13.903480404 | 11.959979853 | 24.534900486 |
| 13.903480404 | 17.956579329 | 24.534900486 |
| 3.439090177 | 2.965079835 | 24.534900486 |
| 3.439090177 | 8.961679579 | 24.534900486 |
| 3.439090177 | 14.958279055 | 24.534900486 |
| 8.671290085 | 2.965079835 | 24.534900486 |
| 8.671290085 | 8.961679579 | 24.534900486 |
| 8.671290085 | 14.958279055 | 24.534900486 |
| 13.903489760 | 2.965079835 | 24.534900486 |
| 13.903489760 | 8.961679579 | 24.534900486 |
| 13.903489760 | 14.958279055 | 24.534900486 |
| 4.335130000 | 1.481930061 | 22.276699543 |
| 4.335130000 | 7.478529939 | 22.276699543 |
| 4.335130000 | 13.475129951 | 22.276699543 |
| 9.567330143 | 1.481930061 | 22.276699543 |
| 9.567330143 | 7.478529939 | 22.276699543 |
| 9.567330143 | 13.475129951 | 22.276699543 |
| 14.799529818 | 1.481930061 | 22.276699543 |
| 14.799529818 | 7.478529939 | 22.276699543 |
| 14.799529818 | 13.475129951 | 22.276699543 |
| 4.335140292 | 4.480229933 | 22.276699543 |
| 4.335140292 | 10.476829677 | 22.276699543 |
| 4.335140292 | 16.473429152 | 22.276699543 |
| 9.567340434 | 4.480229933 | 22.276699543 |
| 9.567340434 | 10.476829677 | 22.276699543 |
| 9.567340434 | 16.473429152 | 22.276699543 |
| 14.799540109 | 4.480229933 | 22.276699543 |
| 14.799540109 | 10.476829677 | 22.276699543 |
| 14.799540109 | 16.473429152 | 22.276699543 |
| 0.822988000 | 1.465929966 | 24.534900486 |
| 0.822988000 | 7.462529978 | 24.534900486 |
| 0.822988000 | 13.459129454 | 24.534900486 |
| 6.055187909 | 1.465929966 | 24.534900486 |
| 6.055187909 | 7.462529978 | 24.534900486 |
| 6.055187909 | 13.459129454 | 24.534900486 |
| 11.287387584 | 1.465929966 | 24.534900486 |
| 11.287387584 | 7.462529978 | 24.534900486 |
| 11.287387584 | 13.459129454 | 24.534900486 |
| 0.822987006 | 4.464229972 | 24.534900486 |

```
 0.822987006        10.460830252        24.534900486
 0.822987006        16.457429728        24.534900486
 6.055186973         4.464229972        24.534900486
 6.055186973        10.460830252        24.534900486
 6.055186973        16.457429728        24.534900486
11.287386648         4.464229972        24.534900486
11.287386648        10.460830252        24.534900486
11.287386648        16.457429728        24.534900486
 2.616090014         1.499139950        20.000000298
 2.616090014         7.495739962        20.000000298
 2.616090014        13.492339974        20.000000298
 7.848290156         1.499139950        20.000000298
 7.848290156         7.495739962        20.000000298
 7.848290156        13.492339974        20.000000298
13.080490299         1.499139950        20.000000298
13.080490299         7.495739962        20.000000298
13.080490299        13.492339974        20.000000298
 2.544110143         1.450860071        26.780399680
 2.544110143         7.447459681        26.780399680
 2.544110143        13.444059693        26.780399680
 7.776310052         1.450860071        26.780399680
 7.776310052         7.447459681        26.780399680
 7.776310052        13.444059693        26.780399680
13.008509727         1.450860071        26.780399680
13.008509727         7.447459681        26.780399680
13.008509727        13.444059693        26.780399680
 2.616090014         4.497419851        20.000000298
 2.616090014        10.494019327        20.000000298
 2.616090014        16.490619875        20.000000298
 7.848290156         4.497419851        20.000000298
 7.848290156        10.494019327        20.000000298
 7.848290156        16.490619875        20.000000298
13.080490299         4.497419851        20.000000298
13.080490299        10.494019327        20.000000298
13.080490299        16.490619875        20.000000298
 2.544110143         4.449159943        26.780399680
 2.544110143        10.445759419        26.780399680
 2.544110143        16.442359967        26.780399680
 7.776310052         4.449159943        26.780399680
 7.776310052        10.445759419        26.780399680
 7.776310052        16.442359967        26.780399680
13.008509727         4.449159943        26.780399680
13.008509727        10.445759419        26.780399680
13.008509727        16.442359967        26.780399680
 3.371540092         2.901329834        30.936500430
```

| | | |
|---|---|---|
| 3.371540092 | 8.897929577 | 30.936500430 |
| 3.371540092 | 14.894529053 | 30.936500430 |
| 8.603740468 | 2.901329834 | 30.936500430 |
| 8.603740468 | 8.897929577 | 30.936500430 |
| 8.603740468 | 14.894529053 | 30.936500430 |
| 13.835940143 | 2.901329834 | 30.936500430 |
| 13.835940143 | 8.897929577 | 30.936500430 |
| 13.835940143 | 14.894529053 | 30.936500430 |
| 3.371540092 | 5.899629839 | 30.936500430 |
| 3.371540092 | 11.896229851 | 30.936500430 |
| 3.371540092 | 17.892829327 | 30.936500430 |
| 8.603740468 | 5.899629839 | 30.936500430 |
| 8.603740468 | 11.896229851 | 30.936500430 |
| 8.603740468 | 17.892829327 | 30.936500430 |
| 13.835940143 | 5.899629839 | 30.936500430 |
| 13.835940143 | 11.896229851 | 30.936500430 |
| 13.835940143 | 17.892829327 | 30.936500430 |
| 0.755415987 | 1.402199936 | 30.936500430 |
| 0.755415987 | 7.398799814 | 30.936500430 |
| 0.755415987 | 13.395399825 | 30.936500430 |
| 5.987615838 | 1.402199936 | 30.936500430 |
| 5.987615838 | 7.398799814 | 30.936500430 |
| 5.987615838 | 13.395399825 | 30.936500430 |
| 11.219815513 | 1.402199936 | 30.936500430 |
| 11.219815513 | 7.398799814 | 30.936500430 |
| 11.219815513 | 13.395399825 | 30.936500430 |
| 0.755415987 | 4.400499808 | 30.936500430 |
| 0.755415987 | 10.397099551 | 30.936500430 |
| 0.755415987 | 16.393699027 | 30.936500430 |
| 5.987615838 | 4.400499808 | 30.936500430 |
| 5.987615838 | 10.397099551 | 30.936500430 |
| 5.987615838 | 16.393699027 | 30.936500430 |
| 11.219815513 | 4.400499808 | 30.936500430 |
| 11.219815513 | 10.397099551 | 30.936500430 |
| 11.219815513 | 16.393699027 | 30.936500430 |
| 5.111970245 | 2.900420008 | 32.579100132 |
| 5.111970245 | 8.897019752 | 32.579100132 |
| 5.111970245 | 14.893619763 | 32.579100132 |
| 10.344170387 | 2.900420008 | 32.579100132 |
| 10.344170387 | 8.897019752 | 32.579100132 |
| 10.344170387 | 14.893619763 | 32.579100132 |
| 15.576370062 | 2.900420008 | 32.579100132 |
| 15.576370062 | 8.897019752 | 32.579100132 |
| 15.576370062 | 14.893619763 | 32.579100132 |
| 5.111970245 | 5.898719477 | 32.579100132 |

| | | |
|---|---|---|
| 5.111970245 | 11.895319489 | 32.579100132 |
| 5.111970245 | 17.891918965 | 32.579100132 |
| 10.344170387 | 5.898719477 | 32.579100132 |
| 10.344170387 | 11.895319489 | 32.579100132 |
| 10.344170387 | 17.891918965 | 32.579100132 |
| 15.576370062 | 5.898719477 | 32.579100132 |
| 15.576370062 | 11.895319489 | 32.579100132 |
| 15.576370062 | 17.891918965 | 32.579100132 |
| 2.495870173 | 1.401149910 | 32.579100132 |
| 2.495870173 | 7.397749520 | 32.579100132 |
| 2.495870173 | 13.394350068 | 32.579100132 |
| 7.728070082 | 1.401149910 | 32.579100132 |
| 7.728070082 | 7.397749520 | 32.579100132 |
| 7.728070082 | 13.394350068 | 32.579100132 |
| 12.960269757 | 1.401149910 | 32.579100132 |
| 12.960269757 | 7.397749520 | 32.579100132 |
| 12.960269757 | 13.394350068 | 32.579100132 |
| 2.495870173 | 4.399449514 | 32.579100132 |
| 2.495870173 | 10.396049794 | 32.579100132 |
| 2.495870173 | 16.392649269 | 32.579100132 |
| 7.728070082 | 4.399449514 | 32.579100132 |
| 7.728070082 | 10.396049794 | 32.579100132 |
| 7.728070082 | 16.392649269 | 32.579100132 |
| 12.960269757 | 4.399449514 | 32.579100132 |
| 12.960269757 | 10.396049794 | 32.579100132 |
| 12.960269757 | 16.392649269 | 32.579100132 |
| 5.113719798 | 2.902349836 | 29.305401444 |
| 5.113719798 | 8.898949848 | 29.305401444 |
| 5.113719798 | 14.895549859 | 29.305401444 |
| 10.345919941 | 2.902349836 | 29.305401444 |
| 10.345919941 | 8.898949848 | 29.305401444 |
| 10.345919941 | 14.895549859 | 29.305401444 |
| 15.578119616 | 2.902349836 | 29.305401444 |
| 15.578119616 | 8.898949848 | 29.305401444 |
| 15.578119616 | 14.895549859 | 29.305401444 |
| 5.113719798 | 5.900649574 | 29.305401444 |
| 5.113719798 | 11.897249585 | 29.305401444 |
| 5.113719798 | 17.893849061 | 29.305401444 |
| 10.345919941 | 5.900649574 | 29.305401444 |
| 10.345919941 | 11.897249585 | 29.305401444 |
| 10.345919941 | 17.893849061 | 29.305401444 |
| 15.578119616 | 5.900649574 | 29.305401444 |
| 15.578119616 | 11.897249585 | 29.305401444 |
| 15.578119616 | 17.893849061 | 29.305401444 |
| 2.497610137 | 1.403179996 | 29.305401444 |

| | | |
|---|---|---|
| 2.497610137 | 7.399779873 | 29.305401444 |
| 2.497610137 | 13.396379885 | 29.305401444 |
| 7.729810280 | 1.403179996 | 29.305401444 |
| 7.729810280 | 7.399779873 | 29.305401444 |
| 7.729810280 | 13.396379885 | 29.305401444 |
| 12.962009955 | 1.403179996 | 29.305401444 |
| 12.962009955 | 7.399779873 | 29.305401444 |
| 12.962009955 | 13.396379885 | 29.305401444 |
| 2.497610137 | 4.401480136 | 29.305401444 |
| 2.497610137 | 10.398079611 | 29.305401444 |
| 2.497610137 | 16.394680159 | 29.305401444 |
| 7.729810280 | 4.401480136 | 29.305401444 |
| 7.729810280 | 10.398079611 | 29.305401444 |
| 7.729810280 | 16.394680159 | 29.305401444 |
| 12.962009955 | 4.401480136 | 29.305401444 |
| 12.962009955 | 10.398079611 | 29.305401444 |
| 12.962009955 | 16.394680159 | 29.305401444 |
| 0.434074003 | 1.187329977 | 35.656398535 |
| 0.434074003 | 7.183929721 | 35.656398535 |
| 0.434074003 | 13.180529733 | 35.656398535 |
| 5.666273795 | 1.187329977 | 35.656398535 |
| 5.666273795 | 7.183929721 | 35.656398535 |
| 5.666273795 | 13.180529733 | 35.656398535 |
| 10.898473938 | 1.187329977 | 35.656398535 |
| 10.898473938 | 7.183929721 | 35.656398535 |
| 10.898473938 | 13.180529733 | 35.656398535 |
| 3.050169981 | 2.665719786 | 35.656398535 |
| 3.050169981 | 8.662319530 | 35.656398535 |
| 3.050169981 | 14.658919006 | 35.656398535 |
| 8.282369890 | 2.665719786 | 35.656398535 |
| 8.282369890 | 8.662319530 | 35.656398535 |
| 8.282369890 | 14.658919006 | 35.656398535 |
| 13.514569565 | 2.665719786 | 35.656398535 |
| 13.514569565 | 8.662319530 | 35.656398535 |
| 13.514569565 | 14.658919006 | 35.656398535 |
| 0.434074003 | 4.185619529 | 35.656398535 |
| 0.434074003 | 10.182219809 | 35.656398535 |
| 0.434074003 | 16.178819284 | 35.656398535 |
| 5.666273795 | 4.185619529 | 35.656398535 |
| 5.666273795 | 10.182219809 | 35.656398535 |
| 5.666273795 | 16.178819284 | 35.656398535 |
| 10.898473938 | 4.185619529 | 35.656398535 |
| 10.898473938 | 10.182219809 | 35.656398535 |
| 10.898473938 | 16.178819284 | 35.656398535 |
| 3.050169981 | 5.664010142 | 35.656398535 |

| | | |
|---|---|---|
| 3.050169981 | 11.660610154 | 35.656398535 |
| 3.050169981 | 17.657209629 | 35.656398535 |
| 8.282369890 | 5.664010142 | 35.656398535 |
| 8.282369890 | 11.660610154 | 35.656398535 |
| 8.282369890 | 17.657209629 | 35.656398535 |
| 13.514569565 | 5.664010142 | 35.656398535 |
| 13.514569565 | 11.660610154 | 35.656398535 |
| 13.514569565 | 17.657209629 | 35.656398535 |
| 1.306100032 | 5.677840353 | 37.427300215 |
| 1.306100032 | 11.674440364 | 37.427300215 |
| 1.306100032 | 17.671039840 | 37.427300215 |
| 6.538300057 | 5.677840353 | 37.427300215 |
| 6.538300057 | 11.674440364 | 37.427300215 |
| 6.538300057 | 17.671039840 | 37.427300215 |
| 11.770500200 | 5.677840353 | 37.427300215 |
| 11.770500200 | 11.674440364 | 37.427300215 |
| 11.770500200 | 17.671039840 | 37.427300215 |
| 3.922199986 | 1.159659905 | 37.427300215 |
| 3.922199986 | 7.156259649 | 37.427300215 |
| 3.922199986 | 13.152859661 | 37.427300215 |
| 9.154399895 | 1.159659905 | 37.427300215 |
| 9.154399895 | 7.156259649 | 37.427300215 |
| 9.154399895 | 13.152859661 | 37.427300215 |
| 14.386599570 | 1.159659905 | 37.427300215 |
| 14.386599570 | 7.156259649 | 37.427300215 |
| 14.386599570 | 13.152859661 | 37.427300215 |
| 1.306100032 | 2.679559916 | 37.427300215 |
| 1.306100032 | 8.676159928 | 37.427300215 |
| 1.306100032 | 14.672759939 | 37.427300215 |
| 6.538300057 | 2.679559916 | 37.427300215 |
| 6.538300057 | 8.676159928 | 37.427300215 |
| 6.538300057 | 14.672759939 | 37.427300215 |
| 11.770500200 | 2.679559916 | 37.427300215 |
| 11.770500200 | 8.676159928 | 37.427300215 |
| 11.770500200 | 14.672759939 | 37.427300215 |
| 3.922199986 | 4.157949993 | 37.427300215 |
| 3.922199986 | 10.154549736 | 37.427300215 |
| 3.922199986 | 16.151149212 | 37.427300215 |
| 9.154399895 | 4.157949993 | 37.427300215 |
| 9.154399895 | 10.154549736 | 37.427300215 |
| 9.154399895 | 16.151149212 | 37.427300215 |
| 14.386599570 | 4.157949993 | 37.427300215 |
| 14.386599570 | 10.154549736 | 37.427300215 |
| 14.386599570 | 16.151149212 | 37.427300215 |
| 1.306100032 | 5.677840353 | 39.805400372 |

```
    1.306100032        11.674440364        39.805400372
    1.306100032        17.671039840        39.805400372
    6.538300057         5.677840353        39.805400372
    6.538300057        11.674440364        39.805400372
    6.538300057        17.671039840        39.805400372
   11.770500200         5.677840353        39.805400372
   11.770500200        11.674440364        39.805400372
   11.770500200        17.671039840        39.805400372
    3.922199986         1.159659905        39.805400372
    3.922199986         7.156259649        39.805400372
    3.922199986        13.152859661        39.805400372
    9.154399895         1.159659905        39.805400372
    9.154399895         7.156259649        39.805400372
    9.154399895        13.152859661        39.805400372
   14.386599570         1.159659905        39.805400372
   14.386599570         7.156259649        39.805400372
   14.386599570        13.152859661        39.805400372
    1.306100032         2.679559916        39.805498719
    1.306100032         8.676159928        39.805498719
    1.306100032        14.672759939        39.805498719
    6.538300057         2.679559916        39.805498719
    6.538300057         8.676159928        39.805498719
    6.538300057        14.672759939        39.805498719
   11.770500200         2.679559916        39.805498719
   11.770500200         8.676159928        39.805498719
   11.770500200        14.672759939        39.805498719
    3.922199986         4.157949993        39.805400372
    3.922199986        10.154549736        39.805400372
    3.922199986        16.151149212        39.805400372
    9.154399895         4.157949993        39.805400372
    9.154399895        10.154549736        39.805400372
    9.154399895        16.151149212        39.805400372
   14.386599570         4.157949993        39.805400372
   14.386599570        10.154549736        39.805400372
   14.386599570        16.151149212        39.805400372


Ag/WSe2/Si
1.0
       15.9893999100         0.0000000000         0.0000000000
        0.0000000000        18.4659004211         0.0000000000
        0.0000000000         0.0000000000        50.0000000000
     Ag    W    Se    Si
    144   36    72   108
Cartesian
      0.000000000         0.000000000        20.000000298
```

| | | |
|---|---|---|
| 0.000000000 | 6.155300324 | 20.000000298 |
| 0.000000000 | 12.310600648 | 20.000000298 |
| 5.329800129 | 0.000000000 | 20.000000298 |
| 5.329800129 | 6.155300324 | 20.000000298 |
| 5.329800129 | 12.310600648 | 20.000000298 |
| 10.659600258 | 0.000000000 | 20.000000298 |
| 10.659600258 | 6.155300324 | 20.000000298 |
| 10.659600258 | 12.310600648 | 20.000000298 |
| 5.290209790 | 6.110600036 | 26.654100418 |
| 5.290209790 | 12.265899810 | 26.654100418 |
| 5.290209790 | 18.421200684 | 26.654100418 |
| 10.620009919 | 6.110600036 | 26.654100418 |
| 10.620009919 | 12.265899810 | 26.654100418 |
| 10.620009919 | 18.421200684 | 26.654100418 |
| 15.949809571 | 6.110600036 | 26.654100418 |
| 15.949809571 | 12.265899810 | 26.654100418 |
| 15.949809571 | 18.421200684 | 26.654100418 |
| 0.000000000 | 3.077650162 | 20.000000298 |
| 0.000000000 | 9.232950211 | 20.000000298 |
| 0.000000000 | 15.388249984 | 20.000000298 |
| 5.329800129 | 3.077650162 | 20.000000298 |
| 5.329800129 | 9.232950211 | 20.000000298 |
| 5.329800129 | 15.388249984 | 20.000000298 |
| 10.659600258 | 3.077650162 | 20.000000298 |
| 10.659600258 | 9.232950211 | 20.000000298 |
| 10.659600258 | 15.388249984 | 20.000000298 |
| 5.290209790 | 3.032949874 | 26.654100418 |
| 5.290209790 | 9.188249923 | 26.654100418 |
| 5.290209790 | 15.343550247 | 26.654100418 |
| 10.620009919 | 3.032949874 | 26.654100418 |
| 10.620009919 | 9.188249923 | 26.654100418 |
| 10.620009919 | 15.343550247 | 26.654100418 |
| 15.949809571 | 3.032949874 | 26.654100418 |
| 15.949809571 | 9.188249923 | 26.654100418 |
| 15.949809571 | 15.343550247 | 26.654100418 |
| 1.762600047 | 6.139490538 | 22.230899334 |
| 1.762600047 | 12.294790862 | 22.230899334 |
| 1.762600047 | 18.450090635 | 22.230899334 |
| 7.092399819 | 6.139490538 | 22.230899334 |
| 7.092399819 | 12.294790862 | 22.230899334 |
| 7.092399819 | 18.450090635 | 22.230899334 |
| 12.422199947 | 6.139490538 | 22.230899334 |
| 12.422199947 | 12.294790862 | 22.230899334 |
| 12.422199947 | 18.450090635 | 22.230899334 |
| 1.762600047 | 3.061840101 | 22.230899334 |

| | | |
|---|---|---|
| 1.762600047 | 9.217140425 | 22.230899334 |
| 1.762600047 | 15.372440198 | 22.230899334 |
| 7.092399819 | 3.061840101 | 22.230899334 |
| 7.092399819 | 9.217140425 | 22.230899334 |
| 7.092399819 | 15.372440198 | 22.230899334 |
| 12.422199947 | 3.061840101 | 22.230899334 |
| 12.422199947 | 9.217140425 | 22.230899334 |
| 12.422199947 | 15.372440198 | 22.230899334 |
| 3.525830055 | 6.124380217 | 24.449199438 |
| 3.525830055 | 12.279679991 | 24.449199438 |
| 3.525830055 | 18.434980865 | 24.449199438 |
| 8.855629946 | 6.124380217 | 24.449199438 |
| 8.855629946 | 12.279679991 | 24.449199438 |
| 8.855629946 | 18.434980865 | 24.449199438 |
| 14.185429598 | 6.124380217 | 24.449199438 |
| 14.185429598 | 12.279679991 | 24.449199438 |
| 14.185429598 | 18.434980865 | 24.449199438 |
| 3.525830055 | 3.046730055 | 24.449199438 |
| 3.525830055 | 9.202030104 | 24.449199438 |
| 3.525830055 | 15.357330428 | 24.449199438 |
| 8.855629946 | 3.046730055 | 24.449199438 |
| 8.855629946 | 9.202030104 | 24.449199438 |
| 8.855629946 | 15.357330428 | 24.449199438 |
| 14.185429598 | 3.046730055 | 24.449199438 |
| 14.185429598 | 9.202030104 | 24.449199438 |
| 14.185429598 | 15.357330428 | 24.449199438 |
| 4.427499516 | 1.523010067 | 22.230899334 |
| 4.427499516 | 7.678310253 | 22.230899334 |
| 4.427499516 | 13.833610027 | 22.230899334 |
| 9.757299645 | 1.523010067 | 22.230899334 |
| 9.757299645 | 7.678310253 | 22.230899334 |
| 9.757299645 | 13.833610027 | 22.230899334 |
| 15.087099297 | 1.523010067 | 22.230899334 |
| 15.087099297 | 7.678310253 | 22.230899334 |
| 15.087099297 | 13.833610027 | 22.230899334 |
| 4.427499516 | 4.600659816 | 22.230899334 |
| 4.427499516 | 10.755959590 | 22.230899334 |
| 4.427499516 | 16.911260464 | 22.230899334 |
| 9.757299645 | 4.600659816 | 22.230899334 |
| 9.757299645 | 10.755959590 | 22.230899334 |
| 9.757299645 | 16.911260464 | 22.230899334 |
| 15.087099297 | 4.600659816 | 22.230899334 |
| 15.087099297 | 10.755959590 | 22.230899334 |
| 15.087099297 | 16.911260464 | 22.230899334 |
| 0.860926953 | 1.507910065 | 24.449199438 |

| | | |
|---|---|---|
| 0.860926953 | 7.663210389 | 24.449199438 |
| 0.860926953 | 13.818510162 | 24.449199438 |
| 6.190726784 | 1.507910065 | 24.449199438 |
| 6.190726784 | 7.663210389 | 24.449199438 |
| 6.190726784 | 13.818510162 | 24.449199438 |
| 11.520526436 | 1.507910065 | 24.449199438 |
| 11.520526436 | 7.663210389 | 24.449199438 |
| 11.520526436 | 13.818510162 | 24.449199438 |
| 0.860926953 | 4.585559952 | 24.449199438 |
| 0.860926953 | 10.740859725 | 24.449199438 |
| 0.860926953 | 16.896160599 | 24.449199438 |
| 6.190726784 | 4.585559952 | 24.449199438 |
| 6.190726784 | 10.740859725 | 24.449199438 |
| 6.190726784 | 16.896160599 | 24.449199438 |
| 11.520526436 | 4.585559952 | 24.449199438 |
| 11.520526436 | 10.740859725 | 24.449199438 |
| 11.520526436 | 16.896160599 | 24.449199438 |
| 2.664880051 | 1.538820128 | 20.000000298 |
| 2.664880051 | 7.694120039 | 20.000000298 |
| 2.664880051 | 13.849420913 | 20.000000298 |
| 7.994679941 | 1.538820128 | 20.000000298 |
| 7.994679941 | 7.694120039 | 20.000000298 |
| 7.994679941 | 13.849420913 | 20.000000298 |
| 13.324479593 | 1.538820128 | 20.000000298 |
| 13.324479593 | 7.694120039 | 20.000000298 |
| 13.324479593 | 13.849420913 | 20.000000298 |
| 2.625309964 | 1.494130021 | 26.654100418 |
| 2.625309964 | 7.649430208 | 26.654100418 |
| 2.625309964 | 13.804729981 | 26.654100418 |
| 7.955110093 | 1.494130021 | 26.654100418 |
| 7.955110093 | 7.649430208 | 26.654100418 |
| 7.955110093 | 13.804729981 | 26.654100418 |
| 13.284910222 | 1.494130021 | 26.654100418 |
| 13.284910222 | 7.649430208 | 26.654100418 |
| 13.284910222 | 13.804729981 | 26.654100418 |
| 2.664880051 | 4.616469877 | 20.000000298 |
| 2.664880051 | 10.771770476 | 20.000000298 |
| 2.664880051 | 16.927070250 | 20.000000298 |
| 7.994679941 | 4.616469877 | 20.000000298 |
| 7.994679941 | 10.771770476 | 20.000000298 |
| 7.994679941 | 16.927070250 | 20.000000298 |
| 13.324479593 | 4.616469877 | 20.000000298 |
| 13.324479593 | 10.771770476 | 20.000000298 |
| 13.324479593 | 16.927070250 | 20.000000298 |
| 2.625309964 | 4.571780046 | 26.654100418 |

| | | |
|---|---|---|
| 2.625309964 | 10.727080645 | 26.654100418 |
| 2.625309964 | 16.882380418 | 26.654100418 |
| 7.955110093 | 4.571780046 | 26.654100418 |
| 7.955110093 | 10.727080645 | 26.654100418 |
| 7.955110093 | 16.882380418 | 26.654100418 |
| 13.284910222 | 4.571780046 | 26.654100418 |
| 13.284910222 | 10.727080645 | 26.654100418 |
| 13.284910222 | 16.882380418 | 26.654100418 |
| 0.764953071 | 1.403700060 | 31.070399284 |
| 0.764953071 | 7.558999971 | 31.070399284 |
| 0.764953071 | 13.714300295 | 31.070399284 |
| 6.094753022 | 1.403700060 | 31.070399284 |
| 6.094753022 | 7.558999971 | 31.070399284 |
| 6.094753022 | 13.714300295 | 31.070399284 |
| 11.424553150 | 1.403700060 | 31.070399284 |
| 11.424553150 | 7.558999971 | 31.070399284 |
| 11.424553150 | 13.714300295 | 31.070399284 |
| 0.764953071 | 4.481350084 | 31.070399284 |
| 0.764953071 | 10.636649858 | 31.070399284 |
| 0.764953071 | 16.791950732 | 31.070399284 |
| 6.094753022 | 4.481350084 | 31.070399284 |
| 6.094753022 | 10.636649858 | 31.070399284 |
| 6.094753022 | 16.791950732 | 31.070399284 |
| 11.424553150 | 4.481350084 | 31.070399284 |
| 11.424553150 | 10.636649858 | 31.070399284 |
| 11.424553150 | 16.791950732 | 31.070399284 |
| 3.429879881 | 6.020160444 | 31.070399284 |
| 3.429879881 | 12.175460218 | 31.070399284 |
| 3.429879881 | 18.330761092 | 31.070399284 |
| 8.759679533 | 6.020160444 | 31.070399284 |
| 8.759679533 | 12.175460218 | 31.070399284 |
| 8.759679533 | 18.330761092 | 31.070399284 |
| 14.089480138 | 6.020160444 | 31.070399284 |
| 14.089480138 | 12.175460218 | 31.070399284 |
| 14.089480138 | 18.330761092 | 31.070399284 |
| 3.429879881 | 2.942510007 | 31.070399284 |
| 3.429879881 | 9.097810331 | 31.070399284 |
| 3.429879881 | 15.253110655 | 31.070399284 |
| 8.759679533 | 2.942510007 | 31.070399284 |
| 8.759679533 | 9.097810331 | 31.070399284 |
| 8.759679533 | 15.253110655 | 31.070399284 |
| 14.089480138 | 2.942510007 | 31.070399284 |
| 14.089480138 | 9.097810331 | 31.070399284 |
| 14.089480138 | 15.253110655 | 31.070399284 |
| 2.540400072 | 1.402289985 | 32.849702239 |

| | | |
|---|---|---|
| 2.540400072 | 7.557590034 | 32.849702239 |
| 2.540400072 | 13.712890358 | 32.849702239 |
| 7.870200201 | 1.402289985 | 32.849702239 |
| 7.870200201 | 7.557590034 | 32.849702239 |
| 7.870200201 | 13.712890358 | 32.849702239 |
| 13.199999853 | 1.402289985 | 32.849702239 |
| 13.199999853 | 7.557590034 | 32.849702239 |
| 13.199999853 | 13.712890358 | 32.849702239 |
| 2.540400072 | 4.479940147 | 32.849702239 |
| 2.540400072 | 10.635239921 | 32.849702239 |
| 2.540400072 | 16.790540795 | 32.849702239 |
| 7.870200201 | 4.479940147 | 32.849702239 |
| 7.870200201 | 10.635239921 | 32.849702239 |
| 7.870200201 | 16.790540795 | 32.849702239 |
| 13.199999853 | 4.479940147 | 32.849702239 |
| 13.199999853 | 10.635239921 | 32.849702239 |
| 13.199999853 | 16.790540795 | 32.849702239 |
| 5.205309905 | 6.018869928 | 32.849702239 |
| 5.205309905 | 12.174170252 | 32.849702239 |
| 5.205309905 | 18.329470025 | 32.849702239 |
| 10.535110034 | 6.018869928 | 32.849702239 |
| 10.535110034 | 12.174170252 | 32.849702239 |
| 10.535110034 | 18.329470025 | 32.849702239 |
| 15.864909686 | 6.018869928 | 32.849702239 |
| 15.864909686 | 12.174170252 | 32.849702239 |
| 15.864909686 | 18.329470025 | 32.849702239 |
| 5.205309905 | 2.941220041 | 32.849702239 |
| 5.205309905 | 9.096520365 | 32.849702239 |
| 5.205309905 | 15.251820689 | 32.849702239 |
| 10.535110034 | 2.941220041 | 32.849702239 |
| 10.535110034 | 9.096520365 | 32.849702239 |
| 10.535110034 | 15.251820689 | 32.849702239 |
| 15.864909686 | 2.941220041 | 32.849702239 |
| 15.864909686 | 9.096520365 | 32.849702239 |
| 15.864909686 | 15.251820689 | 32.849702239 |
| 2.542780057 | 1.404909953 | 29.307401180 |
| 2.542780057 | 7.560210140 | 29.307401180 |
| 2.542780057 | 13.715509913 | 29.307401180 |
| 7.872579948 | 1.404909953 | 29.307401180 |
| 7.872579948 | 7.560210140 | 29.307401180 |
| 7.872579948 | 13.715509913 | 29.307401180 |
| 13.202379600 | 1.404909953 | 29.307401180 |
| 13.202379600 | 7.560210140 | 29.307401180 |
| 13.202379600 | 13.715509913 | 29.307401180 |
| 2.542780057 | 4.482560528 | 29.307401180 |

| | | |
|---|---|---|
| 2.542780057 | 10.637860577 | 29.307401180 |
| 2.542780057 | 16.793160350 | 29.307401180 |
| 7.872579948 | 4.482560528 | 29.307401180 |
| 7.872579948 | 10.637860577 | 29.307401180 |
| 7.872579948 | 16.793160350 | 29.307401180 |
| 13.202379600 | 4.482560528 | 29.307401180 |
| 13.202379600 | 10.637860577 | 29.307401180 |
| 13.202379600 | 16.793160350 | 29.307401180 |
| 5.207680122 | 6.021400330 | 29.307401180 |
| 5.207680122 | 12.176700654 | 29.307401180 |
| 5.207680122 | 18.332000428 | 29.307401180 |
| 10.537480251 | 6.021400330 | 29.307401180 |
| 10.537480251 | 12.176700654 | 29.307401180 |
| 10.537480251 | 18.332000428 | 29.307401180 |
| 15.867279903 | 6.021400330 | 29.307401180 |
| 15.867279903 | 12.176700654 | 29.307401180 |
| 15.867279903 | 18.332000428 | 29.307401180 |
| 5.207680122 | 2.943749893 | 29.307401180 |
| 5.207680122 | 9.099050217 | 29.307401180 |
| 5.207680122 | 15.254349990 | 29.307401180 |
| 10.537480251 | 2.943749893 | 29.307401180 |
| 10.537480251 | 9.099050217 | 29.307401180 |
| 10.537480251 | 15.254349990 | 29.307401180 |
| 15.867279903 | 2.943749893 | 29.307401180 |
| 15.867279903 | 9.099050217 | 29.307401180 |
| 15.867279903 | 15.254349990 | 29.307401180 |
| 0.442168000 | 1.218760015 | 36.044299603 |
| 0.442168000 | 7.374059926 | 36.044299603 |
| 0.442168000 | 13.529360800 | 36.044299603 |
| 5.771968010 | 1.218760015 | 36.044299603 |
| 5.771968010 | 7.374059926 | 36.044299603 |
| 5.771968010 | 13.529360800 | 36.044299603 |
| 11.101768139 | 1.218760015 | 36.044299603 |
| 11.101768139 | 7.374059926 | 36.044299603 |
| 11.101768139 | 13.529360800 | 36.044299603 |
| 3.107039831 | 2.736280023 | 36.044299603 |
| 3.107039831 | 8.891580346 | 36.044299603 |
| 3.107039831 | 15.046880120 | 36.044299603 |
| 8.436840198 | 2.736280023 | 36.044299603 |
| 8.436840198 | 8.891580346 | 36.044299603 |
| 8.436840198 | 15.046880120 | 36.044299603 |
| 13.766639850 | 2.736280023 | 36.044299603 |
| 13.766639850 | 8.891580346 | 36.044299603 |
| 13.766639850 | 15.046880120 | 36.044299603 |
| 0.442168000 | 4.296410314 | 36.044299603 |

| | | |
|---|---|---|
| 0.442168000 | 10.451710363 | 36.044299603 |
| 0.442168000 | 16.607010137 | 36.044299603 |
| 5.771968010 | 4.296410314 | 36.044299603 |
| 5.771968010 | 10.451710363 | 36.044299603 |
| 5.771968010 | 16.607010137 | 36.044299603 |
| 11.101768139 | 4.296410314 | 36.044299603 |
| 11.101768139 | 10.451710363 | 36.044299603 |
| 11.101768139 | 16.607010137 | 36.044299603 |
| 3.107039831 | 5.813930460 | 36.044299603 |
| 3.107039831 | 11.969230783 | 36.044299603 |
| 3.107039831 | 18.124530557 | 36.044299603 |
| 8.436840198 | 5.813930460 | 36.044299603 |
| 8.436840198 | 11.969230783 | 36.044299603 |
| 8.436840198 | 18.124530557 | 36.044299603 |
| 13.766639850 | 5.813930460 | 36.044299603 |
| 13.766639850 | 11.969230783 | 36.044299603 |
| 13.766639850 | 18.124530557 | 36.044299603 |
| 1.330459960 | 5.828139895 | 37.771901488 |
| 1.330459960 | 11.983440219 | 37.771901488 |
| 1.330459960 | 18.138739993 | 37.771901488 |
| 6.660259851 | 5.828139895 | 37.771901488 |
| 6.660259851 | 11.983440219 | 37.771901488 |
| 6.660259851 | 18.138739993 | 37.771901488 |
| 11.990059980 | 5.828139895 | 37.771901488 |
| 11.990059980 | 11.983440219 | 37.771901488 |
| 11.990059980 | 18.138739993 | 37.771901488 |
| 3.995340011 | 1.190359992 | 37.771800160 |
| 3.995340011 | 7.345660316 | 37.771800160 |
| 3.995340011 | 13.500960640 | 37.771800160 |
| 9.325139663 | 1.190359992 | 37.771800160 |
| 9.325139663 | 7.345660316 | 37.771800160 |
| 9.325139663 | 13.500960640 | 37.771800160 |
| 14.654940269 | 1.190359992 | 37.771800160 |
| 14.654940269 | 7.345660316 | 37.771800160 |
| 14.654940269 | 13.500960640 | 37.771800160 |
| 1.330459960 | 2.750490009 | 37.771901488 |
| 1.330459960 | 8.905790333 | 37.771901488 |
| 1.330459960 | 15.061090656 | 37.771901488 |
| 6.660259851 | 2.750490009 | 37.771901488 |
| 6.660259851 | 8.905790333 | 37.771901488 |
| 6.660259851 | 15.061090656 | 37.771901488 |
| 11.990059980 | 2.750490009 | 37.771901488 |
| 11.990059980 | 8.905790333 | 37.771901488 |
| 11.990059980 | 15.061090656 | 37.771901488 |
| 3.995340011 | 4.268010154 | 37.771800160 |

| | | |
|---|---|---|
| 3.995340011 | 10.423310203 | 37.771800160 |
| 3.995340011 | 16.578609976 | 37.771800160 |
| 9.325139663 | 4.268010154 | 37.771800160 |
| 9.325139663 | 10.423310203 | 37.771800160 |
| 9.325139663 | 16.578609976 | 37.771800160 |
| 14.654940269 | 4.268010154 | 37.771800160 |
| 14.654940269 | 10.423310203 | 37.771800160 |
| 14.654940269 | 16.578609976 | 37.771800160 |
| 1.330459960 | 5.828139895 | 40.160998702 |
| 1.330459960 | 11.983440219 | 40.160998702 |
| 1.330459960 | 18.138739993 | 40.160998702 |
| 6.660259851 | 5.828139895 | 40.160998702 |
| 6.660259851 | 11.983440219 | 40.160998702 |
| 6.660259851 | 18.138739993 | 40.160998702 |
| 11.990059980 | 5.828139895 | 40.160998702 |
| 11.990059980 | 11.983440219 | 40.160998702 |
| 11.990059980 | 18.138739993 | 40.160998702 |
| 3.995340011 | 1.190359992 | 40.160998702 |
| 3.995340011 | 7.345660316 | 40.160998702 |
| 3.995340011 | 13.500960640 | 40.160998702 |
| 9.325139663 | 1.190359992 | 40.160998702 |
| 9.325139663 | 7.345660316 | 40.160998702 |
| 9.325139663 | 13.500960640 | 40.160998702 |
| 14.654940269 | 1.190359992 | 40.160998702 |
| 14.654940269 | 7.345660316 | 40.160998702 |
| 14.654940269 | 13.500960640 | 40.160998702 |
| 1.330459960 | 2.750490009 | 40.160998702 |
| 1.330459960 | 8.905790333 | 40.160998702 |
| 1.330459960 | 15.061090656 | 40.160998702 |
| 6.660259851 | 2.750490009 | 40.160998702 |
| 6.660259851 | 8.905790333 | 40.160998702 |
| 6.660259851 | 15.061090656 | 40.160998702 |
| 11.990059980 | 2.750490009 | 40.160998702 |
| 11.990059980 | 8.905790333 | 40.160998702 |
| 11.990059980 | 15.061090656 | 40.160998702 |
| 3.995340011 | 4.268010154 | 40.160998702 |
| 3.995340011 | 10.423310203 | 40.160998702 |
| 3.995340011 | 16.578609976 | 40.160998702 |
| 9.325139663 | 4.268010154 | 40.160998702 |
| 9.325139663 | 10.423310203 | 40.160998702 |
| 9.325139663 | 16.578609976 | 40.160998702 |
| 14.654940269 | 4.268010154 | 40.160998702 |
| 14.654940269 | 10.423310203 | 40.160998702 |
| 14.654940269 | 16.578609976 | 40.160998702 |

# Frictional Properties

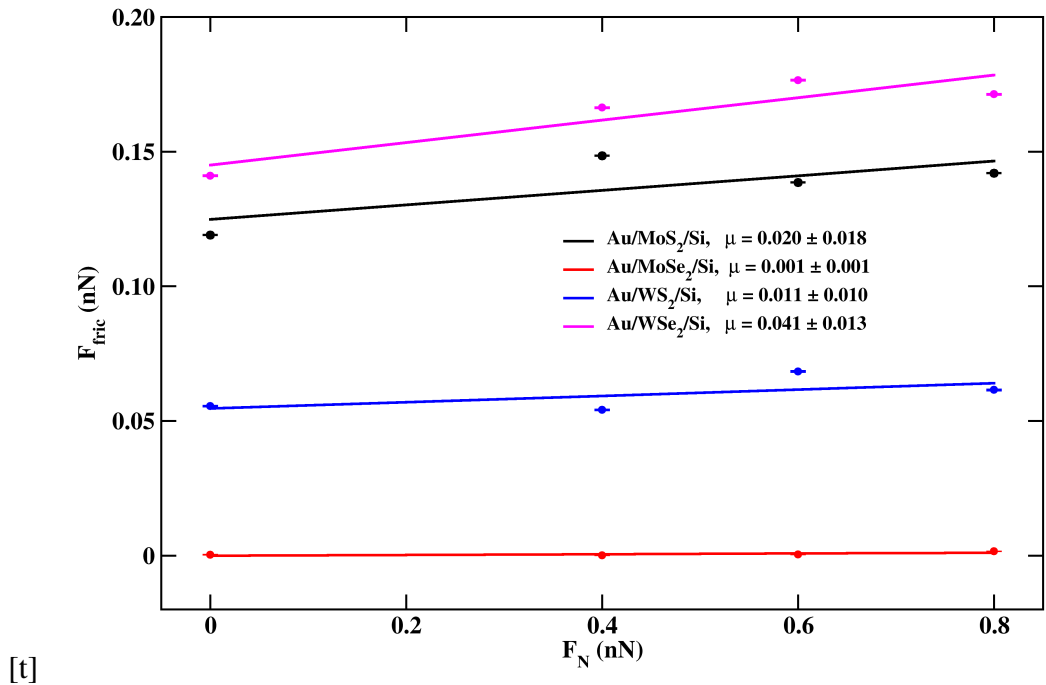

[t]

**Figure S1:** $\mu$ using linear fit function on average friction force as a function of load for all the systems with a tip velocity of 3 m/s.

The Figure S1, Figure S2 and Figure S3 shows friction force as a function of normal load for four Au/MX$_2$/Si systems expect Au/MoS$_2$/Si which is presented in the main manuscript, with linear fits used to extract the coefficient of friction $\mu$. Most of the systems exhibhit nonmonotonic nature with loads and extraction of $\mu$ from a linear fit results in large errors. This same trend is observed in Ag/MX$_2$/Si systems as shown in Figure S6, Figure S7, Figure S8 and Figure S9.

The friction force profile for the Au/MoSe$_2$/Si system is shown in Figure S4. In this case, both lateral slip and zigzag behaviour are suppressed and the tip moves along *y*-direction. This is also observed in the Fourier transform (FT) spectra of Au/MoSe$_2$/Si where the second and third peaks, representing the lateral and ziz-zag motions, are absent (Figure S5).

Figure S10 shows the spatial Fourier transform of the frictional force in the Ag/MX$_2$/Si and Au/MX$_2$/Si systems. In the Au/WSe$_2$/Si case, the first peak dominates than the Ag/WSe$_2$/Si system which leads to higher frictional force.

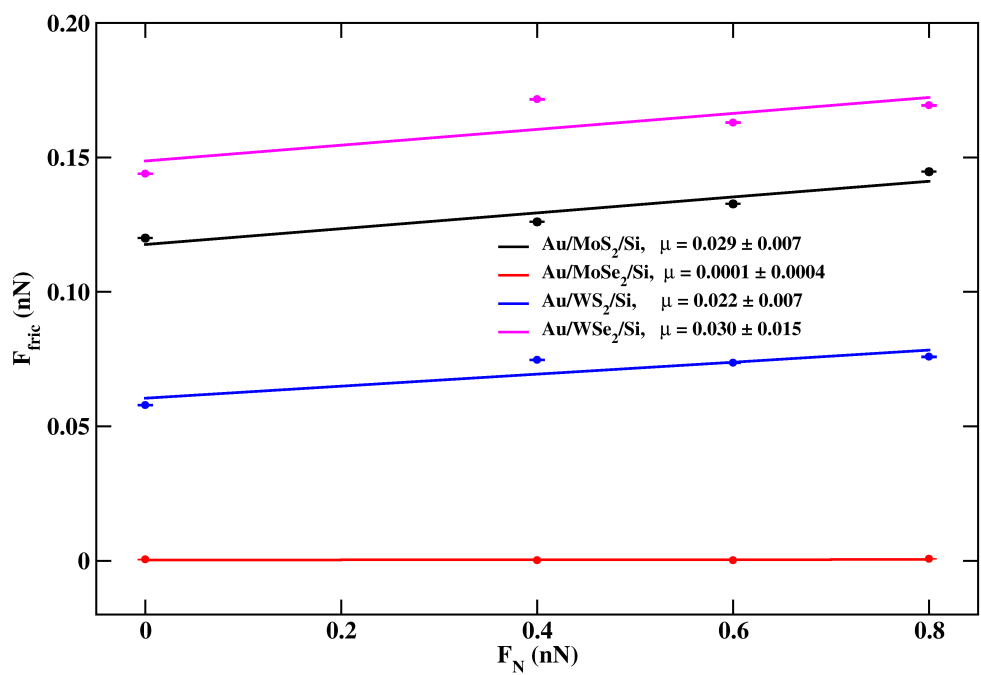

**Figure S2:** $\mu$ using linear fit function on average friction force as a function of load for all the systems with a tip velocity of 4 m/s.

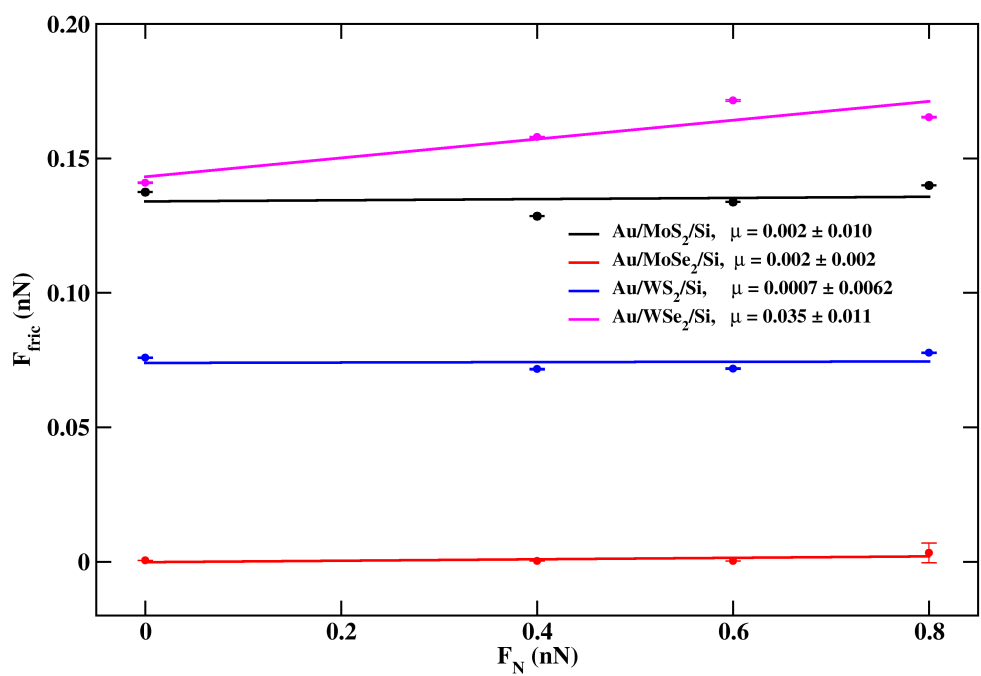

**Figure S3:** $\mu$ using linear fit function on average friction force as a function of load for all the systems with a tip velocity of 5 m/s.

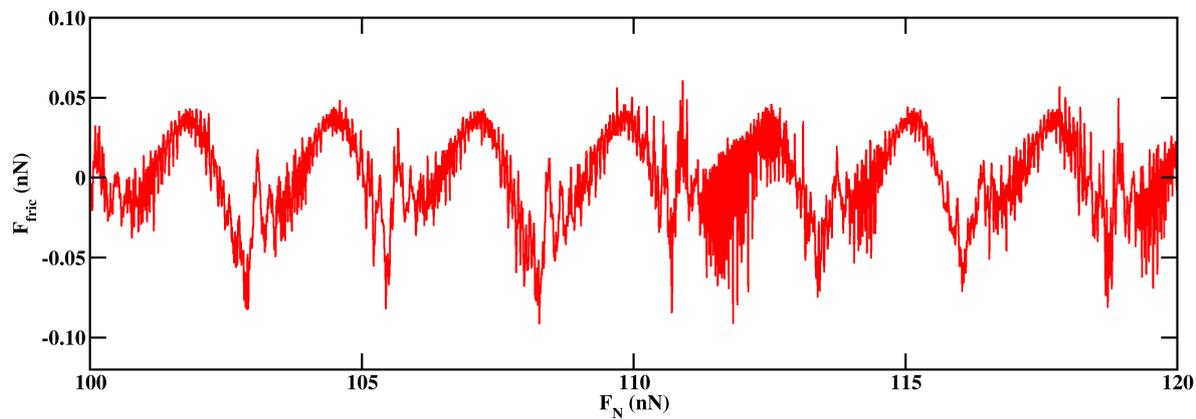

**Figure S4:** Friction force profile of Au/MoSe$_2$/Si where tip moving with velocity 2 m/s for load of 0 nN.

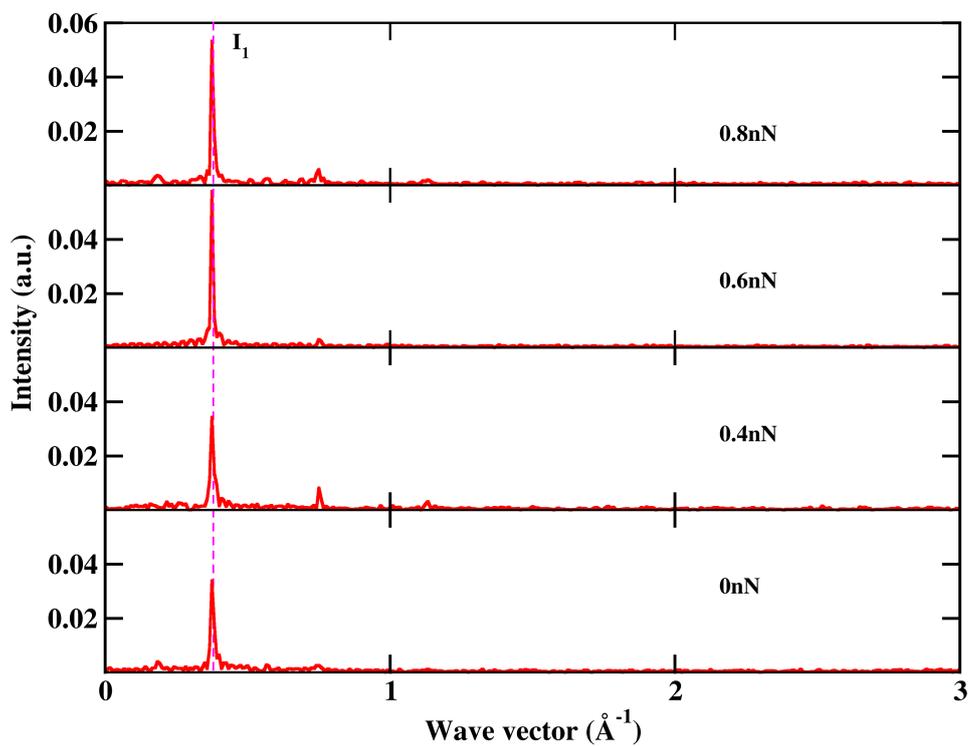

**Figure S5:** Spatial Fourier transform of Au/MoSe$_2$/Si with varying loads at velocity 2 m/s.

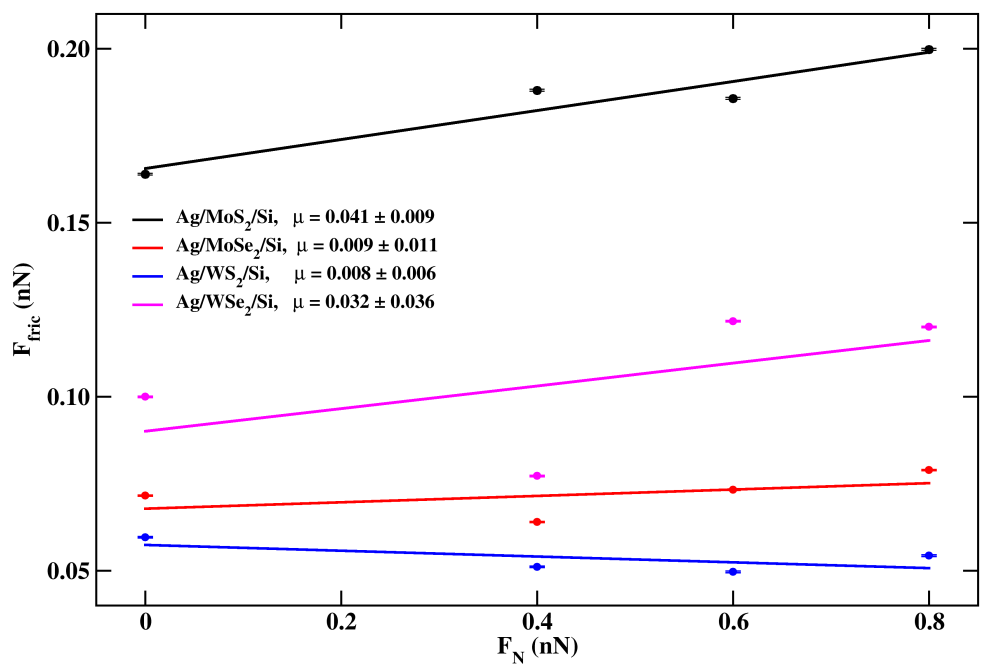

**Figure S6:** $\mu$ extracted from a linear fit on the average friction force as a function of load with a tip velocity of 2 m/s.

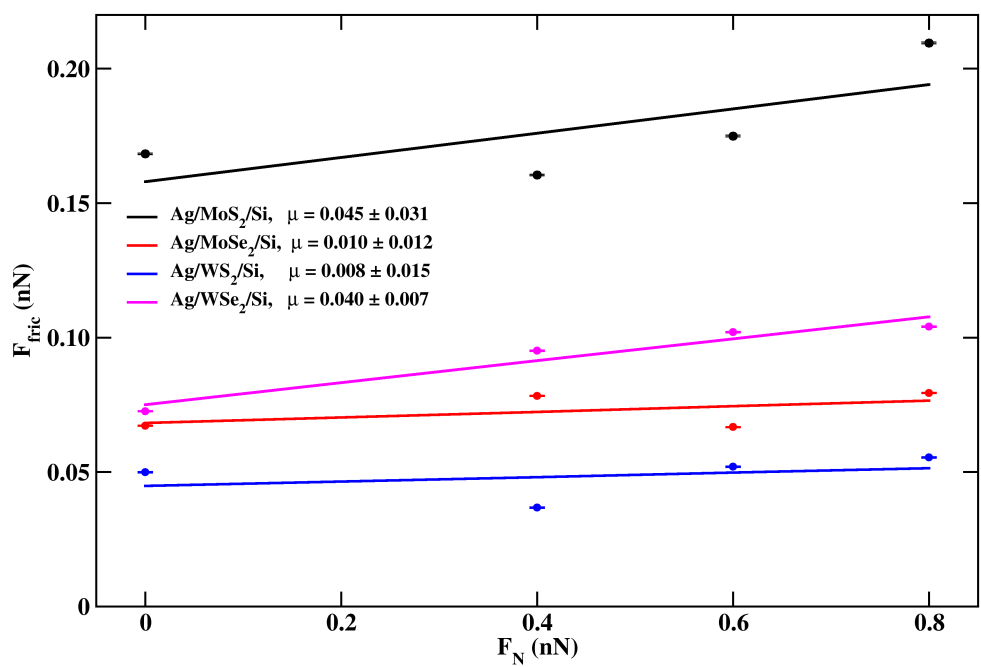

**Figure S7:** $\mu$ extracted from a linear fit function on the average friction force as a function of load with a tip velocity of 3 m/s.

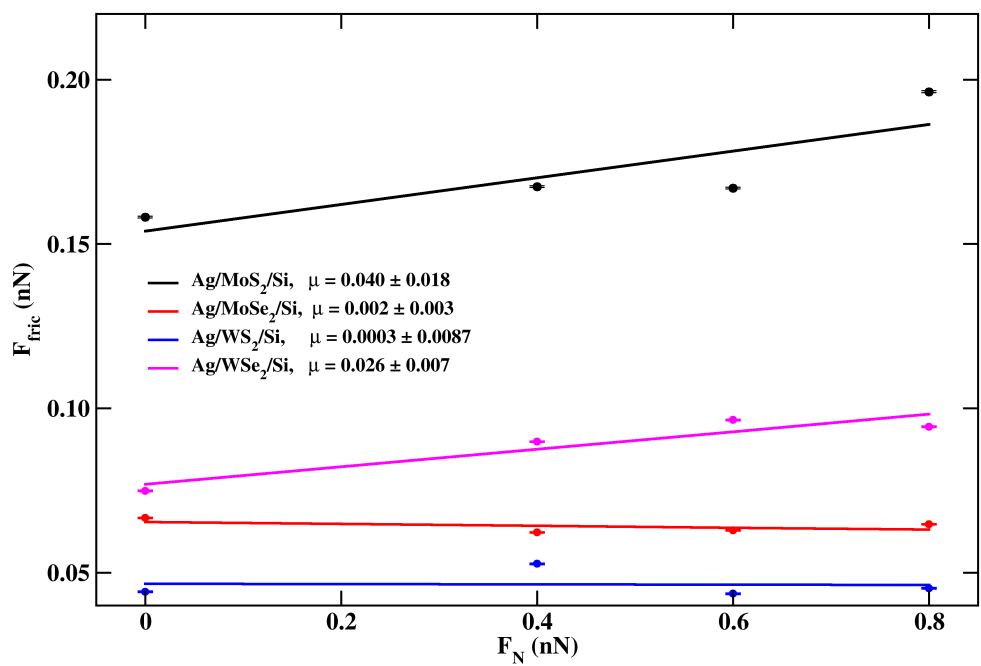

**Figure S8:** $\mu$ extracted from a linear fit function on the average friction force as a function of load with a tip velocity of 4 m/s.

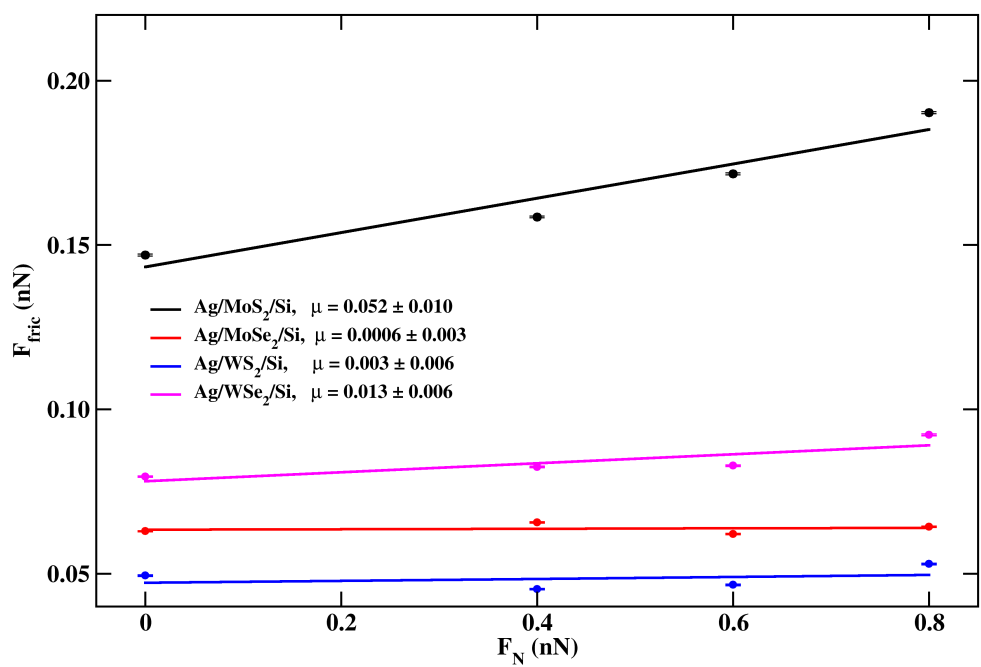

**Figure S9:** $\mu$ extracted from a linear fit function on the average friction force as a function of load with a tip velocity of 5 m/s.

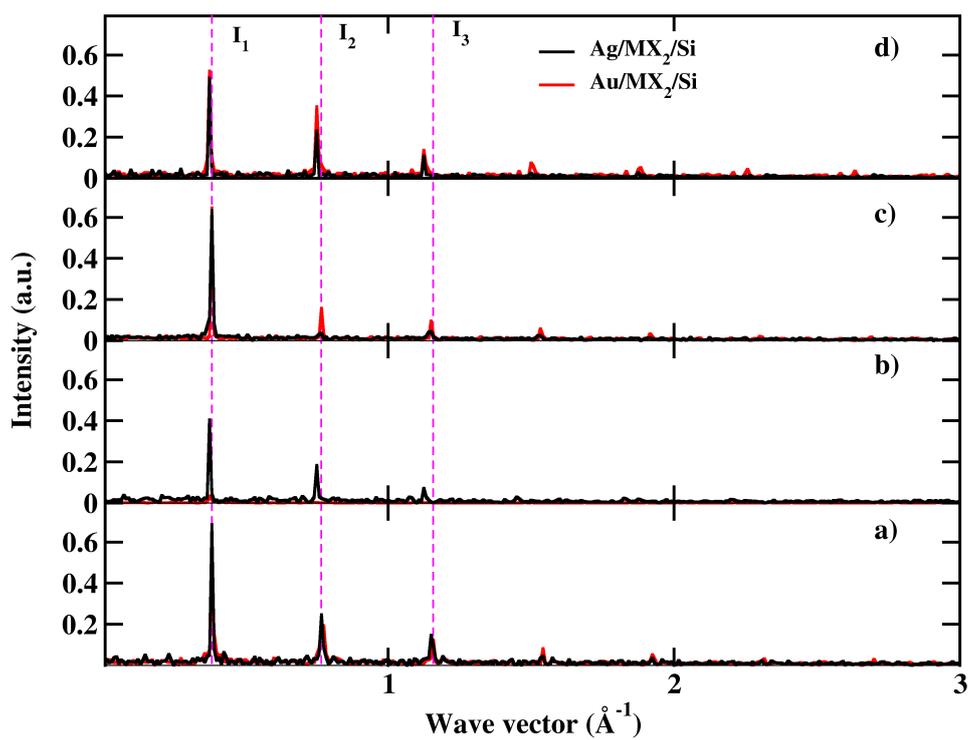

**Figure S10:** Spatial Fourier transform of friction force with monolayer a) $MoS_2$, b) $MoSe_2$, c) $WS_2$ and d) $WSe_2$ at different substrates.



\end{document}